# Exoplanet Mineralogy

**Keith D. Putirka**

*Dept. Earth & Env. Sciences*
*California State University*
*Fresno, California, 93740 U.S.A.*

## INTRODUCTION

Mineral properties hold dominion over planetary behavior as their physical and thermal properties control mantle circulation, plate tectonics, melting behavior, and crustal deformation, as well as global water and $CO_2$ cycles. Estimating exoplanet mineralogy is thus of paramount importance for understanding planetary origins and evolution. Preliminary studies of exoplanets emphasize individual element abundances (e.g., Ca or K contents) or ratios (Mg/Si or Si/Fe), which are a valuable first step. But inter-planet differences in, say, Si or Mg, are only relevant insofar they define a planet's mineralogy.

As will be shown, nothing definitive can yet be said about exoplanet mineralogy, largely because a number of conditions that control mineralogy are unknown, though measurements from the James Webb Space Telescope (JWST) and similar instruments might soon yield the required data for some exoplanets. Differences of opinion are perhaps, then, little more than a contest of speculation; but thoughtful speculation can be profitable, possibly accurate, and in any case essential for progress. The methods noted below will emphasize paths of speculation that minimize underlying assumptions, and several approaches to quantify uncertainties.

To calculate a planet's mineralogy, we will emphasize what geologists refer to as the "major elements" (or "major oxides") i.e., elements that occur at wt. % levels (Si, Ti, Al, Fe, Mn, Mg, Ni, Ca, Na, K, P and O). These elements, along with water, pressure (*P*) and temperature (*T*), control mineral stability. The behavior of minor or trace elements are in turn controlled by mineral assemblage. This chapter will focus on mass balance and some thermodynamic relationships that can be used to estimate mineral assemblages when an exoplanet major element composition is given. Such estimates are usually inferred from star compositions, and some caveats with regard to such approaches will be explored. Mass balance calculations are probably all that we can sensibly do to infer exoplanet mineralogy until vastly better remote-sensing methods are available. But by keeping several caution flags in mind, it should be possible in the meantime to explore possible scenarios regarding how exoplanets form and evolve. Such evolutionary scenarios inevitably reflect back on what we currently comprehend of Earth and its neighboring rocky objects, and reveal the flaws and gaps in such understanding. As we have only one planetary system to use as an analog, exoplanetary studies are tethered to what we know, and still have yet to learn about our own Solar System.

This chapter begins with some basic concepts regarding the structure and mineralogy of rocky planets, how to read and construct ternary diagrams, and why partial melting occurs when plate tectonics is operative. Partial melting is a key concept in that it governs crust and core formation, which in turn control mineralogy. These sections are for astronomers, or geologists new to the study of igneous petrology. From there, computational approaches for estimating planetary mineral assemblages will be introduced. These quantitative methods are simple, consonant with the level of information currently available on exoplanet compositions, and while largely intended for mineralogists, should be accessible to non-specialists as well. Such methods are followed by a study of error when plotting mineral abundances in ternary diagrams, for mineralogists and petrologists who construct such diagrams. The chapter concludes with caveats, and the ways in which exoplanets might surprise us.

## AN OVERVIEW OF PLANETARY STRUCTURE, MINERALOGY & PETROLOGY

### Planetary structure

Rocky objects with radii greater than a few hundred km will undergo partial melting very early in their growth, due to the gravitational energy released upon accretion (Safranov, 1978; Ransford, 1979). In our Solar System, the starting materials that accrete to form the inner planets are thought to be similar to chondritic meteorites (so-named because they contain glass-bearing blebs called "chondrules"), although the match is imprecise (Drake and Righter 2002). The CI chondrite meteorites are of special interest as when volatile elements are ignored, they appear to be very similar to the Sun in composition (Lodders and Fegley 2018) and may approximate the bulk materials that coalesced to form the inner planets (see Jones, this issue). Whatever the parent material, accretionary heating will cause a protoplanet to become "differentiated", into a metallic core and a silicate mantle and crust (Fig. 1). This tripartite planetary structure, of an Fe-Ni alloy metal core, overlain by a silicate mantle and crust (Fig. 1), likely dominates all rocky planets in any planetary system; in our Solar system chondritic meteorites are examples of materials that have escaped such differentiation.

### Mineralogic structure

Table 1 summarizes the minerals that are likely to be the most abundant in these differentiated planetary systems. Mineral assemblages will vary with pressure (*P*) (depth) within a planet. For example, Figure 2 shows the case for Earth, assuming constant composition, where mineral assemblages change drastically across a *P* range of 30 GPa—shifts that are significant for interpreting estimates of exoplanet densities (Dorn et al. 2015; Putirka et al. 2021). It may be challenging to infer crust compositions on exoplanets, but for any planet large enough for the silicate mantle to exceed $P>23$ GPa, such planets are at least likely to have mineralogically distinct "upper" and "lower" mantle layers. Within Earth, the uppermost mantle ($P<15$ GPa) is dominated by olivine and pyroxene, while above 23 GPa, bridgmanite dominates in a region called the "lower mantle" (Figs. 1-2). These two regions are separated by a "transition zone" where olivine is gradually compressed to form the polymorphs wadsleyite and ringwoodite and where Al-rich garnet is compressed to form majorite. Above 100-120 GPa, bridgmanite and ferro-periclase may transform to "post-perovskite" minerals at the base of the mantle, which within Earth might explain the so-called D" layer (Mattern et al. 2005; Wicks and Duffy 2016; Kuwayama et al. 2021; Fig. 1) which has distinct seismic velocities.

Because of the challenges in performing phase equilibrium experiments at very high pressures, we have only a cursory understanding of Earth's lower mantle. And yet this region is crucial for understanding heat loss, mantle circulation and plate tectonics. For example, the core /mantle boundary (CMB) is the likely source of heat that powers mantle plumes—hot upwellings of mantle material that rise to the surface to drive large volume volcanic eruptions. Such plumes are necessarily rooted within the D” layer which might not only be mineralogically distinct, but compositionally distinct as well. The D” layer, and perhaps even larger fractions of the lowermost mantle, are a likely storehouse of subducted crust (e.g., Klein et al. 2017) that might be returned to the upper mantle via mantle plumes (Hofmann and White 1982). Mantle plumes also likely drive plate tectonics (Davies and Richards 1992), provided that the lithosphere (a planet’s rocky outer shell) is sufficiently weak to allow it to break up into tectonic plates (Weller and Lenardic 2018). Absent plate tectonics, mantle plumes might provide the only manifestation of mantle convection, in what is called a “stagnant lid” mode of circulation (Ogawa 1991; Solomotov and Moresi 2000; Stevenson 2002).

**The driving force of planetary differentiation: partial melting**

Planets larger than a few hundred km in radius will have sufficient accretional energy to partially melt. Other sources of heat include the gravitational energy released upon core formation, the latent heat of crystallization as a liquid metallic core crystallizes, and heat produced from radioactive elements. On sufficiently large planets these sources can maintain mantle circulation and partial melting—and thus the possibility of plate tectonics—for billions of years. Plate (and other types of) tectonics will be dealt in greater detail (Putirka, this volume). Here we’ll briefly review partial melting.

***Ternary diagrams*** Most common rocks consist largely of just three minerals, whose abundances are used to define rock type, such as “granite”, which is composed of quartz, plagioclase and alkali feldspar. Most other minerals are “accessory phases” which are used to modify the rock name; a granite that contains biotite and hornblende (and more hornblende than biotite) would be called a “biotite-hornblende granite”. The three dominant minerals can be plotted in a ternary diagram where their abundances are normalized to sum to 1 or 100. And with renormalization, because only two of the abundances can vary independently, ternary diagrams can be plotted in 2-dimensions, such as Figure 3a, which shows the relative abundances of the minerals forsterite (Fo; $Mg_2SiO_4$; a type of olivine), diopside (Di; $CaMgSi_2O_6$; a variety of pyroxene) and anorthite (An; $CaAl_2Si_2O_8$; a variety of plagioclase feldspar). To create a ternary diagram within a 2-dimensional Cartesian coordinate system, the two following equations can be used, which translate the three mineral components, A, B and C, into x-y coordinates:

$$y = C \qquad (1a)$$
$$x = \frac{B}{\cos(30^o)} + C[\tan(30^o] \qquad (1b)$$

Here, *x* is the horizontal Cartesian axis, *y* is the vertical Cartesian axis, *C* is the mineral component that plots at the top of the ternary diagram (An in Fig. 3), *B* is the mineral component that plots at the bottom right apex (Fo in Fig. 3), and the *A* component plots at the origin (0,0). Figure 3a shows a ternary phase diagram that is useful for understanding mantle partial melting, and the process of “igneous differentiation”.

Differentiation occurs because (a) rocks and minerals have different densities and so can be mechanically separated, and (b) because nearly all natural rocks melt incongruently, i.e., when they partially melt, the liquids do not have the same composition as the rock. To illustrate, Figure 3a shows the case of what is called “eutectic” melting, using a ternary eutectic diagram that involves three minerals that are common in the mantles of the inner planets: Fo, Di and An. The eutectic point in this diagram, at 1270°C, and represents the only point in the ternary system where the phases forsterite + diopside + anorthite + liquid coexist in equilibrium. Provided that all three minerals are present, melting will always begin at 1270°C in this eutectic system, regardless of the relative abundances of Fo, Di and An. For example: let's say that a bulk composition falls at the green circle in Fig. 3a, within the Fo + liq field, and so has a composition of approximately 40% An, 22% Di and 38% Fo. Since the rock falls into the Fo + liq field, Fo must be the last phase to be consumed during melting. And since a line drawn from the Fo apex through the bulk composition (green line) intersects the Fo + An cotectic (the dark line that separates the An + liq and the Fo + liq fields), An must be the second to the last phase to melt out of the rock. So upon heating, this rock will begin melting at 1270°C and as heat is added to the system, each of Fo, Di and An will be consumed, in eutectic proportions, and the *T* remains constant at 1270°C until one of the three phases is consumed, in this case Di; with further heating, *T* will now rise, and the liquid will leave the eutectic and migrate up the An + Fo cotectic, consuming both An and Fo, until An is consumed, which will happen at 1300°C. At this point, the only mineral left is Fo and so any further heating will enrich the liquid in Fo, driving the liquid towards the Fo apex (following the green line), which means that the liquid will leave the An + Fo cotectic and head directly towards the bulk composition. Temperature will continue to rise until all of the Fo is consumed, and the last bit of Fo will be consumed and the rock will be completely melted at 1500°C.

Note that in this example, the liquid only matches the bulk composition after melting is completed. The only case where the initial (and all subsequent) liquid(s) will match the bulk composition is the highly unlikely case that the bulk composition matches the eutectic (which in this diagram is at approximately 8% Fo, 43% An and 49% Di). Because rocks consist of at least 8-10 components it is even less likely than it is in Figure 3a that a planetary mantle would have a composition that precisely matches the eutectic (Walker, 1986). However, if the Solar system contains an effectively countless number of rocky objects the odds that any one of them has a eutectic composition is certainly increased, which would have significant consequences for the amount of crust and the rate of heat loss for such planets.

Figure 3b is a closer approximation of partial melting within the inner planets. Melting will begin at the co-saturation point of olivine + clinopyroxene + orthopyroxene ± plagioclase at about 1100-1150°C at 1 atm pressure; the intersection of co-saturation curves here is not a true eutectic (note that the temperature arrows do not all point towards the red dots, which represent co-saturation of Ol, Cpx, Opx and Plag): with continued heating (cooling), there is a reaction relationship as orthopyroxene (olivine) reacts with the liquid to form olivine (orthopyroxene). As *P* increases (i.e., as melting occurs deeper within a given planet), the initial *T* at which melting begins will increase, and the initial liquid compositions that are created move away from the $SiO_2$ apex, becoming enriched in the olivine component—i.e., they contain less $SiO_2$ and more MgO.

Since minerals and liquids are not equal in composition or structure, their physical properties are also unequal (e.g., viscosity, density); in the resulting slurry, there can be a rapid gravitational separation of liquid and solid phases—differentiation. As chondrite-like materials partially melt, Fe is sufficiently abundant relative to O that accretional heating yields a dense metallic liquid that sinks downward through a silicate liquid to form a core. The silicate materials that form the overlying mantle will be rich in pyroxene and olivine, and these in turn can partially melt (Fig. 3b) to yield liquids that are Si-rich and Mg-poor relative to the solid phases, and that are likely buoyant and will rise to form a crust. As can be seen from Figure 3b, the pressure at which mantle melting occurs can have a dramatic influence on the resulting crust composition.

How large must a rocky object be to acquire sufficient energy, either from accretion or radioactive elements, so as to partially melt and become differentiated? Theoretical calculations indicate that by accretional energy alone, any object that reaches a radius (r) >1,200 km should be hot enough to exceed the "peridotite solidus" (the *T* at which peridotite melting begins), which is close to ca. 1,150 °C (Takahashi 1981). But in our own Solar system, Vesta (r = 263 km) and Ceres (r = 476 km) are both differentiated and Vesta appears to have a substantial metallic core. Ceres is also reported to have a distinct core, but its structure is very unclear because the reported densities indicate an ice-rich crust, and density estimates for its core range widely (1950-5150 $kg/m^3$; King et al. 2018), falling short of the density for an Fe-Ni alloy (7874 $kg/m^3$). In any case, when using bulk compositions to model exoplanet silicate mineralogy only the mantle is relevant, except for the smallest of planetary objects. Since for this chapter we are solely concerned with the solid portions of rocky planetary objects, our focus will be on mantle mineralogy.

Fortunately, while Earth is host to >5,000 mineral species, only a small fraction is needed to describe its mantle or crust. The reason is that under the high temperatures that pertain to planetary interiors, reaction rates are rapid and minerals are able to dissolve greater amounts of trace and minor elements. The phase rule also comes into play. The rule is: $f = c - p + 2$, where $c$ is the number of components that are needed to describe $p$ number of phases that exist in an equilibrated system; $f$ describes the "degrees of freedom", or the number of intensive variables (i.e., variables that do not depend upon the size of the system, such as pressure, temperature or elemental concentration) that can be varied independently without changing the total number of phases in a system. The size of $c$ is controlled by the "major elements", since trace elements are usually too low in abundance to dictate what phases can possibly exist. Rearranging then, $p = c - f + 2$, which means that for a given $f$, a small value for $c$ also means a small value for $p$. The end result of all these factors is that most rocks in Earth's mantle and crust are dominated by just a handful of minerals. Table 1 is by no means an exhaustive list, but provides the materials that are likely to be dominant on extrasolar planets, unless, as we note later on, the formation conditions of those planets are very different than those in our inner Solar system.

**Planetary layers**

***Metallic Core*** All planets in our Solar System—even Vesta (a mere 262 km in radius)—contain a metallic core. We infer from the mineralogy of Fe meteorites that these metallic cores consist largely of Fe-Ni alloys. Two minerals are dominant in Fe meteorites, kamacite and taenite, both having the formula (Fe,Ni), with taenite containing more Ni than kamacite. These minerals are probably not relevant to planetary interiors, where in Earth's core for example, the outer core is molten and the Fe and Ni within the inner core are likely contained within a dense, hexagonal closest packed (hcp) mineral structure (Kombayashi et al. 2019). The precise stoichiometry of such inner core phases, though, is not entirely clear (Vocaldo 2010). It has long been noted (Birch 1964) that seismic velocities are too slow within Earth's core to allow it to be composed only of Fe-Ni alloys, and that "light alloying elements", such as S, Si or O likely exist at concentrations of at least 10 wt. % (Wade and Wood 2005). Such light alloying elements might exist at even greater concentrations in the cores of the Moon, Mars or Mercury (Hauck et al. 2013; Helfrich 2017; Terasaki et al. 2019). Because metal cores can represent such a large mass fraction of a planet, the amounts of these light alloying elements, especially Si, can have a significant impact on models of mantle mineralogy.

**Table 1.** *A brief list of abundant minerals in Earth and other rocky objects of the Solar System*

| | Mineral Name | Formula | Where might it dominate? |
|---|---|---|---|
| **Native Elements** | Kamacite, Taenite | (Fe,Ni) | core |
| **Sulfides** | Fe-sulfide (various) | FeS, $Fe_2S$ | crust, inner core |
| **Oxides** | Ferro-Periclase | (Fe,Mg)O | mantle |
| | Spinel | $MgAl_2O_4$ | mantle |
| **Silicates** | Olivine | $(Mg,Fe)_2SiO_4$ | mantle |
| | Wadsleyite | $(Mg,Fe)_2SiO_4$ | Transition zone (mantle) |
| | Ringwoodite | $(Mg,Fe)_2SiO_4$ | Transition zone (mantle) |
| | Serpentine | $(Mg,Fe)_6Si_4O_{10}(OH)_8$ | Crust and mantle |
| | Orthopyroxene | $(Mg,Fe)_2Si_2O_6$ | mantle |
| | Majorite | $(Mg,Fe)_2Si_2O_6$ | Transition zone (mantle) |
| | Bridgmanite | $(Mg,Fe,Al)_2(Si,Al,Fe)_2O_6$ | lower mantle |
| | Clinopyroxene | $Ca(Mg,Fe)Si_2O_6$ | mantle |
| | Ca-perovskite | $CaSiO_3$ | lower mantle |
| | Garnet | $(Mg,Fe,Ca)_3Al_2Si_3O_{12}$ | mantle |
| | Plagioclase Feldspar | $CaAl_2Si_2O_8$-$NaAlSi_3O_8$ | crust |
| | Hornblende | $(Ca,Na)_2(Mg,Fe,Al)_5(Al,Si)_8O_{22}(OH)_2$ | crust |
| | Biotite | $K(Mg,Fe)_3AlSi_3O_{10}(F,OH)_2$ | crust |
| | Alkali Feldspar | $(K,Na)AlSi_3O_8$ | crust |
| | Quartz | $SiO_2$ | crust |
| **Carbonates** | Calcite | $CaCO_3$ | crust |

As will be emphasized later, the first and single most important choice when using bulk compositions to model bulk silicate planet (BSP) mineralogy will be how to handle Fe: how much Fe is partitioned into the core relative to the mantle? What are the total concentrations of light alloying elements in the core? What fraction of those light alloying elements is Si?

***Silicate mantle*** Because only a handful of minerals occur in Earth's mantle and crust, we can use a ternary diagram to classify rocks that occur there (Le Bas and Streckeisen, 1991). For exoplanets, perhaps the only relevant ternary diagram that derives from the study of terrestrial samples would be the ultramafic rock diagram, shown in Figure 4a. Some additional ternary diagrams that are not likely relevant to the inner Solar System, but might be important elsewhere, are described in Putirka and Xu (2021).

The ultramafic rock diagram is used when the sum of the minerals olivine (ol) + clinopyroxene (cpx) + orthopyroxene (opx) is >90% by volume (although mineral proportions are often plotted on a molecular or wt. % basis). Minerals whose abundances are <10% can still be important, and ternary diagrams such as in Figures 3 or 4a, are perhaps best thought of as projections, as illustrated in Figures 4b-c. Figure 4b shows the case for a rock that has a small amount of garnet (gar), and so plots within the three-dimensional quadrilateral ol + cpx + opx + gar, just above the ol + cpx + opx plane. The "projection" of the sample onto the plane ol + cpx + opx is accomplished by renormalizing the sum so that ol + cpx + opx = 1 (or 100). This renormalization has the geometric effect of passing a line that emanates from the garnet apex, through the sample in 3D space, to the ol + cpx + opx plane. When ternary diagrams involve such a projection, it is customary to write (in this case) "+ gar" to indicate the phase from which the samples are being projected, and to show that all the plotted rocks contain the indicated phase. Thus, in the experiments used to produce the 1-atm (solid lines) phase boundaries of Figure 3b, every sample was found to be saturated with plagioclase, which would co-exist with either Ol, Di or Qz, or some combination. The only place where plagioclase would be absent would be at the ternary apices, which would represent case of pure Ol or Di or Qzz. (Note: mineral abbreviations, such as ol, di, qz, are from Whitney and Evans (2010); most abbreviations are lower case; it customary, though hardly required, to label the apices in ternary diagrams using upper case symbols; here, both Ol and ol, will indicate olivine). Most ultramafic rocks that plot into Figure 4a are likely to be saturated with an "aluminous phase", namely either spinel, plagioclase or garnet. It is not customary to note such phases in classification diagrams, but when calculating exoplanet mineralogies, it is probably essential since such projections can involve any number of different minerals.

In Figure 4a, most of Earth's mantle rocks fall into the peridotite field, having >40% ol. There is some discussion that considerable amounts of pyroxenite may exist within Earth's mantle (Sobolev et al. 2005). And there is considerable geochemical evidence that pyroxenite may contribute to lavas erupted above mantle plumes (Prytulak and Elliott, 2007; Putirka et al. 2018). But pyroxenite is not the dominant rock type recovered from Earth's upper mantle (McDonough and Sun 1995) and all estimates "Bulk Silicate Earth" (BSE), as well as estimates of the silicate portions of Mars, Moon, Mercury and Vesta, plot solidly within the peridotite field (Putirka et al. 2021). From here on, we will use Bulk Silicate Planet, or BSP, to refer to bulk planet compositions when their metallic cores are subtracted. For Earth-sized or larger planets the BSP will, in effect, be equivalent to the planet's mantle composition, since except for the smallest of planets, crustal mass fractions are exceedingly small.

Putirka and Rarick (2019) recommend using upper mantle *P-T* conditions as a reference or standard state, to compare planets, something close to 2 GPa and 1350°C. Adopting an upper mantle standard state has many advantages, some practical, some experimental. As to the practical advantages, the *P-T* conditions for rocky exoplanets are not only unknown, are hardly constant, and will vary widely: Unterborn and Panero (2019) show that core-mantle boundary conditions can range to 630 GPa and 5000 K. Most current thermodynamic models, however, barely reach pressures of about 2 GPa and experiments on a limited number of terrestrial compositions extend to only 140 GPa; and because the results of very high *P* experiments are controversial, even the mineralogy of Earth's lower mantle is still a bit of a mystery. On the other hand, any planet that is Moon-sized or larger will attain conditions of ca. 2 GPa and 1350°C at some point in its cooling history. On the experimental side, experiments in the range 1-2 GPa and 1300-1400°C are abundant, have been conducted on a very wide range of compositions, and equilibrate quickly (relative to lower *T*, lower *P* experiments), and are mostly conducted in piston-cylinder apparatus that have low calibration uncertainties (relative to diamond anvil cell experiments). Thermodynamic models, such as MELTS (Ghiorso et al. 1995; Asimow et al. 1998), also readily apply to upper mantle conditions and can be easily updated to handle more exotic compositions as new experiments become available. Only smaller bodies, such as Vesta (whose interior does not exceed 0.4 GPa; Righter and Drake 1996), might require a lower set of *P-T* conditions, perhaps 1 atm and 1000°C, but in a quest for habitable or otherwise Earth-like planets, Vesta-sized objects are of lesser interest.

***Crust: oceanic and continental*** Earth is the only planet to have all of the following: plate tectonics, liquid water at its surface, abundant continental (granitic) crust, and a $N_2$- and $O_2$-rich atmosphere. This combination might not be a coincidence (Campbell and Taylor 1983) and the origin and existence of continental crust is of exceeding importance for understanding planetary evolution. There are also some claims that continent-like crust has been detected outside our Solar System (Zuckerman et al. 2011; Hollands et al. 2018; Hollands et al. 2021); these identifications are often based on the abundance of a single element (e.g., Ca, K, Li) although where more complete compositions are available (Mg, Fe and Si), such a detection is not so clear (Putirka and Xu, 2021).

While Earth's mantle is dominated by olivine and pyroxene, plagioclase is dominant in the overlying crust. Accompanying plagioclase in the lower crust are olivine and pyroxenes (in the very deep crust, garnet, instead of plagioclase, is the Al-rich phase). In the upper crust, plagioclase is joined by quartz, hornblende (a variety of amphibole), alkali feldspar and biotite (a variety of mica). Micas and amphiboles contain significant amounts of water, in the form of $(OH)^-$ and so are "hydrous"; minerals that do not contain $(OH)^-$ as a part of their stoichiometry are referred to as "nominally anhydrous" (as they still can contain trace amounts of $(OH^-)$). During subduction (when a crust-bearing tectonic plate sinks into the mantle), amphiboles, and the mineral serpentine (a mineral that forms when water reacts with olivine), may degrade and recrystallize and so return their water back into the mantle. And when water is added to the mantle, its melting temperature is greatly lowered (Fig. 5), which means that for the same temperature, more melt, and more crust are created. Amphiboles and serpentine are thus an important part of Earth's global water cycle. But as shown by Bell et al. (1995), nominally anhydrous mantle minerals, such as garnet and pyroxene, can also contain enough $(OH)^-$ to hold an ocean's-worth or more of water in Earth's mantle (Bell et al.1995). Guimond et al. (2023) suggest that such nominally anhydrous minerals in the transition zone (Figs. 1-2) may well control the global water cycle.

As noted, a planet's crust is created by partial melting of its mantle. For mantles that consist of peridotite or pyroxenite (e.g., any rocks that fall into the classification of Figure 4), the liquids (and crust) compositions will be basalt (Fig. 6a). Basalts are volcanic rocks that are not fully crystallized, and so like other types of volcanic rocks, they are classified based on their chemical composition, rather than their mineralogy (Fig. 6a). Basalts are an important crust type in that they cover ⅔ of Earth's surface and nearly the entire surfaces of Mercury, Moon, Mars, and Vesta, and most likely Venus as well. Most of Earth's basalt lies beneath the ocean basins, but Earth is unique in having large amounts of "continental crust". The average composition of the continental crust is chemically equivalent to "andesite" (Rudnick and Gao 2014), but perhaps the standout feature of Earth is the large amount of "granitic" rocks ("rhyolite" when it erupts from a volcano; Fig. 6a) that are common within Earth's uppermost crust; these rocks are especially low in MgO and high in $SiO_2$ (and high in $H_2O$, $Na_2O$ and $K_2O$ and other elements that are "incompatible" within mantle minerals; Fig. 6a-b). $SiO_2$, $K_2O$ and water are sufficiently high so as to respectively crystallize large amounts of quartz, hornblende and alkali feldspar.

***Making crust/mantle melting*** Planetary crusts are produced by partial melting of the mantle. Unless melting occurs at very high pressures (e.g., Agee and Walker 1988), such partial melts are quite likely to be less dense than their associated mineral residue, and so will rise towards the surface—where they will remain, unless subduction brings them back into the mantle. The amount of crust that is produced will be proportional to the fraction of melt ($F_m$) that is produced during partial melting. At a given pressure, $F_m$ is dependent upon ambient mantle $T$ relative to the $T$ of solidus (the $T$ at which the mantle begins to melt) and the liquidus (the $T$ at which the mantle is completely melted); the solidus and liquidus curves for Earth, Mercury and Mars are shown in Figure 5 (only for Earth have liquidus and solidus curves been determined at high pressures). Herzberg et al. (2000) and Hirschmann et al. (2000) have calibrated rather precise positions for a solidus of a pyrolite mantle; pyrolite is a primitive mantle composition that is "fertile", meaning it has lots of pyroxene, especially cpx (Fig. 4), and has not had large amounts of melt extracted. Fertile mantle stands in contrast to "depleted mantle" (ol-rich and cpx-poor; Fig. 4), which have had large amounts of melt extracted. Fertile mantle can melt to a greater extent at a given $T$, and so yield large amounts of crust when such melts reach the surface. In general, partial melting will extract $TiO_2$, $Al_2O_2$, CaO, $K_2O$, $Na_2O$ and $H_2O$ out of the mantle, leaving a mantle residue that is depleted in these components (with slightly less $SiO_2$ as well) and enriched in MgO. Much experimental work has been conducted to delimit a peridotite or pyrolite solidus for the uppermost terrestrial mantle. The models of Hirschmann (2000) and Herzberg et al. (2000) are highly precise and accurate but are not calibrated to predict melting conditions in the transition zone or lower mantle. Several somewhat recent studies (Fiquet et al. 2010; Nomura et al. 2014; Kim et al. 2020) have examined partial melting to lower mantle conditions and two (Nomura et al. 2014; Kim et al. 2020) are in remarkable agreement with one another, and also consistent with the lower-$P$ models of the solidus. Due to that agreement, a new curve is fit so as to describe the Herzberg et al. (2000) solidus to about 5 GPa, but then matches the Nomura et a. (2014) and Kim et al. (2020) solidi to 140 GPa. Following Andrault et al. (2011), a modified Simon and Gatzel equation is used, but with an added linear $P$-term, which not only improves the fit at high pressures, and helps to avoid spuriously low $T$ estimates when $P$ approaches 1 atm:

$$T(\,^{o}C)_{solidus} = 1120\left(\frac{P(GPa)}{0.25}+1\right)^{\frac{1}{8.3}} + 6.3[(P(GPa)] \qquad (2a)$$

This solidus is illustrated in Figure 5, and has an error of at least ±30 °C at $P$<5 GPa and an unknown error at $P$>20 GPa, due to the very uncertain uncertainties of diamond anvil experiments. To calculate $F_m$ we also must have an estimate of the liquidus, which was unfortunately not determined by either Nomura et al. (2014) nor Kim et al. (2020). A liquidus curve was determined by Fiquet et al. (2010), but their solidus curve falls to a considerably higher $T$ than the solidi temperatures of either Kim et al. (2020) and Nomura et al. (2014). To calibrate a liquidus curve, we'll assume that while the Fiquet et al. temperatures are too high, that the width of their melting interval is accurate, which allows the liquidus to be constructed when the Fiquet melting interval is combined with Equation 2a. It is also assumed that the 1 atm liquidus $T$ is 1700°C, as in Zhang and Herzberg (1994). The resulting curve is:

$$T(\,^{o}C)_{liquidus} = 1700\left(\frac{P(GPa)}{2.8} + 1\right)^{\frac{1}{4.2}} \qquad (2b)$$

Melting temperatures are greatly lowered in the presence of water; the following equation reproduces the solidus of Till et al. (2007), but fixes the solidus $T$ at $P$>3 GPa to 815°C (Grove and Till 2019):

$$T(\,^{o}C)_{wet\ solidus} = 1120 - 305[\mathrm{erf}\,(0.5P(GPa))] \qquad (2c)$$

Equation 2c is necessarily incomplete in that the temperature of the wet solidus also increases with increasing MgO content (Grove and till 2019; Wang et al. 2020); new experiments at very high pressures are needed to yield improved versions of Equations 2a-c, with compositional parameters being added. Variations of stellar compositions (Putirka and Rarick 2019) appear to indicate some considerable variety in exoplanet bulk compositions. Dry solidus curves have been determined for Mars (Ding et al., 2020), Mercury (Namur et al. 2014), and Allende (a carbonaceous chondrite meteorite; Agee et al. 1995; Asahara et al. 2004). But none of the experimental studies on the dry mantle solidus extend to transition zone depths (15-23 GPa), let alone to the lower mantle (>23 GPa). Andrault et al. (2011) determine the solidus and liquidus of a chondrite meteorite (Fig. 5), but their solidus is oddly higher than that of both bulk silicate Earth (BSE), and of Allende (Fig. 5). So there are no obvious compositional corrections. The lower solidus temperatures obtained by Asahara et al. (2004) and Agee et al. (1995) for Allende are at least intuitively correct in that carbonaceous chondrites are more fertile, having a much lower Mg# (MgO/(MgO+FeO)) compared to BSE. But despite the challenges of DAC experiments, perhaps it would be a mistake to discount the Andrault et al. (2011) experiments entirely, since their chondrite composition is lower in CaO and $Al_2O_3$ compared to BSE. In any case, reproducible experiments are a critical need.

***How much crust can a planet have?*** Besides an $O_2$-rich atmosphere, the identification of continental crust outside our Solar system might provide the single most compelling case for an Earth-like planet. But the potential of such an identification faces a key challenge: the mass fraction of Earth's entire crust is ca. 0.4 wt. % of its total mass, and the continental crust comprises yet a smaller fraction. Some planets have much thicker crusts relative to their planet's radius, but as a mass fraction, crusts are still a very small part of any planet (Table 2). And where the crusts are thickest (Moon, Vesta) they are basaltic, not granitic. In sum, abundant granitic crusts might only evolve on large, Earth-sized planets with plate tectonics, where water is stable on the surface. But in such cases crust mass fractions may be too small to detect. And as Venus plainly shows, Earth-sized planets will not necessarily follow Earth-like evolutionary paths.

For the sake of comparison, Table 2 shows core, mantle and crust mass fractions for the inner planets, Moon and Vesta. They are calculated using reported estimates of layer thicknesses and bulk densities of crusts, mantles and cores (Table 1, notes). These inputs yield bulk densities, (layers weighted by volume fractions) that approach very closely with those obtained from observed volumes and gravity (combining $F = GmM/r^2$ and $F = ma$, where $r$ is the planet radius, $M$ is the mass of the planet, $a$ is the acceleration due to gravity at a planet's surface and $m$ is an arbitrary mass at the surface that cancels with substitution). But this consistency check is no guarantee of accuracy as the densities and thicknesses of the various planetary components are not well known outside of Earth or Moon, where we have seismometers in place to measure seismic velocities (which can be translated readily to density). Despite undoubted errors, there is a clear trend in our Solar System to proportionally greater crust fractions as planet size decreases. A caveat is that the crust of Mercury might have been ablated by late meteorite bombardment Helffrich et al. (2019), and perhaps Vesta has also been significantly ablated as the so-called HED meteorites appear to originate from Vesta (Mittlefehldt, 2015). Earth's crust is a mere 0.4 wt. % of its total mass with the balance being a metallic, Fe-Ni alloy core that comprises 33% of the planet's mass, and remainder being Earth's mantle, which lies between the core and crust (i.e., between 35 km and 2,890 km). Earth's mantle is made of silicate minerals that thus comprise about 67% of Earth's mass (the sum of crust + oceans + atmosphere is insignificant in comparison). At 6.4 x $10^{23}$ kg, Mars is just 10% of Earth's mass (5.97 x $10^{24}$ kg) and has a crust mass fraction of 3%. Moon has 1% Earth's mass and its crust mass fraction is 9%. At 2.6 x $10^{20}$ kg, Vesta (radius = 262 km) is less than $10^{-4}$ Earth masses, yet it is still large enough to be differentiated into a core, mantle and crust; its crust fraction is very uncertain, but appears to be at least 20-30% of its total mass (as calculated from densities of Raymond et al. 2013). Clenet et al. (2014) argue for an even thicker crust (80 km, which implies a 40-50% mass fraction), but such thick crust seems inconsistent with inferred mantle and crust compositions (if Vesta's mantle is peridotite and its crust is basalt, then total melt fractions to produce the crust would be <20%; at >30% melting, the crust would be a very high Mg volcanic rock known as komatiite, rather than basalt). In any case, as we use bulk compositions to examine/model planetary mineralogy, it matters immensely whether the planets are small or large and whether to what extent Fe has been segregated into a metallic core.

**Table 2.** *Some planetary properties of relevance to exoplanet studies*

| **Rocky Objectss[1]** | **Earth** | **Venus** | **Mercury** | **Mars** | **Moon** | **Vesta** | **Phobos** |
|---|---|---|---|---|---|---|---|
| Surface g (m/s$^2$) | 9.806 | 8.872 | 3.7 | 3.728 | 1.625 | 0.25 | 0.0057 |
| Mass (kg) | 5.964 x $10^{24}$ | 4.868 x $10^{24}$ | 3.30 x $10^{23}$ | 6.42 x $10^{23}$ | 7.346 x $10^{22}$ | 2.581 x $10^{20}$ | 1.05 x $10^{16}$ |
| Mass rel. To Earth | 1 | 0.816 | 0.06 | 0.107 | 1.23 x $10^{-2}$ | 4.3 x $10^{-5}$ | 1.8 x $10^{-9}$ |
| Distance (AU) | 1 | 0.7233 | 3871 | 1.5236 | 1 | 2.36 | 1.5326 |
| Radius (km) | 6371 | 6051.4 | 2437.6 | 3389.9 | 1737.1 | 262.5 | 11.1 |
| Core mass % | 32.5 | 42 | 72 | 26 | 1.9 | 16 | Undifferentiated |
| Mantle mass % | 67 | 57 | 26 | 71 | 92 | 52 | |
| Crust mass % | 0.5 | 0.8 | 1.9 | 3.0 | 5.4 | 32 | |

1. Data for rocky bodies are from Lodders and Fegley (1998) and Szurgot (2015), with crust and mantle mass fractions calculated using: Mercury (Hauck et al. 2013; Sori 2018; Knibbe et al. 2021); Venus (Aitta 2012; O'Neil 2021); Earth (Lodders and Fegley 1998; Peterson and DePaolo, 2007); Moon (Viswantahan et al. 2019); Mars (Knapmeyer-Endrun et al. 2021); Vesta (Raymond et al. 2013); 2. Radius calculated for equivalent volume as a sphere.

## GEOCHEMISTRY OF EARTH & THE OTHER INNER PLANETS

If we set aside volatile elements (H, He, C, O, S), which are depleted in the inner planets, the Sun and inner planets are made of what Geologists refer to as the "major elements" (or more commonly expressed as the "major oxides", since O is the dominant anion). These elements are, in order of decreasing abundance by weight in our Solar system: Fe, Si, Mg, Ni, Ca, Al, Na, Cr, Mn, P, K, and Ti. These are also the dominant elements in nearby stars. Just three elements, Mg + Fe + Si, account for >85% of the total in 99% of stars in the Hypatia catalog, and adding Ni, Ca and Al, we account for 96% of all cations by weight in >99% of Hypatia stars. Table 4 shows the bulk compositions of the Sun, the inner planets, Moon and Vesta, and a CI chondrite, when S and O are included, as well as when these compositions are renormalized to a S-free, O-free basis. Bulk inner planet compositions are obtained via mass balance:

$$C_i^{BP} = X^{BSP}C_i^{BSP} + X^{MC}C_i^{MC} \quad (3)$$

where the superscript BP is the "bulk planet", and $X$ is the mass fraction of either a metallic core (MC) or the bulk silicate planet (BSP). Core mass fractions are from Szurgot (2015) and to these are added the median of published estimates of the silicate portions of the planets (see note to Table 4).

In astronomy, it's typical to list elements in order of atomic number. In geology, we are less rigorous, listing the rock-forming elements roughly (often, very roughly), in order of decreasing field strength (charge divided by effective cation radius), and almost always as oxides, tacking on minor and trace elements at the end. As this chapter is designed in part to enroll geologists into the study of exoplanets, we'll follow the geological tradition, listing the oxides as: $SiO_2$, $TiO_2$, $Al_2O_3$, $Cr_2O_3$, NiO, FeO, MgO, MnO, CaO, $Na_2O$, $K_2O$, $P_2O_5$, and the elements in a similar order.

Later sections will introduce some approaches to mass balance to transform chemical compositions into mineral abundances, the latter of which are controlled by the major oxides. As noted, Earth's rocky mantle is dominated by the rock type peridotite (Fig. 4); the proportion of the various minerals in this rock type are controlled by $SiO_2$, $Al_2O_3$, FeO, MgO and CaO. All other elements occur in too low a concentration (see Putirka and Rarick 2019) to greatly affect exoplanet mantle mineralogy, either because their abundances in are low (e.g., Ni) or because the elements are incompatible within olivine, pyroxenes and garnet (e.g., Ti, P) and are concentrated in the crust. Although peridotites are defined by their mineralogy, they are also distinct chemically, having high MgO (ca. 36±3 wt. %) and low $SiO_2$ (42±2 wt. %) (Fig. 6b). Basalts, which are produced by partial melting of peridotite, contain higher $SiO_2$ (45-52 wt. %, by definition; Le Bas et al. 1986) and much lower MgO (usually < 8 wt. %) (Fig. 6a). Basalt also contains higher $Al_2O_3$, CaO and $Na_2O$ than peridotite—sufficient for basaltic liquids to precipitate plagioclase feldspar, in addition to olivine and pyroxenes.

Earth's continental crust (Fig. 6b) has higher $SiO_2$ (on average 60.6 wt. %; Rudnick and Gao 2014) and much lower MgO (4.7 wt. %) placing it in the category of "andesite" (52-63 wt. % by definition; Le Bas et al. 1986) (Fig. 6a). So-called "granitic crust" has even higher $SiO_2$ (often >70 wt. %) and very low MgO (< 1 wt. %) and is chemically equivalent to "dacite" and "rhyolite" in Figure 6a. These granitic rock types appear to only be abundant on Earth. However, studies of Mercury (Vander Kaaden and McCubbin 2016; Vander Kaaden et al. 2017) and an ancient meteorite (Barrat et al. 2021) provide some evidence of andesite crust (albeit with much higher MgO) forming early and elsewhere.

**Table 3.** *Bulk compositions of the planets, a CI chondrite and the Sun (wt. %)*[1]

| | Solar | Mercury | Venus | Earth | Moon | Mars | Vesta | Orgueil |
|---|---|---|---|---|---|---|---|---|
| Si | 7.81 | 7.63 | 16.27 | 15.32 | 20.88 | 16.25 | 19.27 | 15.68 |
| Ti | 0.03 | 0.04 | 0.08 | 0.08 | 0.12 | 0.06 | 0.07 | 0.08 |
| Al | 0.52 | 0.50 | 1.34 | 1.57 | 2.06 | 1.20 | 1.31 | 1.24 |
| Cr | 0.20 | 0.01 | 0.37 | 0.18 | 0.25 | 0.41 | 0.43 | 0.35 |
| Fe | 12.63 | 52.84 | 32.67 | 32.78 | 11.27 | 27.50 | 18.06 | 27.00 |
| Mn | 0.13 | 0.00 | 0.00 | 0.07 | 0.12 | 0.25 | 0.35 | 0.27 |
| Mg | 6.17 | 6.43 | 15.25 | 15.41 | 20.09 | 14.39 | 16.64 | 14.00 |
| Ni | 0.73 | 2.92 | 1.70 | 1.76 | 0.14 | 1.89 | 1.51 | 1.67 |
| Ca | 0.67 | 0.46 | 1.57 | 1.67 | 2.22 | 1.34 | 1.45 | 1.33 |
| Na | 0.32 | 0.28 | 0.14 | 0.18 | 0.04 | 0.33 | 0.04 | 0.74 |
| K | 0.04 | 0.03 | 0.01 | 0.02 | 0.00 | 0.03 | 0.01 | 0.16 |
| P | 0.06 | 0.00 | 0.01 | 0.01 | 0.00 | 0.05 | 0.10 | 0.00 |
| S | 3.31 | 15.17 | 0.00 | 0.55 | 0.11 | 3.06 | 0.00 | |
| O | 67.37 | 13.69 | 30.60 | 30.40 | 42.70 | 33.23 | 40.74 | 37.49 |
| Total | 100 | 100 | 100 | 100 | 100 | 100 | 100 | 100 |

*Bulk Planets (wt. %) on a Sulfur- and Oxygen-free basis*

| | Solar | Mercury | Venus | Earth | Moon | Mars | Vesta | Orgueil |
|---|---|---|---|---|---|---|---|---|
| Si | 26.66 | 10.72 | 23.44 | 22.19 | 36.52 | 25.51 | 32.53 | 25.09 |
| Ti | 0.10 | 0.05 | 0.11 | 0.12 | 0.20 | 0.10 | 0.12 | 0.12 |
| Al | 1.77 | 0.70 | 1.93 | 2.28 | 3.60 | 1.88 | 2.21 | 1.99 |
| Cr | 0.70 | 0.01 | 0.53 | 0.26 | 0.44 | 0.64 | 0.73 | 0.55 |
| Fe | 43.09 | 74.27 | 47.08 | 47.47 | 19.71 | 43.16 | 30.48 | 43.19 |
| Mn | 0.44 | 0.00 | 0.00 | 0.10 | 0.22 | 0.40 | 0.59 | 0.43 |
| Mg | 21.04 | 9.04 | 21.98 | 22.32 | 35.13 | 22.59 | 28.08 | 22.40 |
| Ni | 2.49 | 4.11 | 2.44 | 2.55 | 0.24 | 2.97 | 2.55 | 2.67 |
| Ca | 2.29 | 0.65 | 2.26 | 2.42 | 3.88 | 2.11 | 2.45 | 2.13 |
| Na | 1.09 | 0.40 | 0.20 | 0.26 | 0.06 | 0.51 | 0.07 | 1.18 |
| K | 0.13 | 0.04 | 0.01 | 0.02 | 0.01 | 0.04 | 0.01 | 0.26 |
| P | 0.19 | 0.00 | 0.01 | 0.01 | 0.00 | 0.09 | 0.17 | 0.00 |
| Total | 100 | 100 | 100 | 100 | 100 | 100 | 100 | 100 |

1. All compositions are on a H- and He-free basis. Solar bulk composition is from Magg et al. (2022); planet bulk compositions are from McDonough and Yoshizaki (2022), renormalized so as to add minor elements from the the sources shown in Table 4. Bulk compositions of Mercury and especially Venus are highly uncertain.

## Converting elements to oxides, and oxides to elements

Because O is far and away the dominant anion in the inner planets (Table 4), and because in the mantle, nearly all cations are bound to anions, it is typical that BSPs for the inner planets are reported as wt. % oxides, as are the rock compositions on which most inner planet BSP estimates are based. Table 4 shows BSP compositions for the inner planets and Vesta, taken as median values from a range of published estimates. Estimates of the Sun, in contrast, are most often reported as atomic proportions of individual elements, while star compositions are reported in the same way, but often on a log scale relative to the Sun (see Hinkel, this issue). Given that star compositions provide a first-order (and sometimes our only) estimate of exoplanet compositions (e.g., Hinkel and Unterborn 2018), it is clearly critical to convert back-and-forth between oxides and elemental weight %; for convenience, Table 5 provides multiplicative conversion factors for such conversions.

**Table 4.** *Bulk Silicate Planet Compositions of the Inner Planets and Vesta*

| | Mercury[1] | Venus[2] | Earth[3] | Moon[4] | Mars[5] | Vesta[6] |
|---|---|---|---|---|---|---|
| **$SiO_2$** | 50.99 | 44.93 | 44.86 | 45.53 | 44.41 | 41.24 |
| **$TiO_2$** | 0.19 | 0.20 | 0.20 | 0.20 | 0.13 | 0.12 |
| **$Al_2O_3$** | 2.95 | 3.72 | 4.41 | 4.06 | 2.90 | 2.48 |
| **$Cr_2O_3$** | 0.04 | 0.79 | 0.38 | 0.44 | 0.76 | 0.63 |
| **FeOt** | 0.00 | 9.62 | 8.03 | 12.57 | 17.69 | 23.24 |
| **MnO** | 0.00 | 0.00 | 0.13 | 0.15 | 0.42 | 0.46 |
| **MgO** | 33.32 | 37.20 | 37.87 | 33.80 | 30.48 | 27.59 |
| **NiO** | 0.00 | 0.00 | 0.25 | 0.00 | 0.05 | 1.92 |
| **CaO** | 2.02 | 3.23 | 3.46 | 3.19 | 2.40 | 2.03 |
| **$Na_2O$** | 1.20 | 0.27 | 0.36 | 0.05 | 0.56 | 0.06 |
| **$K_2O$** | 0.12 | 0.02 | 0.03 | 0.01 | 0.04 | 0.01 |
| **$P_2O_5$** | 0.00 | 0.02 | 0.02 | 0.00 | 0.16 | 0.23 |
| **S** | 9.16 | 0.00 | 0.00 | 0.00 | 0.00 | 0.00 |
| **Total** | 100.00 | 100.00 | 100.00 | 100.00 | 100.00 | 100.00 |

1. Nittler et al. (2018); 2. Shah et al. (2022), Lodders and Fegley (1998); 3.McDonough and Sun (1995), Workman and Hart (2005), Lyubetska and Korenaga (2007), Salters and Stracke (2003), Khan et al. (2003), Palme and O'Neil (2003); 4. Longhi (2006), Jones and Delano (1989), Kuskov et al. (2019), Khan et al. (2013), Togashi et al. (2017); 5. Yoshizaki and McDonough (2020), Khan and Connoloy (2008), Wanke and Dreibus (1994), Sanloup et al. (1999), Taylor (2013). 6. Steenstra et al. (2016).

**Table 5.** *Conversion factors for wt. % Element-to-Oxide and Oxide-to-Element*

| | **$SiO_2$** | **$TiO_2$** | **$Al_2O_3$** | **$Cr_2O_3$** | **FeOt** | **MnO** | **MgO** | **NiO** | **CaO** | **$Na_2O$** | **$K_2O$** | **$P_2O_5$** |
|---|---|---|---|---|---|---|---|---|---|---|---|---|
| (a) Oxide weights (O) | 60.08 | 79.88 | 101.96 | 152.00 | 71.85 | 70.94 | 40.30 | 74.69 | 56.0774 | 61.98 | 94.20 | 141.94 |
| (b) Cation atomic wt. (C) | 28.09 | 47.87 | 26.98 | 52.00 | 55.85 | 54.94 | 24.31 | 58.69 | 40.078 | 22.99 | 39.10 | 30.97 |
| (c) Conv. Elem to oxide (O/nC) | 2.14 | 1.67 | 1.89 | 1.46 | 1.29 | 1.29 | 1.66 | 1.27 | 1.40 | 1.35 | 1.20 | 2.29 |
| (d) Conv. Ox to elem (nC/W) | 0.47 | 0.60 | 0.53 | 0.68 | 0.78 | 0.77 | 0.60 | 0.79 | 0.71 | 0.74 | 0.83 | 0.44 |

There has been speculation that some rocky material that once orbited white dwarf stars might have contained large fractions of granitic-like crust. Table 6a provides the unique compositions that comprise Earth's crust to aid such discussions. Rudnick and Gao (2014) provide the most commonly accepted values for Earth's bulk continental crust (CC in Table 6a) while Gale et al. (2013) provide the same for the Mid-Ocean Ridge Basalt (MORB) crust that covers the floor of the ocean basins. As can be seen from Table 6 and Figure 6, the continental crust is distinctly Si-rich compared to oceanic crust and would be classified as andesite, rather than basalt. But the composition of Earth's continental crust ranges much more widely than MORB (which varies quite little) and contains three compositionally distinct layers, an upper, middle and lower crust, each also characterized by Rudnick and Gao (2014) and whose upper crust composition (UC) is shown in Table 6a. For many non-geologists, Earth's crust is perhaps commonly associated with compositions similar to the $SiO_2$-rich upper crust, or the granitic rocks that dominate many high mountain ranges, such as the Sierra Nevada, which are even more enriched in $SiO_2$ (Table 6a, columns b, c). As published compositions are often presented as oxides (Table 6a, columns a-e), one can multiply these by row (d) in Table 5 to obtain elemental concentrations (Table 6a, f-j). The conversion factors of Table 5 account for the weight % of the cations of the indicated compositions; for systems where O is the only or overwhelmingly dominant anion (and in these systems it is) the O weight % can be obtained as O (wt. %) = 100 - sum of cations (wt. %). Be aware that these are minimum estimates of O since Fe is known to also occur in the $Fe^{3+}$ state, occurring as $Fe_2O_3$, as well as FeO. The calculation can be made more precise by substituting estimates of total Fe as FeO (FeOt) with estimates of FeO and $Fe_2O_3$, adjusting the contents of Tables 5 and 6a accordingly.

Mantle rocks are Si-poor and highly Mg-enriched compared to continental crust, as can be ascertained from Figure 6, and Table 6b (column a), which provides the most-cited estimate of Earth's mantle (effectively, bulk silicate Earth or BSE, also referred to as "pyrolite") from McDonough and Sun (1995). The pyrolite, or BSE composition in Table 6b column (a), is re-normalized to 100% in column (b), and then multiplied by row (d) of Table 5 to obtain elemental weight % of cations (Table 6b, column c).

**Table 6a.** *Types of crust on Earth (CC = bulk continental crust; UCC = upper continental crust; MORB = Mid-ocean ridge basalt = oceanic crust; Plume = hot spot at Samoa)*

| | (a) CC | (b) UCC | (c) Granite | (d) MORB | (e) Plume | | (f) CC | (g) UCC | (h) Granite | (i) MORB | (j) Plume |
|---|---|---|---|---|---|---|---|---|---|---|---|
| $SiO_2$ | 60.60 | 66.60 | 72.04 | 50.47 | 45.30 | Si | 28.33 | 31.13 | 33.67 | 23.59 | 21.17 |
| $TiO_2$ | 0.70 | 0.64 | 0.42 | 1.68 | 2.50 | Ti | 0.42 | 0.38 | 0.25 | 1.01 | 1.50 |
| $Al_2O_3$ | 15.90 | 15.40 | 14.32 | 14.70 | 9.20 | Al | 8.42 | 8.15 | 7.58 | 7.78 | 4.87 |
| FeOt | 6.70 | 5.04 | 2.88 | 10.43 | 12.00 | Fe | 5.21 | 3.92 | 2.24 | 8.11 | 9.33 |
| MnO | 0.10 | 0.10 | 0.05 | 0.18 | 0.20 | Mn | 0.08 | 0.08 | 0.04 | 0.14 | 0.15 |
| MgO | 4.70 | 2.48 | 0.66 | 7.58 | 20.50 | Mg | 2.83 | 1.50 | 0.40 | 4.57 | 12.36 |
| CaO | 6.40 | 3.59 | 1.88 | 11.39 | 8.00 | Ca | 4.57 | 2.57 | 1.34 | 8.14 | 5.72 |
| $Na_2O$ | 3.10 | 3.27 | 3.56 | 2.79 | 1.80 | Na | 2.30 | 2.43 | 2.64 | 2.07 | 1.34 |
| $K_2O$ | 1.80 | 2.80 | 4.04 | 0.16 | 0.70 | K | 1.49 | 2.32 | 3.35 | 0.13 | 0.58 |
| $P_2O_5$ | 0.10 | 0.15 | 0.18 | 0.18 | | P | 0.04 | 0.07 | 0.08 | 0.08 | 0.00 |
| | | | | | | O | 46.31 | 47.46 | 48.40 | 44.38 | 42.98 |

| | | | | | Total | | | | | |
|---|---|---|---|---|---|---|---|---|---|---|
| | 100.10 | 100.07 | 100.03 | 99.57 | 100.20 | **Total** | 100.00 | 100.00 | 100.00 | 100.00 | 100.00 |

**Table 6b.** *Convert BSE (Pyrolite from McDonough and Sun 1995) from oxide to elemental weight %*

| | **(a) Earth's Mantle (McDonough & Sun 1995)** | **(b) Earth's Mantle (McDonough & Sun 1995) renormalized** | | **(c) Earth's Mantle (McDonoug h & Sun 1995)** | **(d) Earth's Core[1]** | **(e) Earth bulk comp[1]** | **(f) Earth bulk comp[1]** |
|---|---|---|---|---|---|---|---|
| **$SiO_2$** | 45 | 44.90 | **Si** | 20.99 | 3.6 | 15.34 | 22.16 |
| **$TiO_2$** | 0.201 | 0.20 | **Ti** | 0.12 | 0 | 0.08 | 0.12 |
| **$Al_2O_3$** | 4.45 | 4.44 | **Al** | 2.35 | 0 | 1.59 | 2.29 |
| **$Cr_2O_3$** | 0.384 | 0.38 | **Cr** | 0.26 | 0 | 0.18 | 0.26 |
| **FeOt** | 8.05 | 8.03 | **Fe** | 6.24 | 87.9 | 32.78 | 47.36 |
| **MnO** | 0.135 | 0.13 | **Mn** | 0.10 | 0 | 0.07 | 0.10 |
| **MgO** | 37.8 | 37.71 | **Mg** | 22.74 | 0 | 15.35 | 22.18 |
| **NiO** | 0.25 | 0.25 | **Ni** | 0.20 | 5.5 | 1.92 | 2.77 |
| **CaO** | 3.55 | 3.54 | **Ca** | 2.53 | 0 | 1.71 | 2.47 |
| **$Na_2O$** | 0.36 | 0.36 | **Na** | 0.27 | 0 | 0.18 | 0.26 |
| **$K_2O$** | 0.029 | 0.03 | **K** | 0.02 | 0 | 0.02 | 0.02 |
| **$P_2O_5$** | 0.021 | 0.02 | **P** | 0.01 | 0 | 0.01 | 0.01 |
| | | | **O** | 44.17 | 3 | 30.79 | |
| | 100.23 | 100 | | 100 | 100 | 100 | |

1. For a core composition we use the composition reported by Tronnes et al. (2019) but since they do not report S in Earth's core, we adobe the value of 1.7 wt. % S in the core from Wood et al. (2014), reducing Ni to 5 wt. % and Si to 1.8 wt. %. We assume a core mass fraction ($M_c$) of 0.325. And as in Table 6a, O (wt. %) is obtained by the difference between 100 and the sum of the cation wt. %, in this case, 100 - 55.83 = 44.17 wt. % O. To obtain the composition of bulk Earth we add to the BSP a core composition modified from Tronnes et al. (2019) in that we allow for 1.7 wt. % S in the core (Wood et al. 2014), as the Tronnes et al. (2019) core is curiously S-free. Assuming that mass fraction of Earth's core, $M_c$, is 0.325 then the composition of Bulk Earth, or BE is: $BE = M_c(C_{core}) + (1 - M_c)(C_{BSE}) = 0.325(C_{core}) + 0.675(C_{BSE})$, the results of which are provided in column (e) of Table 6b. Column (f) of Table 6b gives Earth's bulk composition renormalized to an O-free basis. Two notes: First, if an O-free composition is the end goal (and it is often useful to compare cations on a O-free basis) it may be tempting to renormalize BSE on an O-free basis before conducting a weighted addition of the core and mantle compositions. That procedure is fine provided that one adjusts the mass proportion of the mantle downwards to account for the loss of mass by removing O, which occurs upon such a renormalization; the result is not trivial as O is the single most abundant element by mass in Earth's mantle (Table 6, column C). Second, it is not a simple matter to use row (c) of Table 5 and either columns (e) or (f) of Table 6b to express bulk Earth as a set of oxides. For the case of Table 6a, multiplication of columns f-j by row (c) of Table 5 simply leads one back to the values in Table 6a columns (a-e). But if one multiplies the cations of column (e) in Table 6b by row (c) of Table 5, the sum will be >100%; the reason for this is because the calculation assumes that all of the elements in (e) (or f) are fully oxidized to the extent indicated by their formulas (i.e., Fe as FeO), and of course, in bulk Earth, most of the Fe occurs as $Fe^0$. Finally, and as noted for the contents of Table 6a, the estimates of O wt % in Table 6b are also minimum values since within Earth's mantle at least some Fe exists as $Fe_2O_3$ as well as FeO.

## WORKING WITH STELLAR COMPOSITIONS

Two methods of estimating mantle mineralogy will be presented. The first, using matrices, has the advantage of being very easy to code into a spreadsheet or other software and very simple to edit, so as to rapidly test a range of mineralogical models. The second method is somewhat more complex, but has the advantage of yielding only positive values for mineral abundance estimates, even for rather exotic compositions. Hinkel et al. (2022) provide a thorough review of computations and notational conventions. Some notation common in the astronomical literature will be briefly reviewed, but the reader is highly recommended to read Hinkel et al. (2022) for further details as the focus here will be on converting star compositions to wt. % elements and oxides, so as to compare stars to rocks and minerals.

Star compositions are often reported as atomic abundances of a given element (*A(El)*) on a log scale, referenced to H (or sometimes to He for some white dwarfs): *A(El)* = log(*Z/Y*) = log*[n(Z)/n(Y)]* where *n(Z)* is the number of atoms of the element of interest, Z, and in the denominator, *n(Y)* is the number of atoms of either H or He, depending upon which dominates a stellar atmosphere. Star compositions are also often reported as numbers of atoms relative to the Sun, on a log basis. Astronomers use "dex" to denote the $\log_{10}$ scale; so the amount of Fe in a star might be reported as Fe/H, where the dex value of -3.2. This means that the Fe/H ratio is $10^{-3.2} = 6.31 \times 10^{-4}$.

1. *For the case of H- or He-normalized log atomic abundances that are NOT referenced to the Solar composition*: the conversion to elemental wt. % is straightforward: since *Y* is the same for all reported elements for any given star, cation fractions can be obtained by renormalization irrespective of *Y*, so that $10^{[Mg/Y]} + 10^{[Si/Y]} + 10^{[Ca/Y]} + 10^{[Fe/Y]} = 100\% = X_{Mg} + X_{Si} + X_{Ca} + X_{Fe}$, where $X_i$ are the atomic % values of element i. With these atomic proportions one can multiply $X_i$ by atomic weights and renormalize to 100 to obtain elemental weight %.
2. *For the case of H- or He-normalized log atomic abundances that ARE referenced to the Solar composition*: Stars in the immensely useful Hypatia Catalog (Hinkel et al. 2014) are further normalized to the Lodders et al. (2009) Solar composition, and reported as a difference in log concentrations relative to the Sun (as [*Z/H*]). It is thus essential to multiply the Hypatia compositions by the Lodders (2009) Solar composition (Table 8a).

**Table 7.** *Sample calculation converting a stellar abundance (Magg et al. 2022) to element wt. %*

| | Atomic No. | Atomic wt. (AW) | A(El) Magg 2022 | No. of Atoms n(Z) = $10^{A(El)}$ | n(Z) x AW | Wt. % element |
|---|---|---|---|---|---|---|
| **C** | 6 | 12.0107 | 8.56 | 363078054.8 | 4360821592 | 23.7722888 |
| **O** | 8 | 15.9994 | 8.77 | 588843655.4 | 9421145179 | 51.3577955 |
| **Na** | 11 | 22.98977 | 6.29 | 1949844.6 | 44826478.88 | 0.24436404 |
| **Mg** | 12 | 24.305 | 7.55 | 35481338.92 | 862373942.5 | 4.70108716 |
| **Al** | 13 | 26.981538 | 6.43 | 2691534.804 | 72621748.59 | 0.3958853 |
| **Si** | 14 | 28.0855 | 7.59 | 38904514.5 | 1092652742 | 5.95641347 |
| **P** | 15 | 30.97361 | 5.41 | 257039.5783 | 7961443.652 | 0.04340048 |
| **S** | 16 | 32.065 | 7.16 | 14454397.71 | 463480262.5 | 2.52658505 |
| **K** | 19 | 39.0983 | 5.14 | 138038.4265 | 5397067.809 | 0.02942121 |
| **Ca** | 20 | 40.078 | 6.37 | 2344228.815 | 93952002.46 | 0.51216361 |
| **Ti** | 22 | 47.867 | 4.94 | 87096.359 | 4169041.416 | 0.02272683 |
| **Cr** | 24 | 51.9961 | 5.74 | 549540.8739 | 28573982.23 | 0.15576628 |
| **Mn** | 25 | 54.938049 | 5.52 | 331131.1215 | 18191697.78 | 0.09916899 |
| **Fe** | 26 | 55.845 | 7.5 | 31622776.6 | 1765973959 | 9.62691135 |
| **Ni** | 28 | 58.6934 | 6.24 | 1737800.829 | 101997439.2 | 0.55602196 |
| | | | | total | 18344138580 | 100 |

**Table 8a.** *Solar compositions in dex, converted to element wt. %*

| **Reported ratio** | | **(a) Solar Phot.[1] $A(El)_{Sol}$** | **(b) Solar Phot. $n(Z)_{Sol} = 10^{A(El)}$** | **(c) Solar Phot.[2] AW x $10^{A(El)}$** | **(d) Solar Phot. wt. %[3]** |
|---|---|---|---|---|---|
| **C/H** | C | 8.39 | 245470891.6 | 2948277237 | 18.79 |
| **O/H** | O | 8.73 | 537031796.4 | 8592186523 | 54.75 |
| **Na/H** | Na | 6.3 | 1995262.315 | 45870621.71 | 0.29 |
| **Mg/H** | Mg | 7.54 | 34673685.05 | 842743915 | 5.37 |
| **Al/H** | Al | 6.47 | 2951209.227 | 79628163.9 | 0.51 |
| **Si/H** | Si | 7.52 | 33113112.15 | 929998311.2 | 5.93 |
| **P/H** | P | 5.46 | 288403.1503 | 8932886.701 | 0.06 |
| **S/H** | S | 7.14 | 13803842.65 | 442620214.4 | 2.82 |
| **K/H** | K | 5.12 | 131825.6739 | 5154159.744 | 0.03 |
| **Ca/H** | Ca | 6.33 | 2137962.09 | 85685244.62 | 0.55 |
| **Ti/H** | Ti | 4.9 | 79432.82347 | 3802210.961 | 0.02 |
| **Cr/H** | Cr | 5.64 | 436515.8322 | 22697120.86 | 0.14 |
| **Mn/H** | Mn | 5.37 | 13803842.65 | 12878735.75 | 0.08 |
| **Fe/H** | Fe | 7.45 | 28183829.31 | 1573925948 | 10.03 |
| **Ni/H** | Ni | 6.23 | 1698243.652 | 99675693.99 | 0.64 |
| | | | Total: | 15694076987 | 100 |

1. The Solar photosphere composition from Lodders et al. (2009, their Table 4). 2. Product of atomic weight (AW) and $10^{A(El)}$. 3. Column (c) renormalized to 100%.

**Table 8b.** *Star composition from the Hypatia Catalog (Hip 32768), converted to element wt. %*

| **Reported ratio** | | **(e) Hip 32768 Reported dex values $A(X)_{32768}$** | **(f) dex relative to Solar $10^{A(X)}$** | **(g) $n(Z)_{32768} = 10^{A(X)} \times n(Z)_{Sol}$** | **(h) AW x $n(Z)_{32768}$** | **(i) Element wt. %** |
|---|---|---|---|---|---|---|
| **C/H** | C | -0.14 | 0.72 | 177827941 | 2135838051 | 12.63 |
| **O/H** | O | -0.06 | 0.87 | 467735141.3 | 7483481620 | 44.26 |
| **Na/H** | Na | 0.3 | 2.00 | 3981071.706 | 91523922.86 | 0.54 |
| **Mg/H** | Mg | 0.21 | 1.62 | 56234132.52 | 1366770591 | 8.08 |
| **Al/H** | Al | 0.23 | 1.70 | 5011872.336 | 135228023.9 | 0.80 |
| **Si/H** | Si | 0.27 | 1.86 | 61659500.19 | 1731737892 | 10.24 |
| **P/H** | P | | | | | |
| **S/H** | S | 0.67 | 4.68 | 64565422.9 | 2070290285 | 12.24 |
| **K/H** | K | | | | | |
| **Ca/H** | Ca | 0.1 | 1.26 | 2691534.804 | 107871331.9 | 0.64 |
| **Ti/H** | Ti | | | | 0 | 0.00 |
| **Cr/H** | Cr | 0.08 | 1.20 | 524807.4602 | 27287941.18 | 0.16 |
| **Mn/H** | Mn | | | | | |
| **Fe/H** | Fe | 0.02 | 1.05 | 29512092.27 | 1648102793 | 9.75 |
| **Ni/H** | Ni | 0.04 | 1.10 | 1862087.137 | 109292225.1 | 0.65 |

| | | | |
|---|---|---|---|
| | Total | 16907424677 | 100 |

It is common in the recent literature to compare star and planet compositions using elemental ratios; while useful, comparisons on the basis of % abundances, via Tables 5-8 provide advantages. Firstly, two different objects might have the same Mg/Si ratio, for example, but only by comparing % values can one test whether they have the same Mg and Si concentrations. If the concentrations are not the same, imputed common properties or evolutionary histories may be entirely mistaken. In addition, when compositions are expressed on a % basis, mass balance calculations are facilitated, as are comparisons of exoplanetary materials to meteorites and rock and mineral compositions analyzed from samples of the inner planets. And of course, one can also estimate mineral abundances. Absolute concentrations thus provide paths for increasing our understanding of exoplanetary systems, and serve as an insurance policy against concluding that two things are similar when they are really quite different.

## CALCULATING MINERALOGY AND ROCK TYPE: METHODS AND UNCERTAINTIES

Comparisons of elemental concentrations can be instructive, but to truly understand any planet, whether contained within the Solar System or elsewhere, the mineralogy of the silicate fraction is essential. In this section we'll examine how to calculate bulk silicate planet (BSP) composition using a star's bulk composition as input, and then introduce two different methods for converting a BSP into mineral abundances and rock types. Our examples will make considerable use of perhaps an otherwise obscure set of stars, known as "polluted white dwarfs". These stars are special because their atmospheres may record the remnants of rocky planetary debris (see Xu et al. this issue). Caveats accompany the use of these methods, which are detailed in a later section.

### Star compositions, chondrites and planetary bulk compositions

To estimate mantle mineralogy of an exoplanet, more often than not, we will use a star's composition as a nominal planetary bulk composition, taking the refractory elements on a volatile-free basis. The foundation for such a comparison stems from within our Solar system: there is a remarkably strong 1-to-1 correlation between the concentrations of refractory elements in the Sun and a special class of meteorites, the CI carbonaceous chondrites (Lodders et al. 2009; Lodders and Fegley 2018). These meteorites, as might be expected, are C-rich; they also contain "chondrules"—glassy blebs of material that appear to be relics of the earliest stages of Solar System development (Connolly and Jones, 2018). Some caveats about this foundation are worth noting. First, the meteorite database of Nittler et al (2004) contains compositions from 647 distinctly named chondrite meteorites, but for the CI chondrites, we have compositions from only 4 named samples, and most analyses of these derive from a single meteorite, Orgueil, as it is the only CI chondrite with sufficient mass to allow multiple analyses (Palme and Zipfel, 2022). The Sun-CI chondrite connection thus hinges on the observation that 0.6% of all chondrites are similar to the Sun with respect to their refractory elements. On this basis, it is very commonly assumed that most of the rocky bodies of the inner Solar System have a CI chondrite bulk composition (Sanloup et al. 1999; Rubie et al. 2011; Elardo et al. 2019). However, many studies now clearly show that Earth is not chondritic with respect to its trace elements and isotopes (McDonough and Sun 1995; Drake and Righter 2002; Campbell et al. 2012; Mezger et al. 2020) and may even have a non-chondritic major element content (Putirka et al. 2021). This recognition has led to the proposal that enstatite chondrites are a better fit for bulk Earth (e.g., Javoy et al. 2010; Boyet et al. 2018) since they have Earth-like ratios for many isotopes. But enstatite meteorites do not match Si isotopes (e.g. Fitoussi and Bourdon 2012) and they are not a match for various trace and major element compositions (Baedecker and Wasson 1975; Mezger et al. 2020; Putirka et al. 2021). We've known about this Earth-meteorite mismatch issue for some time, as McDonough and Sun (1995) concluded that no meteorite subclass provides a match to Earth's bulk composition. Mezger et al. (2020) provide intriguing evidence that the component that is missing from the meteorite database but that comprises Earth, may be found more heavily concentrated in the compositions of Venus and Mercury. In any case, and noting these caveats, we will proceed on the assumption that rocky planets at least roughly mimic the non-volatile compositions of the stars they orbit, which may indeed be true for the inner planets (Putirka et al. 2021, their Fig. 4).

**Models for core formation (a 1st-order control on mantle mineralogy)**

The amount of Fe that partitions into the core relative to the mantle has a tremendous influence on mineralogy, in large part because Fe readily partitions into silicate minerals. The recent focus on exoplanetary Mg/Si ratios is not entirely misplaced. But the ratio (Mg+Fe)/Si provides a more precise classification scheme, separating peridotites from pyroxenites, and determining when a mantle assemblage might be nominally saturated with quartz or ferro-periclase (Putirka and Rarick 2019). The mantle (Mg+Fe)/Si ratio is, of course, dependent upon the fraction of Fe that is sequestered into the core. And to obtain mantle composition from a star composition, that amount of Fe that will partition into the core will be the first decision that must be made (though in a later section we'll introduce a method to calculate Fe in the core).

There are many models of core formation but from a purely geochemical standpoint, oxygen abundance (often measured as a "fictive" oxygen fugacity, or $fO_2$, i.e., the partial pressure of $O_2$ in a vapor phase, if that vapor were in equilibrium with the core) is likely a key thermodynamic variable. And that variable will be essentially unknown. The problem is this: oxygen is so abundant in FGMK-type stars (the Sun is a G-type star) that all the planet-building major elements Si, Al, Fe, Mg, Ni and Ca are likely to be fully oxidized (to form $SiO_2$, $Al_2O_3$, FeO, MgO, NiO and CaO), with O leftover (Unterborn and Panero 2017; Putirka and Rarick 2019). However, every differentiated planetary object in the inner Solar system contains a metallic Fe-Ni alloy core that is at least roughly devoid of O, and so the inner planets are effectively O-depleted relative to the Sun. In calculating an exoplanet's mantle mineralogy we must decide, perhaps independent of star composition, the fraction of Fe that occurs in the core, as $Fe^0$, and what fraction occurs in the mantle, as $Fe^{2+}$ (in mineral such as $(Mg,Fe)_2SiO_4$, and $(Mg,Fe)SiO_3$). But the actual fictive $fO_2$ in a planetary mantle may indeed be a function of the final calculated mineralogy: Guimond et al. (2023) show that, because pyroxenes can accommodate $Fe^{3+}$, but olivine cannot, mantle mineralogy can control the total amount of O that can be dissolved into the mantle, and hence control the $fO_2$ of the mantle, rather than our usual way of thinking, i.e. that $fO_2$ controls mineral proportions. N.B.: it is common in petrologic studies to describe mantle rocks as following a given "$fO_2$ buffer", mostly because their $FeO/Fe_2O_3$ ratios can mimic such trends (e.g., Ni-NiO; Rhodes and Vollinger 2005); in such discussions it is understood that no such buffers actually exist and that $fO_2$ values are, indeed, "fictive", in that the rocks in question have not equilibrated with a vapor phase; $fO_2$ is useful as it can be controlled in experiments, but in this work, all calculations deal with a defined O content, not a defined $fO_2$). In this work, we'll ignore $Fe^{3+}$ (usually expressed as $Fe_2O_3$, which accounts for as much as 10% of all Fe in Earth's upper mantle; Rhodes and Vollinger 2008) and treat all Fe as if it were $Fe^{2+}$ (usually expressed as FeO). To be clear that we are accounting for all Fe in the mantle (though perhaps not all O), independent of the oxidation state of Fe, we'll refer to this "total Fe" as FeOt. Since O contents effectively unknown for exoplanets, Putirka and Rarick (2019) recommend use of $\alpha_{Fe} = Fe^{BSP}/Fe^{BP}$, where $Fe^{BSP}$ is the cation fraction of Fe in the bulk silicate planet (crust + mantle), and $Fe^{BP}$ is the cation fraction of Fe in the bulk planet (crust + mantle + core). This method effectively constrains the bulk mantle O content (and a bulk fictive $fO_2$, although $fO_2$ will vary widely across a planet, even within a well-mixed mantle, e.g., Stolper et al. 2020). If we assume that exoplanetary O contents are roughly Earth-like (Doyle et al. 2019), $\alpha_{Fe}$ is uncertain but may be close to 0.27-0.30 (see Putirka and Rarick 2019). In such a model, high-Fe planets will both have larger cores but also slightly higher FeOt in their BSP compositions. Using our Solar system as a model of how O might vary within another planetary system, Putirka and Rarick (2019) recommend exploring $\alpha_{Fe}$ values that range from at least 0 (Mercury-like) to 0.54 (Mars-like), but of course, our Solar System might not define the limits of $\alpha_{Fe}$ and exoplanetary systems might range to greater $\alpha_{Fe}$.

**Calculating mantle mineralogy**

***Normative abundances*** In 1902, four geologists, Cross, Iddings, Prison and Washington, published a method to classify igneous rocks (Cross et al. 1902). Their method involves using rock compositions to quantify the minerals that are "capable of crystallizing from a magma" of the same composition. Their approach would come to be known as the "CIPW norm". A "norm" is a theoretical mineral assemblage, to be put into contrast with a "mode" which would refer to the actual minerals identified in a given rock. Cross et al. (1902) suggested their norms could constitute a "standard mineralogy" by which different rocks could be compared. The use of such a method is that a volcanic rock, which is incompletely crystallized, could then be compared on a mineralogical basis to a plutonic rock (an igneous rock that has fully crystallized, at some depth below the surface). The norm, or standard mineralogy, is also helpful for comparing rocks that have crystallized under very different *P-T* conditions. This standard mineralogy is effectively a "what if' calculation: what if a volcanic rock were to crystallize completely? What if a rock that formed deep in the crust were to re-crystallize at 1 atm pressure? What minerals might form and how would those minerals differ from, say, that rock over there?

***Cross and his collaborators were not fools*** Cross et al. (1902) recognized that their standard mineralogy was not a substitute for mineral identification. On page 558 they write "The standard mineral composition of a rock is called its norm, and this may be *quite different from its actual mineral composition*, or mode" (emphasis altered from the original). As an example, if their method calculates that a rock is corundum (Cn; $Al_2O_3$) normative, that rock might not contain corundum; the *P-T* conditions might not be conducive, or Al might be distributed differently than as calculated, or the rock contains components (e.g., water, carbon, etc.) that go unaccounted in the method. If there is error in the attempt, why make the normative estimate in the first place? Because, as Cross et al. (1902) hypothesized, the normative classification can group rocks together that follow a similar evolutionary path. Our hypothetical Cn-normative system might not contain corundum, but it's likely on a path towards corundum saturation, and knowing that path can tell us about a rock's evolutionary history (Cawthorn and Brown, 1976) and its source materials (Chappell and White, 1992).

In summary Cross et al. (1902) proposed a classification scheme. Their system does not substitute for examining rocks in the field or under a microscope, but it does provide insights into a system's origin and evolutionary path. Exoplanets can similarly benefit from the method. The dozens of minerals calculated in a traditional CIPW norm (Verma et al. 2003; Buckle et al. 2023), however, are largely irrelevant to our current studies of exoplanets. The abundances of Si, Mg, and Fe utterly dominate over all other cations, and our current emphasis is to estimate mantle mineral assemblages; the numbers of minerals that will control exoplanetary thermal and mechanical properties of a planet are few. The two methods described below are thus abbreviated for the specific task of estimating mineral abundances for exoplanetary mantles, and as we learn more about how S, C, O and H vary in planetary bodies, these methods will quite likely require further refinement.

***Method 1: the J.B. Thompson approach (Supplement 1)*** Thompson (1982) proposed a method based on linear algebra that is quite different from the CIPW norm that has been classically used by geologists. But Thompson's (1982) approach has the virtue of allowing a very wide range of different mineral assemblages to be tested very quickly, which can be useful for exoplanets since we must at least hold the possibility that some may have mineral assemblages that are exotic to the inner Solar System (See Supplementary file 1 for example calculations of Method 1).

In the Thompson (1982) approach, we can calculate a normative mineral assemblage using a BSP (e.g., as in Table 4) as input. For our example, we'll consider a case where the compositions are expected to plot into Figure 4a (as peridotite or pyroxenite), so the desired outputs are the fractions of olivine ($F_{Ol}$), clinopyroxene ($F_{Cpx}$), and orthopyroxene ($F_{Opx}$). Some exoplanets might have sufficiently high Fe to allow wüstite to form, which we can test by calculating $F_{Wus}$. As input, we need only the mole fractions of $SiO_2$, FeO, MgO and CaO, renormalized to sum to 1. For the calculation, we must also decide on specific mineral compositions (Table 9a), which for the case of Models 1 and 2 in Table 9a are obtained from experiments on peridotite compositions conducted at upper mantle *P-T* conditions (ca. 2 GPa and 1300-1400 °C. Kinzler 1997; Walter 1998).

Taking olivine as an example, both observations and experiments show that Earth's upper mantle typically has olivine that is close to 90% forsterite (Fo; $Mg_2SiO_4$) and 10% fayalite (Fa; $Fe_2SiO_4$), expressed as Fo90. The specific formula for olivine is thus $Mg_{1.8}Fe_{0.2}SiO_4$, which requires 1 unit of $SiO_2$, 1.8 units of MgO and 0.2 units of FeO, which are the table entries for olivine in Table 9a, Model 1. Model 1 entries for clinopyroxene and orthopyroxene similarly provide the compositions expected for a peridotite mantle, equilibrated at upper mantle conditions.

**Table 9a.** *Mineral compositions*

| ***Model 1*** | $SiO_2$ | FeO | MgO | CaO |
|---|---|---|---|---|

| | | | | | |
|---|---|---|---|---|---|
| Olivine | $(Mg,Fe)_2SiO_4$ | 1 | 0.2 | 1.8 | 0 |
| Clinopyroxene | $Ca(Mg,Fe)Si_2O_6$ | 2 | 0.2 | 1.8 | 1 |
| Orthopyroxene | $(Mg,Fe)_2Si2O_6$ | 2 | 0.2 | 1.8 | 0 |
| Wüstite | FeO | 0 | 1 | 0 | 0 |
| | | | | | |
| ***Model 2*** | | $SiO_2$ | $Al_2O_3$ | FmO | CaO |
| Olivine | $(Mg,Fe)_2SiO_4$ | 1 | 0 | 2 | 0 |
| Clinopyroxene | $Ca(Mg,Fe)Si_2O_6$ | 1.8 | 0.3 | 1.3 | 0.6 |
| Orthopyroxene | $(Mg,Fe)_2Si2O_6$ | 1.8 | 0.3 | 1.8 | 0.1 |
| Garnet | $(Mg,Fe,Ca)_3Al_2Si_3O_{12}$ | 3 | 2 | 2.7 | 0.3 |
| | | | | | |
| ***Model 3*** | | $SiO_2$ | $Al_2O_3$ | FmO | CaO |
| Olivine | $(Mg,Fe)_2SiO_4$ | 1 | 0 | 2 | 0 |
| Clinopyroxene | $Ca(Mg,Fe)Si_2O_6$ | 1.75 | 0.25 | 1.25 | 0.75 |
| Orthopyroxene | $(Mg,Fe)_2Si2O_6$ | 1.9 | 0.3 | 1.75 | 0.05 |
| Garnet | $(Mg,Fe,Ca)_3Al_2Si_3O_{12}$ | 3 | 2 | 2.4 | 0.6 |
| | | | | | |
| ***Model 4*** | | Si | Fe | Mg | Ca |
| Olivine | $(Mg,Fe)_2SiO_4$ | 1.33 | 0.267 | 2.4 | 0 |
| Cpx + Opx | $Ca_{0.5}(Mg,Fe)_{1.5}Si_2O_6$ | 2 | 0.2 | 1.3 | 0.5 |
| Fe Metal | Fe | 0 | 1 | 0 | 0 |
| Quartz | $SiO_2$ | 1 | 0 | 0 | 0 |

**Table 9b**. *Transformation matrix*

| | Olivine | Clinopyroxene | Orthopyroxene | Wüstite |
|---|---|---|---|---|
| $SiO_2$ | -1 | 0 | 1 | 0 |
| FeO | 0 | 0 | 0 | 1 |
| MgO | 1.11 | 0 | -0.56 | -0.11 |
| CaO | 0 | 1 | -1 | 0 |

Note: each of the models in Table 9a, however simple, are thermodynamic models: a set of *P* and *T* conditions are assumed to apply (close to ca. 2 GPa and 1400°C), and the minerals in any given model are the phases that are presumed to have the lowest Gibbs Free Energy at such *P*-*T* conditions. The models in Table 9a also presume an anhydrous and largely C- and S-free bulk composition, with an Earth-like O content (e.g., fictive $fO_2$ close to the Ni-NiO buffer; Rhodes and Vollinger 2008). To the extent that we choose to compare exoplanets using such a standard mineralogy as Models 1 or 2 in Table 9a, these conditions act as a *de facto* standard state.

As to the calculation, in a spreadsheet, it is convenient to handle large amounts of data using row matrices, and so we can employ Model 1 using the equation:

$$[F_{Ol}\ F_{Cpx}\ F_{Opx}\ F_{Wus}] = [X^{BSP}_{SiO_2}\ X^{BSP}_{FeO}\ X^{BSP}_{MgO}\ X^{BSP}_{CaO}] \times \begin{bmatrix} X^{Ol}_{SiO_2} & X^{Ol}_{FeO} & X^{Ol}_{MgO} & X^{Ol}_{CaO} \\ X^{Cpx}_{SiO_2} & X^{Cpx}_{FeO} & X^{Cpx}_{MgO} & X^{Cpx}_{CaO} \\ X^{Opx}_{SiO_2} & X^{Opx}_{FeO} & X^{Opx}_{MgO} & X^{Opx}_{CaO} \\ X^{Wus}_{SiO_2} & X^{Wus}_{FeO} & X^{Wus}_{MgO} & X^{Wus}_{CaO} \end{bmatrix}^{-1}$$

(4)

In Equation 4, the row matrix on the left-hand side is the output, which is the mole fractions of the minerals indicated in a matrix such as those in Table 9a. On the right-hand side, the row matrix contains the mole fractions of the oxides i for the bulk silicate planet, $X_i^{BSP}$; in our example in Model 1 (Table 9a) we require the mole fractions of $SiO_2$, MgO, FeO or CaO, where $X_{SiO2}+X_{MgO}+X_{FeO}+X_{CaO} = 1$. And the square matrix on the right-hand is the inverse of the matrix of Model 1, which is given in Table 9b. If we have a BSP where the mole fractions of $SiO_2$, FeO, MgO and CaO are respectively, 0.4, 0.09, 0.49 and 0.02, we have:

$$[0.144\ \ 0.020\ \ 0.108\ 0.036] = [0.4\ \ 0.09\ 0.49\ \ 0.02] \times \begin{bmatrix} -1 & 0 & 1 & 0 \\ 0 & 0 & 0 & 1 \\ 1.11 & 0 & -0.56 & -0.11 \\ 0 & 1 & -1 & 0 \end{bmatrix} \quad (5)$$

And the mineral proportions on a molecular basis, obtained from Equations 4-5, of Ol, Cpx, Opx and Wus are respectively 0.144, 0.020, 0.108 and 0.036, which upon renormalization to 100 yields a mineral assemblage that is 46% olivine 6% clinopyroxene, 35% orthopyroxene and 12 % orthopyroxene. For a ternary diagram where Ol, Cpx and Opx lie at the apices, it is necessary to renormalize again, so that Ol + Cpx + Opx = 100 (or 1), so we have ca. 53% Ol, 7% Cpx and 40% Opx on a molecular basis. There is no reason not to use molecular proportions in ternary diagrams such as Figure 4a, and interestingly, terrestrial mantle rocks plot in effectively the same field in Figure 4a regardless of whether molecular or weight proportions are employed. However, individual compositions can plot quite differently, especially when the minerals at the apices have very different molecular weights. In any case, to obtain weight % values each mole fraction or % value would then be multiplied by the molecular formula weight of the mineral with further renormalization to 100%. Figure 4a was developed on the basis of nominal "% volume" (though really "% area", since most "volume" measurements are determined from 2D thin sections); mineral densities could be used to convert weights to volume proportions, but this additional step is not recommended.

If one prefers column matrices, then instead of Equation (4) we have:

$$\begin{bmatrix} F_{Ol} \\ F_{Cpx} \\ F_{Opx} \\ F_{Wus} \end{bmatrix} = \left[\begin{bmatrix} X_{SiO_2}^{Ol} & X_{FeO}^{Ol} & X_{MgO}^{Ol} & X_{CaO}^{Ol} \\ X_{SiO_2}^{Cpx} & X_{FeO}^{Cpx} & X_{MgO}^{Cpx} & X_{CaO}^{Cpx} \\ X_{SiO_2}^{Opx} & X_{FeO}^{Opx} & X_{MgO}^{Opx} & X_{CaO}^{Opx} \\ X_{SiO_2}^{Wus} & X_{FeO}^{Wus} & X_{MgO}^{Wus} & X_{CaO}^{Wus} \end{bmatrix}^T\right]^{-1} \times \begin{bmatrix} X_{SiO_2}^{BSP} \\ X_{FeO}^{BSP} \\ X_{MgO}^{BSP} \\ X_{CaO}^{BSP} \end{bmatrix} \quad (6)$$

Equation 6 contains two critical adjustments to the right-hand side of Equation 4, namely we must take the transpose of the mineral composition matrix (Model 1, Table 9a) before calculating its inverse, and we must reverse the order of the multiplication (matrix multiplication is non-commutative).

Equations 4-6 demonstrate the method, but do the very simple thermodynamic models of Table 9a have any chance of accurately predicting mineral abundances? Models 1, 2 and 3 all assume that mineral compositions are constant across all bulk compositions where the model is applied—this is highly unlikely, but apparently not fatal. Figures 7a-b and Figure 8 show the case where Model 2 is applied to predict mineral abundances for peridotites (Warren 2016) and pyroxenites (Bodinier et al. 2008), where mineral abundances are either measured to high precision or can be calculated from measured mineral and bulk compositions. The results of our tests (Figure 7) show that Model 2 can yield estimates that are precise to ±10% and perhaps as precise as ± 3%, provided the bulk compositions used as input fall within the ultramafic rock diagram of Figure 4; it might even be useful for terrestrial peridotites. Model 2 is clearly less accurate for pyroxenites (Fig. 7b), capturing just 65% of the variation of natural samples compared to 99% for peridotites (compare $R^2$ values). But by adjusting the coefficients so as to obtain Model 3 (Table 9a), one can quickly devise a new Model that increases precision and accuracy for such compositions (Fig. 7c). This approach, of adjusting such matrices so as to better predict cases of measured or well-constrained mineral abundances, is highly recommendd.

Note that Models 1-3 in Table 9a are only appropriate when olivine and the pyroxenes dominate, such as the rocky mantles of planetary objects; the models are not appropriate for plotting bulk planet compositions, as these are likely to have large metallic cores. Model 4 in Table 9a shows how one might address the issue: here, due to their identical stoichiometry, clinopyroxene (Cpx) and orthopyroxene (Opx) are combined to form a single pyroxene component (Pyx) and a pure-Fe metal component is added. This particular matrix describes 97% of the core mass % ($M_c$) values of the inner planets and Vesta (Figure 7d), using Table 3 and bulk planet estimates from McDonough and Yoshizaki (2022) as input (measured values of $M_c$ are from Szurgot 2015). (N.B.: the bulk compositions of Venus, and even Mercury, are highly uncertain; confidence in their compositions is low, especially for Venus).

As can be seen from these examples, the Thompson (1982) approach has advantages and disadvantages: it is ideal in quickly calculating mineral abundances for a large number of samples, and testing a wide range of possible mineral assemblages. Matrices can also be quickly edited so as to optimize performance in predicting mineral abundances in test data sets. But predicated mineral abundances can be negative, which is acceptable for classification (e.g., Thompson 1982), but not helpful if one wants to perform experiments on possible exoplanet compositions. Negative estimates show that a model does not yield a plausible mineral assemblage; this is, of course, a signal to find a new set of minerals or mineral compositions. The model also requires square matrices for transformation, which limits the range of minerals that can be explored for a given set of bulk composition and hence also limits the range of compositions and mineral assemblages for which accurate mineral abundances may be obtained.

Method 2 illustrates a new approach, analogous to the original CIPW norm, that yields only positive mineral abundances regardless of bulk composition.

***Method 2: a CIPW norm analog for exoplanets (Supplement 2)*** The method is fashioned in the spirit of Cross et al. (1902), using a cation accounting system, but is specifically designed for exoplanet studies. Method 2 allows for (a) calculation of Fe metal (core) mass fractions, based on Fe-Si mass balance, (b) a single set of fixed, end-member mineral compositions that, without adjustment, yield positive mineral sums for any exoplanet composition, including highly exotic ones, and (c) a determination whether a system might be saturated in ferrosilite ($Fe_2Si_2O6$)—a mineral that is rare (because it is thermodynamically unstable) in the inner planets, but has been putatively detected in white dwarf dust disks (Reach et al. 2009). The method is as follows (See also Supplementary File 2, a spreadsheet that provides the calculations for Method 2, below):

Step 1: For input, obtain/convert a composition consisting of any of Si, Ti, Cr, Al, Fe, Mg, Ni, Ca, Na, K and P, to elemental mole % (the calculation assumes an anhydrous bulk composition; O is determined from charge balance on the cations). Any of the listed elements may be undetermined (set to 0), but useful output requires estimates for Si, Fe, and Mg at a minimum. Note: for a correct accounting, anytime a component calculates as < 0, assign a value of 0, since a negative value indicates an insufficiency of an essential ingredient for the component to form.

Step 2: Assign Cr to chromite, Chr ($FeCr_2O_4$) and eskolaite, Esk ($Cr_2O_3$):

(a) If either of Fe or Cr = 0, then Chr = 0
if Fe and Cr are >0, then
if (Cr/Fe)>2, Chr = Fe/(1/3), else Chr = Cr/(2/3).

(b) If Cr > Fe, then excess Cr = Cr1 = Cr – (2/3)($FeCr_2O_4$) and Cr1 is assigned to Esk, where Esk = (Cr1)/2.

Step 3: Excess Fe, Fe1 = Fe1 – (1/3)$FeCr_2O_4$.

Step 4: Assign Ti to Ilmenite, Ilm ($FeTiO_3$) and rutile, Rt ($TiO_2$):

(a) If Ti or Fe1 = 0, then Ilm = 0
if Ti > 0 and Fe1 > 0, then
if Ti < Fe1, Ilm = Ti/(1/2), else Ilm = Fe1/(1/2)

(b) Excess Ti (Ti1) is assigned to the mineral rutile, where Ti1= Ti – (1/2)$FeTiO_3$ and Rt = Ti1

Step 5: Excess Fe, Fe2: = Fe2 – (1/2)$FeTiO_3$

Step 6: Assign Mg# (= Mg/([Mg+Fe]) to silicate minerals (0-1 possible; Earth like values are 0.80 to 0.90) and a fraction of Ni to silicates, $F_{Ni}$ (0-1 possible; Earth-like value is ca. 0.05): The dominant silicate minerals in the mantles of the inner planets (olivine, pyroxenes, garnet) are dominantly solid solutions of Mg and Fe, with substantial amounts of Ni. For simplification, the choice of Mg# and $F_{Ni}$ at this step will set the Mg# and $F_{Ni}$ values for all silicates that contain such elements. The real import of this step, and Steps 7-9, is to determine the amount of Fe + Ni that will be segregated into a metallic core and how much is retained by silicate minerals in the mantle; the larger Mg# and $F_c$ are the less Fe and Ni are available to form a metallic core.

Step 7: Calculate PFNS, the potential amount of Ni and Fe that can partition into silicate phases: if Si > (Fe2 + Mg + Ni), then PFNS = Fe2 + Ni, else PFNS = ($F_{Ni}$ )(Ni) + (Mg - Mg#[Mg])/Mg#.

Step 8: Assign Fe and Ni to silicate phases (FNS) and determine any excess Fe and Ni (FeNi) relative to Si:

(a) if (Fe2 + Ni) > PFNS, then FNS = PFNS, else FNS = Fe2 + Ni.

(b) Excess Fe and Ni are assigned to FeNi as follows: if (Fe2+Ni) > PFNS, FeNi = Fe2 + Ni - PFNS, else FeNi = 0. FeNi represents the first step in determining the amount of Fe and Ni that will partition into a metallic phase.

Step 9: Assign Fe, Ni and Mg to Silicates as Fm: Fm = FNS + Mg. (Note to astronomers: “Fm” is a shorthand for “ferro-magnesian”, which indicates minerals rich in Fe and Mg, such as olivine and the pyroxenes). Here, Fm includes Ni since quantifying Ni is important to quantifying bulk planet and planetary core compositions.

Step 10: Assign P to apatite, Ap ($Ca_5(PO_4)_3(OH)$) and calculate excess Ca and P:

(a) if both P > 0 and Ca > 0, then Ap = P/(3/8), else Ap = 0.
(b) Excess Ca, Ca1 = Ca – (5/8)Ap.
(c) Excess P, if any, is assigned to phosphorous oxide in the final tally where P1 = P – (3/8)Ap and phosphorous oxide = P1/2.

Step 11: Assign Na, K, Al and Si to alkali feldspar, Afs (Na,K)$AlSi_3O_8$), and nepheline, Ne (Na,K)$AlSiO_4$:

(a) Establish Si/Al ratio, SAR: If Al and Si > 0, then SAR = Si/Al, else SAR = 0
(b) Establish the amount of Si in feldspar, SiF: if SAR > 3 then SiF = 1; if SAR < 1, then SiF = 0. If 3 > SAR > 1, then SiF = (SAR – 1)/2
(c) Establish SiN, the amount of Si assigned to nepheline: SiN = 1 – SiF
(d) Calculate TA, total alkalis: TA = Na + K
(e) Calculate amount of Afs:
    if either TA or SiF are not > 0, then Afs = 0;
    if TA > 0 and SiF > 0, then
        if both TA < Al and 3(TA) < Si then Afs = (SiF)(TA)/(1/5), else
        if either TA is not < Al or 3(TA) is not < Si, then
            if 3(Al) < Si then Afs = (SiF)(Al)/(1/5) else Afs = (SiF)(Si)/(3/5)
(f) Calculate TA1, excess total alkalis: TA1 = TA – (1/5)Afs
(g) Calculate Al1, excess Al: Al1 = Al – (1/5)Afs
(h) Calculate Si1, excess Si: Si1 = Si – (3/5)Afs
(i) Calculate Ne, nepheline: if either of TA1 = 0 or Al1 = 0, then Ne = 0; if both TA1 > 0 and Al1 > 0 then:
        if both TA1< Al1, and TA1 < Si1 then Ne = (SiN)(TA1)/(1/3)
            if either TA1 > Al1 or TA1 > Si1, then
                if Si1 > Al1, then Ne = (SiN)(Al1)/(1/3)
                    else Ne = (SiN)(Si1)/(1/3).
(j) Calculate excess Al: Al2 = Al1 – (1/3)Ne
(k) Calculate excess Si: Si2 = Si1 – (1/3)Si5
(l) Calculate excess alkalis: TA2 = TA1 – (1/3)Ne

Step 12: Assign Al and Fm to garnet, PyAl (PyAl is the sum of pyrope ($Mg_3Al_2Si_3O_{12}$) and almandine ($Fe_3Al_2Si_3O_{12}$) garnet):

(a) If 3Al2 < 2Si2 and 3Al2 < 2Fm, then PyAl = Al/(1/4)
    if either 3Al2 > 2(Si2) or 3Al2 > 2Fm, then
        if Si2 < Fm, then PyAl = Si/(3/8), else
            if Si2 > Fm, PyAl = Fm/(3/8).
(b) Calculate excess Al: Al3 = Al2 – (1/4)PyAl. This excess Al in the final tally is assigned to corundum (Cn; $Al_2O_3$), where Cm= (Al3)/2.
(c) Calculate excess Fm: Fm1 = Fm – (3/8)PyAl
(d) Calculate excess Si: Si3 = Si2 – (3/8)PyAl

Step 13: Assign excess Ca to Diopside + Hedenbergite, DiHd (Ca(Mg,Fe)$Si_2O_6$):

(a) If Ca1 < (Si3/2) and Ca1 < Fm1, then DiHd = Ca1/(1/4)
    if either Ca1 > (Si3/2) or Ca1 > Fm1 then
        if Fm1 < (Si3)/2 then DiHd = Fm1/(1/4), else DiHd = (Si3)/(1/2)
(b) Calculate excess Ca: Ca2 is: Ca2 = Ca1 – (1/4)Di
(c) Calculate excess Fm: Fm2 = Fm1 – (1/4)Di

(d) Calculate excess Si: Si4 = Si3 – (1/2)Di

Step 14: Determine the potential fractions of olivine (Fol) and orthopyroxene (Fopx) (the steps that follow apportion the remaining amounts of Fm and Si between the phases olivine and orthopyroxene, based on Fm/Si ratios):

(a) If either Fm2 or Si4 = 0, then Fol = 0;
if both Fm2 >0 and Si4 > 0, then
if (Fm2/Si4)/2 >1 then Fol = 1, else
if Fm2/Si4 < 1, then Fol = 0 else
Fol = Fm2/Si4 – 1

(b) Fopx = 1 – Fol; (note: later calculations will take care of the case that either or both of Fm2 and Si2 = 0, and where Fol = 0 and Fopx = 1).

(c) Determine Si assigned to olivine (SiOl): if Fm2 > 0 and Si4 > 0 then SiOl = (Fol)(Si4), else SiOl = 0.

(d) Determine Si assigned to orthopyroxene (SiOpx): if Fm2 > 0 and Si4 > 0, then SiOpx = (Fopx)(Si4), else SiOpx = 0.

Step 15: Determine amounts of olivine (FoFa) and orthopyroxene (EnFs)

(a) FoFa = SiOl/(1/3)
(b) if Fm2 < SiOpx, then EnFs = Fm2/(1/2), else EnFs = SiOpx/(1/2)
(c) Calculate excess Fm: Fm3 = Fm2 - (2/3)FoFa – (1/2)EnFs
(d) Calculate excess Si: if Si4 < Fm2, then Si5 = 0, else
if Si4 > Fm2, then Si5 = Si4 – (1/3)FoFa – (1/2)EnFs.

Step 16: Assign remaining Fm to ferro-periclase, FmO: If Fm3 > 0, FmO = Fm3, else FmO = 0.

Step 17: Assign any excess FeNi and Si to ferrosilite (Fs; $FeSiO_3$):

(a) if Si5 > FeNi then Fs = FeNi/(1/2) else Fs = Si3/(1/2)
(b) Si6 = Si5 – (1/2)Fs
(c) FeNi1 = FeNi - (1/2)Fs

Step 18: Assign remaining oxides to any excess Al (corundum; Cm), Si (quartz; Qz) and alkalis (alkali oxides, AO):

(a) Corundum, Cm = Al3
(b) Quartz , Qz = Si6
(c) Alkali Oxides, AO = TA2

The above procedure yields the following estimates of mineral abundances:

| | |
|---|---|
| Chromite | = Chr from Step 2a |
| Eskolaite | = Step 2b |
| Ilmenite | = Ilm from Step 4a |
| Rutile | = Rt in Step 4b |
| Apatite | = Ap from Step 10a |
| Phosphorous oxide | = P1 from Step 10c |
| Alkali feldspar | = Afs from Step 11e |
| Nepheline | = Ne from Step 11i |
| Pyrope + Almandine (Gar) | = PyAl from Step 12a |
| Diopside + Hedenbergite (Cpx) | = DiHd from Step 13a |
| Calcium Oxide (CaO) | = Ca2 from Step 13b |
| Forsterite + Fayalite (Ol) | = FoFa from Step 15a |
| Enstatite + Ferrosilite (Opx) | = EnFs from Step 15b |
| Ferropericlase | = FmO from Step 16 |
| Ferrosilite | = Fs from Step 17 |
| Corundum | = Cm from Step 18a |
| Quartz | = Qz from Step 18b |
| Alkali Oxides | = AO from Step 18c |

These minerals should sum to precisely 100%; many may be 0 but none should be negative. An example calculation is provided in Table 10, using a Solar bulk composition.

The procedure is tested on nearly 9,000 rock compositions that include peridotites (McDonough and Sun 1995), pyroxenites (GEORORC) and meteorites of all types (Nittler et al. 2004), including irons, pallasites and mesosiderites; all yield mineral sums of precisely 100%, with predicted mineral assemblages that yield little to no Fe for silicate rock samples, and significant metallic Fe for all pallasites and most mesosiderites. Sums higher or lower than 100 indicate that elements/components are either being double- or under-counted.

How well does it work? For predicting mineral modes, Method 2 is less accurate than Method 1. But as with the CIPW norm, it is designed for classification; accurate estimates of mineral abundances are possible, but only under a set of caveats and assumptions, noted in a later section. But for a small cost in accuracy we gain flexibility: in a single model, Method 2 yields positive mineral abundances for any composition, estimates of metallic core mass fractions and predicts saturation in minor phases for potentially any exotic composition. Method 2 could be refined to allow some fraction of Al and/or Na in pyroxenes, by adding an Al partitioning function, just as we added Mg-Fe-Ni partitioning in Step 6, although it's not clear that in any immediately recognizable future we will be quite ready to discuss the Ca-Tschermak's components in exoplanet clinopyroxenes. In any case, Method 2 yields a "standard mineralogy" that can describe any exoplanetary mantle, no matter how extreme or bizarre the composition, and like Method 1 (Fig. 7d), describes 98% of the variations in $M_c$, with a precision of ± 3 wt. %.

**Table 10.** *Results from Method 2, using an anoxic, Solar bulk composition*

| Si | Ti | Cr | Al | Fe | Ni | Mg | Ca | Na | K | P |
|---|---|---|---|---|---|---|---|---|---|---|
| | | | | ***Solar elemental wt. %*** | | | | | | |
| 26.78 | 0.10 | 0.70 | 1.78 | 43.29 | 2.44 | 21.15 | 2.30 | 1.10 | 0.12 | 0.23 |
| | | | | ***Mole proportions*** | | | | | | |
| 0.95 | 0.00 | 0.01 | 0.07 | 0.78 | 0.04 | 0.87 | 0.06 | 0.05 | 0.00 | 0.01 |
| | | | | ***Mole %*** | | | | | | |
| 33.60 | 0.08 | 0.47 | 2.32 | 27.31 | 1.47 | 30.66 | 2.02 | 1.68 | 0.11 | 0.26 |
| ***Step 2*** | ***Step 2*** | ***Step 3*** | ***Step 4*** | ***Step 4*** | ***Step 5*** | ***Step 6*** | ***Step 6*** | | | |
| ***Chromite*** | ***Excess Cr*** | ***Fe1*** | ***Ilmenite*** | ***Excess Ti*** | ***Fe2*** | ***Ni #*** | ***Mg#*** | | | |
| 0.71 | 0.00 | 27.08 | 0.15 | 0.00 | 27.00 | 0.05 | 0.90 | | | |
| ***Step 7*** | ***Step 8*** | ***Step 8*** | | ***Step 9*** | | | | | | |
| ***PNFS*** | ***FNS*** | ***FeNi*** | ***Total Fe + Ni*** | ***Fm*** | | | | | | |
| 3.49 | 3.49 | 24.98 | 28.47 | 34.15 | | | | | | |
| ***Step 10*** | ***Step 10*** | ***Step 10*** | ***Step 11*** | ***Step 11*** | ***Step 11*** | ***Step 11*** | ***Step 11*** | | | |
| ***Apatite*** | ***Ca1*** | ***Excess P*** | ***SAR*** | ***SiF*** | ***SiN*** | ***Afs*** | ***TA1*** | | | |
| 0.68 | 1.60 | 0.00 | 14.45 | 1.00 | 0.00 | 8.98 | 0.00 | | | |
| ***Step 11 cont.*** | ***Step 11*** | ***Step 11*** | ***Step 11*** | ***Step 11*** | ***Step 11*** | | | | | |
| ***Al1*** | ***Si1*** | ***Ne*** | ***TA2*** | ***Al2*** | ***Si2*** | | | | | |
| 0.53 | 28.21 | 0.00 | 0.00 | 0.53 | 28.21 | | | | | |
| ***Step 12*** | ***Step 12*** | ***Step 12*** | ***Step 12*** | ***Step 13*** | ***Step 13*** | ***Step 13*** | ***Step 13*** | | | |
| ***PyAl*** | ***Al3*** | ***Fm1*** | ***Si3*** | ***DiHd*** | ***Ca2*** | ***Fm2*** | ***Si4*** | | | |
| 2.11 | 0.00 | 33.36 | 27.42 | 6.39 | 0.00 | 31.76 | 24.23 | | | |
| ***Step 14*** | ***Step 14*** | ***Step 14*** | ***Step 14*** | ***Step 15*** | ***Step 15*** | ***Step 15*** | ***Step 15*** | | | |
| ***Fol*** | ***Fopx*** | ***SiOl*** | ***SiOpx*** | ***FoFa*** | ***EnFs*** | ***Fm3*** | ***Si5*** | | | |
| 0.31 | 0.69 | 7.53 | 16.69 | 22.60 | 33.39 | 0.00 | 0.00 | | | |
| ***Step 16*** | ***Step 17*** | ***Step 17*** | ***Step 17*** | | | | | | | |
| ***Ferropericlase*** | ***Fs*** | ***Si6*** | ***FeNi1*** | | | | | | | |
| 0.00 | 0.00 | 0.00 | 24.98 | | | | | | | |
| Final Mineral % | | | | | | | | | | |
| ***Chromite*** | ***Eskolaite*** | ***Ilmenite*** | ***Rutile*** | ***Apatite*** | ***$P_2$O5*** | ***Alk. Fspar*** | ***Nepheline*** | ***Garnet*** | ***Clinopyroxene*** | |
| 0.71 | 0.00 | 0.15 | 0.00 | 0.68 | 0.00 | 8.98 | 0.00 | 2.11 | 6.39 | |
| ***CaO*** | ***Olivine*** | ***Orthopyroxene*** | ***Ferropericlase*** | ***Ferrosilite*** | ***Fe Metal*** | ***Corundum*** | ***Quartz*** | ***Alkali Ox.*** | | ***Total*** |
| 0.00 | 22.60 | 33.39 | 0.00 | 0.00 | 24.98 | 0.00 | 0.00 | 0.00 | | 100.00 |

## ERROR ANALYSIS IN TERNARY PROJECTIONS (USING POLLUTED WHITE DWARFS AS EXAMPLES)

Much discussion about exoplanet compositions derives from using star compositions to predict the compositions of exoplanets (e.g., Hinkel et al. 2016). The reasoning is that the inner planets have relative Mg, Si and Fe contents that might closely approximate Solar proportions (Putirka et al. 2021; their Fig. 4). However, as is evident from Table 3, the inner planets do not precisely mimic the Sun. McDonough and Yoshizaki (2022) make a highly compelling case that Fe decreases with heliocentric distance, arguing that the shift is controlled by the Sun's magnetic field. Clearly, we need better models of how and why the inner planets vary in composition so that we might apply such understanding to other stars, to better predict exoplanet compositions. On top of this, the uncertainties reported for star compositions have their own challenges (Hinkel et al. 2016; Rogers et al. 2024). Repeat observations of given stellar objects are another highly useful task to improve estimates of error.

In the meantime, uncertain uncertainties are unlikely to halt our thinking about plausible exoplanet compositions—and neither should they. To be sure, the less we understand about error, the more our calculations are rooted in the soil of speculation rather than rising into the atmosphere of improved understanding. Such challenges lead Treirweiler et al. (2023) to conclude that errors are "hopelessly large" and that, in effect, when it comes to estimating mantle mineralogy: it's too hard; don't try. That approach obviates all efforts that stem from mineralogy, such as studies of mantle circulation, plate tectonics, and the origin of life. Rather than give up entirely however, another option is to think more carefully about error, which provides a very precise and accurate guide to future work.

### What are Polluted white dwarfs? (possibilities and nominal problems)

As described by Xu et al. (2024, this volume) and by Xu and Bonsor (2021), polluted white dwarfs (PWDs) provide exceptional opportunities to examine the bulk geochemical and mineralogical characteristics of exoplanets. These stars represent a very late stage in stellar evolution, where a Sun-like star first expands to a Red Giant, obliterating planetary objects within a distance equivalent to the orbit of Mars, and then collapses to form a white dwarf, that is not much larger than Earth. Any planetary debris that survives the Red Giant phase (so the planetary analogs are, in effect, mostly outer Solar system materials) might later be absorbed into a white dwarf, yielding a "polluted" atmosphere, hence the name "polluted white dwarf". Because white dwarf atmospheres typically consist only of H or He (heavier elements sink rapidly to the stellar core), any elements that are heavier than He are likely to represent remnants of planetary debris that once orbited such stars and have very recently polluted their atmospheres. Putirka and Xu (2021) estimate mineral abundances for 23 PWDs, which represents a subset of the hundreds of white dwarfs that have been analyzed, where Si, Fe, Mg and Ca are all reported.

Putirka and Xu's (2021) apply Method 1 to calculate mineral abundances for 23 PWDs and speculate that some might have mineral assemblages that are exotic to our Solar system. They note that their mineral abundance estimates are by no means definitive because, as also emphasized below, current thermodynamic models are calibrated almost solely on Earth-like compositions. Putirka and Xu (2021) thus proposed a provisional mineral classification system, to describe exoplanet mantles that fall outside the ultramafic rock triangle (Fig. 4a). Trierweiler et al. (2023) test those results and conclude that most exoplanets are like CI chondrites. How do the same data yield such differing conclusions? It might stem from a misunderstanding of certain statistical tests. Figures 9-11 will use PWDs as a case study, to examine errors more closely and to also illustrate how certain choices when plotting ternary diagrams affect plotted uncertainties.

**Monte Carlo simulations**

Monte Carlo simulations are especially useful when data sets are small. In geology, they are often not needed, but it's no less critical to understand how to interpret them. First let's consider the need. If geologists have a large outcrop, that outcrop can re-sample and re-analyze (say, for $SiO_2$) many times over, perhaps using different analytical instruments. If the mean and standard deviation of repeat measurements are respectively 52.4 wt. % and ±1.2 wt. %, we would report the $SiO_2$ as 52.4±1.2 wt. %. This means that there is a 68% (±1σ) chance that the $SiO_2$ content falls within the range 51.2-53.6 wt. %. We might also report a smaller "standard error", which accounts for the sample size, *n*. The standard error is smaller than the standard deviation, and decreases with increasing *n*. The standard error is useful if we think that the errors are random and only the mean value is important. For exoplanets, or PWDs, obtaining repeated observations is not as readily feasible and so a Monte Carlo simulation can be employed to "simulate" the distribution. The Monte Carlo approach simulates the case that many samples are available. The problem is that a mean and standard deviation are required as input, and with limited data, such inputs are uncertain, which is why Hinkel et al. (2016) report the "spread" of their dex values (e.g. for Fe/H). The hope, in any case, is that measurement errors are random and reported "spreads" are closer to a range than a standard deviation, although they might not be. Fortunately, systematic errors have not yet been documented in the astronomical literature (e.g., that Ca/H is always high relative to Fe/H, or that small stars have lower Si/H than large ones), but if such errors are ever discovered they will need to be accounted for.

With these assumptions in place, Figure 9a shows 1,000 simulations, of the PWD known as Ton 345. Figure 9a, projects PWDs into the Fe-free system: $Mg_2SiO_4$-$Mg_2Si_2O_6$-$CaMgSi_2O_6$, which effectively assumes that the mantle is Fe-free. This approach yields errors much larger than shown in the ternary projection of Putirka and Xu (2021), who instead use repeat analyses and independent models of certain PWD compositions (by UV and optical spectra, discussed below).

Figure 9a is a "what if" calculation: "What if we were to re-analyze Ton 345 1,000 times? Given the reported error, what range of compositions might result?". The calculations thus take the place of the analyses that we would otherwise perform if the analyses were easier to obtain.

To interpret a set of Monte Carlo simulations, we must also choose a desired level of certainty: we know that a plotted composition might plot differently if we were to re-analyze it. So what range of compositions should we consider when assessing possible mineral assemblages? The range depends upon how confident one wants to be when discussing such possibilities. The larger the level of confidence we require, the larger the range of compositions we must consider as plausibly real. In Figure 9a, the Monte Carlo results for Ton 345 are contoured at the ±1σ (68%), ±2σ (95%) and ±3σ (99.7%) confidence intervals. If we desire a 68% (±1σ) level of certainty, then we have decided to accept that Ton 345 may have a mineralogy represented by any point that falls within the 68% (red) contour. If we instead decide that only a ±3σ confidence level will do, then we must accept that Ton 345 could have a mineral assemblage that falls anywhere within the much larger 99.7% contour. Note, however, that Fig. 9a says nothing about whether Ton 345 is statistically the same as any other PWD. For that test, we need a new Monte Carlo simulation for a second PWD, to test whether contours might intersect.

A critical aspect of Figure 9a is that the error inherent to Method 1 is trivial in comparison to compositional (input) error. Figure 9a shows a blue field that encompasses the $\pm 3\sigma$ (99.7%) uncertainty region when Model 1 of Table 9 is used to predict mineral abundances when such abundances are known (Figure 7). Treirweiler et al. (2023) imply that Method 1 (Table 9) is the source of the error but the large ellipses in Fig. 9 result from the error on the input compositions. This mis-match in error is the main reason why very simple thermodynamic and mineralogic models are advocated here. Compositional precision is not yet so precise that there is any use in debating the extent to which small amounts of Al partition into Opx, or whether Ti partitions into Ol.

In any case, within the present context, every Monte Carlo simulation requires two inputs and a decision: a composition, an estimate of its uncertainty in the form of a standard deviation and a choice about what level of certainty is acceptable. For all subsequent Monte Carlo tests (Figs. 9b-i) the input compositions represent either a particular PWD bulk composition, or the mean of the 23 PWD compositions used by Putirka and Xu (2021), which are input as dex values (dex units are $\log_{10}$ units, so if a star has Fe/H reported as -3.2 dex, then the Fe/H ratio of the star is $10^{-3.2} = 6.31 \times 10^{-4}$). The errors are then propagated from dex inputs through to the calculation of mineral abundances. As will be shown, the shape and magnitude of the errors plotted within a ternary depend upon other items besides the input standard deviations. As the intent here is to illustrate how certain choices in constructing a ternary diagram affect the plotted errors, every Monte Carlo simulation in Figure 9 thus uses a single set of standard deviations (an average of published uncertainties across 23 PWDs), which in dex notation are: ±0.12 for Fe and Ca, ±0.14 for Mg and ±0.13 for Si (the average of the uncertainties for the 23 PWDs of Putirka and Xu, 2021). Note that these errors lose their symmetry when uncertainties on the dex (log) scale are converted to a linear scale. For example, the contours of the Monte Carlo results for PWD Ton 345 (Fig. 9a) appear symmetric. But the lower panel of Fig. 9a isolates the distribution of olivine, which are skewed to higher abundances. The distribution of olivine is very well fit by a SHASH, or sinh-arcsinh, distribution (Jones and Pewsey (2009), and less so by a normal distribution (which is a special case of the SASH distribution, when $\gamma = 0$ and $\delta = 1$). Until many more PWD compositions are obtained, it's not yet clear that there is much to exploit in understanding these distributions, except to note that for any of the distributions shown in Fig. 9, high, positive olivine abundances are more likely than negative abundances, which fall under a very low probability tail (and similarly, highly positive Opx and Cpx values are under similarly under low-probability tails). Also, the greater the reported dex uncertainty, the greater the asymmetry in the probability density function.

**Quirks of ternary projections, and their plotted errors**

Errors within a ternary diagram are dependent on some not-so-obvious choices about how a given ternary is constructed. Figures 9b-e show the errors on PWD bulk compositions when they are projected into the ternary system using Model 4 in Table 9. This model allows estimates of metallic Fe (Figs. 9b-e) as well as silicate mineral abundances in the mantle, which it does by collapsing the two pyroxenes, Opx and Cpx, into a single pyroxene component (Pyx).

Several results are notable, especially for the very high-Fe PWD known as PG0843+517 which contains 87% Fe in its bulk composition. Errors are small for PG0843+517 because the remaining components are too low to have any great impact on error. For example, in the 1,000 Monte Carlo simulations of PG0843+517, Ca ranges between just 0-3 wt. %; the standard deviation on Ca must therefore be some value less than the range, or < ±3%, even though the error on Ca is identical to that of Fe (±0.12 dex). In short, errors are weighted by their respective elemental abundances. Error on PG0843+517b is also particularly small because one of the ternary phases (Fe metal) matches one of the input components (wt. % Fe). Since Ca controls Cpx, is it possible that the error in Figure 9a is narrow perpendicular to the Cpx apex because the errors on Ca are low? This is a tempting interpretation (Trierweiler t al. 2023) but is incorrect; in every panel in Figures 9a-e, the error assigned to Ca is precisely the same as that assigned to Fe and the errors on the dex scale are not much different than for Mg or Si. Figure 9a shows a narrow distribution perpendicular to the Cpx because Ton 345, like PG0843+517, has low Ca. Incidentally, note also that mole and weight % projections have different magnitudes of uncertainty (Figs. 9b-e) and their contours have different geometries.

> *Tenet 1: In ternary projections, errors are weighted by the relative abundances of the elements, and thus depend upon the input composition, even when differing compositions have the same analytical errors; this also means that extreme compositions can have lower plotted ternary errors in some projections, especially when one of the input components matches a projected mineral component.*

> *Tenet 2: Error depends upon the units used to describe the various projected phases (wt. %, mole % or volume %).*

Note also that errors are elongated in the direction of the Ol-Opx tie-line (Trierweiler et al. 2023) in Figure 9a. This issue is not intrinsic to the method. To test for such error, one can apply the standard deviations of Figure 7 to a Monte Carlo simulation. Figure 7 compares the mineral abundances calculated by Model 1 to the mineral abundances of actual rocks, where bulk and mineral compositions are known and where mineral proportions can be obtained by mass balance. The blue field in Figure 9a shows that Ol/Opx and Ol/Pyx ratios can be determined with comparative accuracy relative to the spread of Ol/Opx in the PWD data set. We can safely conclude that the range in the calculated Ol/Opx ratios among PWDs is not a result of the method applied to determine the PWD mineral abundances. Figures 9b-e also show how compositional errors can be compressed along the Ol-Pyx tie-line, especially in the mole % projection (Figs. 9b-e).

> *Tenet 3: Errors in ternary projections reflect a competition between the plotted mineral components, and the geometry of error ellipses depend strongly on the choice of minerals being projected.*

By collapsing Cpx and Opx into a single Pyx component (Figs. 9b-e), however, we've lost some critical mineralogic information. Experimental studies, for example, show that Cpx-rich systems melt at lower temperatures, and so produce greater amounts of crust at given temperature (Lambert et al. 2016). To regain such information, Figures 9f-g show projections of BSPs for the PWDs of Putirka and Xu (2021), using Model 1 of table 9. Bulk silicate compositions are also calculated using $\alpha_{Fe} = 0.27$. But in this projection, the positions of the plotted points are not affected by the choice of $\alpha_{Fe}$ values (Figs. 9f-g) because Wus (FeO) is used as the apex from which the compositions are being projected. By projecting from Wus, the positions of the plotted points are affected only by the Fe assigned to the minerals in the matrix of Model 1. Any other changes in Fe, e.g., via an assumed $\alpha_{Fe}$ value, will only move points along the projection line (as in Figures 4b-c), with no impact on the positions of the points in the projected (Fe-bearing) plane, Ol-Cpx-Opx.

*Tenet 4: The choice of mineral from which a ternary plot is projected affects whether a given compositional parameter affects the final plotted position.*

Errors in Figure 9g are much smaller than in Fig. 9a for two reasons: first, Fig. 9a is an Fe-free system, whereas, by virtue of applying Model 1 in Table 9a, our mineral compositions in Fig. 9g are Fe-bearing. In addition, we plot our compositions in mole %. The issue with regard to Fe is parallel to that of Figures 9b-e: by allowing Fe into the calculations, there is less room for variations in Mg and Si, especially for Fe-rich compositions. To illustrate, Figure 9g shows 1$\sigma$ errors for PWDs with high Fe (PG0843+517; 24 wt. %), low Fe (NLTT 43806; 4.6 wt. %), and intermediate Fe (PWD G241-6; 7.5 wt. % Fe), the latter of which also has high Mg (56 wt. %). Also plotted is error for the average of all 23 PWD compositions (11 wt. % Fe) of Putirka and Xu (2021). Plotted errors for the high-Fe case are smaller because even at just 24 wt. % Fe in the BSP, the Fe content is still high enough to restrict variation in the remaining three elements (Si, Mg and Ca). The high-Mg case also has less error in the Ol-Opx direction, mostly by virtue of its higher Mg content. A key issue is that errors are thus dependent upon choices made regarding mineral compositions and style of projection.

One final quirk, and one that may be explored to some advantage, is that ternary errors are also affected by the choice of formulas used to describe a given phase (Figs. 9h-i). Pyroxenes are commonly expressed on the basis of 3-oxygens instead of the 6-oxygen basis used in Figs. 9f-g. Figure 9h shows the results of that choice. Errors on the mean and low-Fe compositions are considerably larger than in projections that use pyroxene formulas expressed on the basis of 6 oxygens (Figs. 9f-g). But a more unconventional approach is shown in Figure 9i that reduces error, by expressing olivine compositions on the basis of 6 oxygens instead of the usual 4. This choice allows the molecular weight of olivine to more closely match that of the pyroxenes (so the formula for forsterite, usually expressed as $Mg_2SiO_4$, becomes $Mg_3Si_{1.5}O_6$). Error on PG0843+517 is smaller, for example, but without the concomitant amplification of error elsewhere in the ternary as occurs in Figure 9h.

*Tenet 5: Error can be minimized in some projections when stoichiometric mineral formulas are written so as to equalize weight or molar proportions.*

None of the plots of Figures 9a-i are intrinsically better or worse than any other, nor does any single plot provide a definitive view of error. Thus, none need be preferred— except insofar as they allow one to minimize error, better describe a system, or improve tests of hypotheses.

**Are any exoplanets exotic?**

Understanding how to interpret the uncertainties plotted within Fig. 9 is essential to considering possible exoplanet mineralogies. For example, if we require a ±3σ, or 99.7% certainty (e.g., as in Trierweiler et al. 2023), Figures 9a,f show that a wide range of mineralogies are possible: *our choice of a 3σ confidence interval requires us to consider any mineral assemblage encompassed within the 99.7% ellipse as possibly real*. One could, instead, choose a smaller level of confidence, say 1σ, but even at that level of confidence some PWDs plot outside the ternary.

The above approach treats error as if only one PWD observation is available. But more than one PWD composition has been measured. Should we average all the observed PWDs and treat them as one composition? Or are there multiple PWD compositions that we must consider? Three different tests will be presented.

First, we can compare PWDs to one another, as Figs. 10a-b. Figures 10a-b show mineral abundances for 23 PWD compositions obtained using both Method 1 (Fig. 10a) and Method 2 (Fig. 10b), along with1,000 Monte Carlo results (as in Fig. 9) using the mean PWD composition as input. If the 23 PWD compositions represent a random sampling of effectively equivalent compositions, then the distribution of the observed 23 PWDs compositions should mimic the distribution from the Monte Carlo simulations. Significantly, a large fraction of PWDs plot where the Monte Carlo simulations predict a low density of observations (Figs. 10a-b). And only four of the PWDs overlap with inner Solar System compositions, regardless of whether Method 1 or 2 is employed (Figs. 10a-b). The same result can also be seen in the Fe-free system of Fig. 9a: about half of PWDs there do not fall within that ternary diagram and many plot where the Monte Carlo simulations also predict that few compositions should occur, if we are randomly sampling only one composition (note how adding Fe shifts the positions of the data compared to the Fe-free case of Fig. 9a). It is thus appears that by sampling multiple PWDs we have probably sampled different PWD compositions.

A second test is more quantitative, and can be used to calculate the probability that any PWDs fall outside the ternary diagrams of Figs. 10a-b. The pertinent probability calculation involves the contours of every PWD composition (not shown in Figs. 10a-b), not just one. The simplest case is to calculate the probability that *at least one of two PWDs fall outside the ternary*, (call them A and B), where we can use the equation:

$$P(A \text{ or } B) = P(A) + P(B) - P(A \,\&\, B) \qquad (7a)$$

Here, P(A or B) is the probability that either A or B fall outside the ternary, P(A) is the probability that A falls outside the ternary, P(B) is the probability that B falls outside the ternary, and P(A & B) = P(A)P(B), is the probability that both A and B fall outside the ternary. For the case of whether A, or B or C fall outside the ternary, we have:

$$P(A \text{ or } B \text{ or } C) = P(A) + P(B) + P(C) + P(A \,\&\, B \,\&\, C) - P(A \,\&\, B) - P(B \,\&\, C) - P(A \,\&\, C) \qquad (7b)$$

Equation 7b can, of course, be expanded to cover any number of observations, using terms analogous to those in Equation 7a. If the Monte Carlo simulations for A, B and C indicate that P(A) = 0.7, P(B) = 0.5 and P(C) = 0.3, then the resulting probabilities that at least one PWD falls outside the ternary are: P(A or B) = 0.85, and P(A or B or C) = 0.9. As should be expected, *the* *more PWDs that are observed to fall outside the ternary, the higher the probability that at least one PWD falls outside the ternary*.

A third test is to compare "probability density functions" (or pdf); for illustrative purposes, we'll apply a Gaussian distribution generated from a Monte Carlo simulation, using the global mean and standard deviations as parameters (Figs. 10c-f). We could also use a SHASH distribution (Fig. 9a lower panel), or other distributions, but the Central Limit Theorem states that a random sample from any distribution is likely to approximate a Gaussian distribution, so a Gaussian pdf is a good place to start, especially when the true distribution is unknown. This test examines whether the distribution of observed PWDs (Figs. 10c-f; green histogram) mimics the distribution one would expect if we have a random sample of what is, effectively, a single compositions, or a very narrow range of similar compositions (Figs. 10c-f; orange, Gaussian curve). The gray shaded areas of the Gaussian pdfs (Figs. 10d-f) show the region where mineral abundances would fall within the ultramafic ternary (Fig. 4a); blue bars (Figs. 10c-f) mark values for Earth (or nearly equivalently, CI chondrites). The mean values of the PWD compositions (green values) are not too different from Earth, but not all PWDs are Earth-like, let alone well described by the Gaussian distribution. The poorest fits are for Fe metal, when bulk PWDs are used as input (Fig. 10c) and for Clinopyroxene when BSP compositions are considered (Fig. 10f). In both cases, the observed values extend well outside even the 99.7% confidence intervals. If we applied a SHASH distribution, the observed high Cpx and high Opx cases would be predicted to be even less probable. In any case, the observed ranges for bulk Fe, Opx and Cpx might be real, since the Gaussian pdf suggests that these are very low probability events if error in random. Orthopyroxene is perhaps especially interesting as it shows an observed mode at -20 wt. % Opx (so nominally ferro-periclase normative; Putirka and Xu 2021), which is far below the predicted mean value of +34 wt. %. Olivine abundances are better predicted by a Guassian distribution, as one mode closely approaches the predicted mean; but a second mode at >100% Ol (also nominally ferro-periclase normative) could indicate a bi-modal distribution, or an entirely different mean that is unlike Earth.

The interpretation of Figures 10c-f is highly challenged by the small number of PWD observations; it is conceivable that new observations could well shift the observed abundances so as to better mimic a single distribution, Gaussian or otherwise. But we must interpret the data we have, not the data we wished we had. Regardless, multiple, independent observations are needed.

**Comparing multiple observations of individual PWDs**

Some good news is that multiple and independent analyses of PWDs are available, which indicate vastly less uncertainty than the Monte Carlo simulations based on a single observation (Fig. 9a). The independent compositional estimates of certain PWDs stem from the fact that planet-building elements, Fe, Si, Mg and Ca, can be detected from both the UV and optical portions of the electromagnetic spectrum (e.g., Jura et al. 2012; Klein et al. 2012; Xu et al. 2017, 2019). Optical and UV spectra can also yield different estimates of H and He (as H is determined either by Lyman alpha lines in the UV, or Balmer lines in the optical part of the spectrum), and estimates of H and He in turn affect estimates of the planet-building elements (see Xu et al. 2017). Optical and UV spectra thus provide highly independent tests of reproducibility. These tests are more than just differences in curves fitted to different parts of the electromagnetic spectrum. As explained by Xu et al. (2019), optical and UV results also rely on different models of white dwarf structure, as the UV and optical lines emanate from different depths in PWD atmospheres. Figure 11 shows that in some cases, though by no mean all, UV and Optical data (and some repeat observations by one or the other method) are in remarkable agreement.

It should be noted that some astronomers (e.g., Rogers et al. 2024) are greatly concerned about order-of-magnitude differences between UV and optical spectra for some elemental ratios (e.g., Si/H). But such contrasts matter less than at first appearances: while Fe/H and Si/H might differ greatly, Fe/Si ratios might still be very similar (Xu et al. 2017, 2019). The PWD PG0843+517 is illustrative: UV spectra (Gansicke et al. 2012) yield Fe/H = -4.6 dex, while optical spectra (Xu et al. 2019) yield Fe/H = -3.84 dex, and astronomers are naturally concerned about this nearly order-of-magnitude difference in Fe/H. But when estimating mineral abundances, it's not the absolute dex values of Fe/H, or any other given element, that matter so much as the relative values of Fe/H, Mg/H and Si/H. For PG 0843+517, the optical spectra not only yield higher Fe/H but also higher Mg/H and Si/H; as a result, the weight % values of Fe are not very different: 87.9 wt. % from the optical data and 73.5 wt. % from UV. Both estimates of Fe for PG0843+517 are well above the mean Fe (42.4 ± 22.3 wt. %) of the 23 PWDs examined by Putirka and Xu (2021). Averaging the two estimates for PG 0843+517 yields 80.7 ± 10.2 wt. %. The order-of-magnitude dex-scale difference in Fe/H thus collapses to a <20% difference in Fe on a wt. % basis, and a ±10.2 wt. % uncertainty for the Fe content of PG 0843+517. On the basis of this averaging, one can reasonably conclude that PG 0843+517 is very likely enriched in Fe relative to other PWDs. The repeat observations of PG 0843-517 clearly bolster this view.

Figure 11 shows mineral abundance estimates for PG 0843+517 and six other PWDs where multiple published abundances are available. Except for PG 1015, the range of reported compositions from independent measurements is considerably less than the errors indicated in Figs. 9a. The mean and median errors shown in Fig. 11 are indeed greater than the uncertainties shown in similar plots in Putirka and Xu (2021), where a similar analysis, but using only PG 1225 and HS 2253, was applied. However, even in this expanded comparison, the mean and median error ranges, which include the very large errors in PG 1015, are less than even the 1$\sigma$ errors from the Monte Carlo simulations (Figs. 9a-i), which implies that mean values obtained via one method (UV) can be faithfully duplicated by another (Optical). (Note that PG 0843+517 exhibits a wider mineralogical range than most, but this is not due to its <20% error on Fe, but rather the 400% difference in Mg: 4 % vs. 16%, respectively for UV and optical results).

Figure 11 also provides a test of whether any PWDs are exotic, i.e., whether they have mantles that fall outside the ultramafic ternary. PG0843 was considered as exotic by Putirka and Xu (2021), but taking the mean of the two observations, it looks to fall within the ternary. But PG1225 plotted within the ternary in Putirka and Xu (2021), but the additional estimates here indicate that it might be exotic. This example harkens back to a key point about the Monte Carlo simulations: we must be prepared to accept that some PWDs that fall within the ternary, might actually be exotic, once more analyses are obtained. In any case, four of the seven PWDs (WD 1425, GD 40, G 241-6 and HS 2253) with multiple composition estimates fall entirely outside the ultramafic ternary.

The above arguments can only be controversial to the extent that reproducibility is an unreliable measurement strategy. For the future of exoplanet research, reproducibility should, of course, be a top priority.

**Do Methods 1 and 2 agree in predicting PWD mineralogy?**

A final aspect of reproducibility is to test whether independent methods, such as the two introduced in this chapter, are capable of yielding similar results. And they do. Of the 23 PWDs, eight fall entirely within the ultramafic ternary using Method 1, and for these 8 compositions, the molar abundances calculated by Methods 1 and 2 correlate with $R^2$ values of 0.8, 0.95 and 0.93 for Ol, Cpx and Opx are respectively. Although well correlated, Method 2 yields systematically higher Cpx and lower Opx values than Method 1.

Of the 23 PWDs in Figures 9a-i, five are predicted to have Ol<0 when Method 1 (Model 1 of Table 9a) is applied. For agreement, Method 2 should then yield Ol = 0 for these cases and it does so for four of the five (80%). Putirka and Xu (2021) hypothesized that such systems could be Qz-saturated (not a necessary result of Method 1) and Method 2 also predicts Qz>0 for two of the five cases (where Ol<0 by Method 1). But Method 2 allows any excess Si (step 15) to react with excess Fe (from step 8) to form Ferrosilite (Fs; step 17), and Fs >0 for the other two cases where Ol = 0 by Method 2. If Fs were unstable for any reason (or removed from the calculation), then all 4 cases of Ol = 0 would indeed yield Qz>0 by Method 2 (and any excess Fe from step 8 would partition into a metallic Fe phase, i.e., a metal core). Another test involves the 10 PWDs that are predicted to have Opx<0 by Method 1. Method 2 should predict Opx = 0, and it does so for 80% of these cases. Putirka and Xu (2021) predicted that when Opx<0 by Method 1 such systems would be saturated with ferro-periclase, and Method 2 predicts ferro-periclase saturation for all eight cases where Opx = 0.

Methods 1 and 2 are thus likely to yield a similar classifications of exoplanets, even when such planets have compositions that are exotic to the ultramafic ternary diagram (Figure 4).

## THERMODYNAMIC CONCEPTS & RELATED CAVEATS

The goal of Putirka and Xu (2021) was to show, in part, that some exoplanets may well fall outside the ultramafic ternary and that conclusion seems to be well-founded, at least thus far. But like Cross et al. (1902) we are under no illusion that our published mineral abundances represent precise estimates of exoplanetary rock types. The reasons that warrant caution are several, including the simplicity of our thermodynamic models, the potential that currently unknown phases might be stable in exotic compositions, and because even in non-exotic cases, other components, such as O, S and C, might occur in sufficient abundances to require new phases or altogether novel mineral assemblages.

### Solid solution ("*All Models are wrong; some are useful*" – George Box)

Minerals are defined as naturally occurring substances that have an ordered atomic arrangement and a definite chemical composition. However, "definite" does not mean "fixed". To take a contrary example, a liquid has no fixed stoichiometry, or rather no fixed ratios of certain elements, so there are no definite formulas that we use to describe them. Liquids then, are not minerals. In contrast, materials such as quartz ($SiO_2$), corundum ($Al_2O_3$) or forsterite olivine ($Mg_2SiO_4$) have very definite ratios of, respectively, Si/O, Al/O and Mg/Si, among others. The formulas we write are, in effect, recipes, needed to make a given mineral. But there is some flexibility in most recipes. This variability is referred to as "solid solution".

Olivine is a useful example as it is the dominant mineral in the mantles of the inner planets and its chemistry is not terribly complex. Olivine has two common end-members, the Fe-end member, fayalite (Fa; $Fe_2SiO_4$) and the just-mentioned Mg-end member, forsterite (Fo). For Fa, its stoichiometry indicates that Fe is in the $Fe^{2+}$ state. Both $Fe^{2+}$ and $Mg^{2+}$ fill similar sites within the olivine structure, which are referred to as "M1" and "M2" (for "metal 1" and "metal 2"; Fig. 12a). The M1 and M2 sites are called "octahedral" sites because Mg and Fe are bonded to 6 oxygens that together create an 8-sided octahedron when we draw lines connecting the oxygens. Magnesium and Fe happily partition into these octahedral sites, as do other elements that have the same charge and similar ionic radii, i.e., Mn, Ni, and to a lesser extent, Ca. The smaller $Si^{4+}$ atom occurs within a tetrahedral (or "T") site (Fig. 12a), where Si is bonded to 4 oxygens that create a 4-sided polyhedron (Fig. 12a). (Many other substitutions are possible; for example, very small amounts of $P^{5+}$ can partition into the T site and $Al^{3+}$ can enter both the T and M1 sites, with charge balance being taken up by vacancies or $Fe^{3+}$ in the M1; see Shea et al. 2019 and references therein). The M1 site is a tad smaller than M2 and (so the larger $Ca^{2+}$ will favor M2 while the smaller $Ni^{2+}$ will favor M1), but for the case of $Fe^{2+}$ and $Mg^{2+}$ their radii are sufficiently similar so as to yield no energetic preference for either Fe or Mg into the M1 or M2. This ease of partitioning of Mg and $Fe^{2+}$ allows olivine to exhibit "complete solid solution" between the Fo and Fa end-members, meaning that a natural olivine can have any composition, from pure Fo to pure Fa, or any mixture in between.

Perhaps the most glaring simplification of Method 1 and the projections of Figures 9-10 is that minerals are assigned a constant composition, regardless of the bulk composition with which they are equilibrated. So long as the magnitude of compositional error so greatly outweighs the error associated with calculating mineral abundances (Fig. 9a), the simplification will do. But if compositional error decreases, the limitations of this approach may be exposed. To give just one example, experiments by Zhao et al. (2018) indicate that the viscosity of Mars' Fe-rich mantle may be 5 times less viscous than the relatively Fe-poor terrestrial mantle—and such differences in viscosity can greatly affect whether or not we predict a planet to exhibit terrestrial-like plate tectonics (e.g., Weller and Lenardic, 2018). So when applying either of Methods 1 or 2, it might be critical to use a separate model—adjusting Fe contents in the matrix of Method 1 or in the Mg# of Method 2—for each bulk composition, especially if mineralogical outputs are used as inputs into numerical models that simulate mantle circulation.

The amounts of Fe and Mg in olivine are very sensitive to temperature, as well as liquid composition, these two variables being intimately connected. And for exoplanets that equilibrate at different temperatures, *T*-dependent models could also be useful. Figure 12b shows a series of diagrams that compare Gibbs Free Energy (*G*) to composition (*X*), where *X* can range from pure Fa on the left, to pure Fo on the right. In the topmost panel, which shows the case for a high T, noted as T1, the liquid has a lower free energy at every possible value of *X*, and so the liquid is stable for all compositions. In the *T-X* phase diagram in the bottom panel, the region in red represents a range of *T-X* where only liquid is stable and olivine will not crystallize. As *T* decreases (from T2 down to T4 in Fig. 12b), the free energy curve for olivine decreases at a faster rate than for liquid, and it intersects that of the liquid at some *T* between *T1* and *T2*. The compositions of the liquid and equilibrium olivine can be found from the common tangent of the liquid and olivine *G-X* curves (panel *T2*), which are projected downward onto the *T-X* diagram to show the equilibrium liquid and olivine compositions in *T-X* space. With further decreases in *T* (*T3*), the positions of the common tangent points shift towards more Fe-rich compositions. And when *T* is low enough (*T4*), olivine has a lower Gibbs Free Energy than the liquid at all compositions and only olivine is stable, as in the green area of the *T-X* diagram in Fig. 10b.

The *G-X* diagrams illustrate how the *T-X*, or binary phase, diagram of Figure 12b is produced, but the *T-X* diagram (Fig. 12c) is sufficient to describe the equilibrium state of any system. For example, the bulk composition (*BC*) in Figure 12c, is 60% Fo (and thus 40% Fa). Upon heating, it would begin melting at 1480ºC, and the first bit of liquid to be produced would have a composition of 22% Fo (or Fo22). At 1600ºC, the system is partially melted, with the liquid having a composition of Fo33, which is in equilibrium with crystals of Fo74. With further heating, the *BC* would be completed melted at 1850ºC and the liquid would be identical in composition to *BC*, while the last bit of solid to dissolve (melt) into the liquid would have a composition of Fo97.

Other minerals besides olivine, but still common to Earth's mantle, exhibit solid solution. For exoplanet studies it's probably safe to ignore the minor element solid solution series and concern ourselves only with the overwhelmingly dominant Fe-Mg solid solutions that occur within olivine, the pyroxenes and garnet, but other potentially important solid solution series are shown in Table 11.

**Table 11.** *Solid Solution Series*

| Mineral Series | Mineral Names | Abbreviations | End Member Formulas | | Exchange Reaction |
|---|---|---|---|---|---|
| Olivine | Forsterite - Fayalite | Fo-Fa | $Mg_2SiO_4$ | Fe2SiO4 | $Mg^{2+} \Leftrightarrow Fe^{2+}$ |
| Orthopyroxene | Enstatite - Ferrosilite | En-Fs | $Mg_2Si_2O_6$ | $Fe_2Si_2O_6$ | $Mg^{2+} \Leftrightarrow Fe^{2+}$ |
| Clinopyroxene | Diopside - Hedenbergite | Di-Hd | $CaMgSi_2O_6$ | $CaFeSi_2O_6$ | $Mg^{2+} \Leftrightarrow Fe^{2+}$ |
| | Diopside - Ca Tschermak | Di-CaTs | $CaMgSi_2O_6$ | $CaAl_2SiO_6$ | $Mg^{2+}Si^{4+} \Leftrightarrow 2Al^{3+}$ |
| | Diopside - Jadeite | Di-Jd | $CaMgSi_2O_6$ | $NaAlSi_2O_6$ | $Ca^{2+}Mg^{2+} \Leftrightarrow Na^{1+}Al^{3+}$ |
| Plagioclase Feldspar | Anorthite - Albite | An-Ab | $CaAl_2Si_2O_8$ | $NAlSi_3O_8$ | $Ca^{2+}Al^{3+} \Leftrightarrow Na^{1+}Si^{4+}$ |
| Alkali Feldspar | Albite - Orthoclase | Ab-Or | $NaAlSi_3O_8$ | $KAlSi_3O_8$ | $Na^{1+} \Leftrightarrow K^{1+}$ |
| Garnet | Pyrope - Almandine | Py-Al | $Mg_3Al_2Si_3O_{12}$ | $Fe_3Al_2Si_3O_{12}$ | $Mg^{2+} \Leftrightarrow Fe^{2+}$ |
| | Pyrope - Grossular | Py-Gr | $Mg_3Al_2Si_3O_{12}$ | $Ca_3Al_2Si_3O_{12}$ | $Mg^{2+} \Leftrightarrow Ca^{2+}$ |
| Oxides | Periclase - Wüstite | Per-Wus | MgO | FeO | $Mg^{2+} \Leftrightarrow Fe^{2+}$ |

Although Methods 1 and 2 appear useful (Figs. 7-8, 9a), some significant fraction of the predictive error (Figs. 7-8) is assuredly due to the lack of treatment of solid solution. For example, Cpx abundances are under-predicted by both models, probably because insufficient Al is assigned to Cpx. Again, it seems vastly premature to speculate on the amounts of Al that may be partitioned into exoplanetary Cpx given the large input errors of Fig. 9a. But if the need arises, the CIPW norm-like approach of Method 2 can and should be modified to accommodate such solid solution effects. If we are ever in a position to evaluate crust and mantle compositions on exoplanets, accounting for *T*-sensitive partitioning may be crucial and the kind of modelling that is conducted within MELTS (Ghiorso et al. 1995; Asimow et al. 1998), if properly tested against known cases, should improve predictions of mineral abundances and melting behavior.

## Gibbs-Duhem equation

The Gibbs-Duhem equation provides a caution flag when comparing the mineralogy of planets of different bulk compositions. The equation is a mathematical statement that all intensive parameters (pressure, temperature and chemical activities) are interconnected—a sort of geochemical Indra's Net. The equation is often expressed as a function solely of composition. For a system at equilibrium and so at constant *P* and *T*, the Gibbs-Duhem equation is:

$$\sum_i n_i d\mu_i = 0 \quad (8)$$

where $n_i$ is the number of moles of component i in a given phase and $\mu_i$ is the chemical potential of i. The conceptual import of equation 8 is that with any change in the number of moles of any component in a system there must be consequent changes in chemical potentials, so that the sum remains 0 for the system to approach equilibrium. Since a Gibbs-Duhem equation applies to every phase in a given equilibrium assemblage, the equation can be used to derive the phase rule, which provides the degrees of freedom ($f$), for a system of $c$ components and $p$ phases; $f$ is the number of intensive variables ($P$, $T$, mole fraction of a given component)) that can be varied independently without affecting the number of phases, $p$, in the system:

$$f = c - p + 2 \quad (9)$$

The Gibbs-Duhem equation thus applies to both homogenous (a single phase) and heterogenous (multi-phase) systems. Since Gibbs Free Energy can be expressed as:

$$dG = VdP - SdT + \sum_i \mu_i dn_i \quad (10)$$

several substitutions lead to the following equation that expressly illustrates the interdependencies of $P$, $T$ and $\mu_i$ (see Denbigh 1981):

$$VdP - SdT = \sum_i n_i d\mu_i \quad (11)$$

and thus if an equilibrated system is perturbed in any way, by a shift in $T$, $P$ or composition ($d\mu_i$) there must be a reactive perturbation in any or all these variables to maintain the equality of Eqn. 11.

Equations 10-11 are as a reminder that if one exoplanet has a different composition than another, then even at equivalent $P$-$T$ conditions, we can be certain that there are mineralogical differences also. The recommended standard state (i.e., $P$-$T$ conditions of Earth's upper mantle, or ca. 2 GPa and 1350$^{o}$C) is a recommendation to focus on the compositional—and consequently mineralogical—implications.

Small compositional shifts might be accommodated entirely by solid solution. But small differences also have the potential to precipitate a wholesale change in phase assemblage. To illustrate, Figures 13a-b show a hypothetical case where Si and Mg vary within a range that might describe Earth or the inner planets. Throughout the ranges in MgO and $SiO_2$ in Earth's mantle Ol and Opx are stable. Less certain (until we conduct experiments to find out) is whether other phases may be lurking in the near compositional distance. For example, at $SiO_2$ greater than observed in Earth's mantle (Fig. 13c) we might calculate that Qz is stable, when instead another phase (x) may have a lower Gibbs Free Energy and so would displace Qz from the equilibrium assemblage. Similarly, we might calculate that periclase is stable as MgO contents increase, but another phase (y), might replace Per (Fig. 13d). The same issue applies to any of the planet-building elements, which means that mineral assemblages could look quite different than as calculated by Methods 1 and 2.

## O, C, S, H & other unknowns

Oxygen is far and away the dominant anion in the rocky materials that comprise the inner Solar System. Its abundance also provides a first-order control on the fraction of Fe that will be partitioned into a planet's metal core. But H, C, and S, are also highly abundant in stellar atmospheres and large concentrations of any of H, C, O, or S could play havoc with estimates of exoplanet mineralogy.

***Oxygen (and Fe)*** Despite its importance, O is rarely if ever measured in rock or mineral samples. The reason is because in X-ray fluorescence (XRF) spectroscopy (the most common method of rock analysis) samples are bombarded with so-called "primary" X-rays that interact with the inner electron shells of target elements to induce "secondary" X-rays, via fluorescence. For light elements, such as O, the energies of the secondary X-rays are very low and more readily absorbed by the sample, leaving lesser amounts of energy to reach a detector. Since O is the dominant anion in nearly all rocks and minerals, though, total O contents can be calculated by charge balance, once all the major cations have been quantified. In geological contexts, O is thus largely known by calculation, the only prominent exception being partial melting experiments conducted at 1-atm pressure, where an experimenter can dictate the $O_2$ content of the atmosphere. In experimental studies, the amount of $O_2$ is a system is often expressed as "oxygen fugacity" or $fO_2$, which is the partial pressure of $O_2$ in the atmosphere that is in equilibrium with a given system—usually a magma. The two quantities, $fO_2$ and total O content, are by no means identical, but when a system is in contact with an atmosphere of a given O content, they are certainly related to one another.

Because many systems of interest (e.g., Earth's mantle) are not in equilibrium with a vapor phase, nearly all discussions of oxygen fugacity in Geology are fictive, and hinge upon measurements of Fe. Iron occurs in two different oxidation states, $Fe^{2+}$ and $Fe^{3+}$, in silicate rocks and minerals. Oxygen contents, or $fO_2$, are thus most commonly inferred from $Fe_2O_3$/FeO ratios (based on experiments where the systems are equilibrated with an atmosphere of known $fO_2$). Several publications provide excellent summaries of the fictive $fO_2$ of bulk Earth and Earth's upper mantle (e.g., Wade and Wood 2005; Corgne et al. 2008; Frost and McCammon 2008), including relevant thermodynamic theory, and Stolper et al. (2022) and Guimond et al. (2023) illustrate how mantle mineralogy and $fO_2$ are intimately related to one another. Doyle et al. (2019) apply this theory to interpret measurements of O contents of several white dwarfs. Interested readers are also highly encouraged to read the very accessible discussion of fugacity by its inventor, Gilbert Lewis (Lewis 1901), who devised the concept to link chemical transfer processes to Gibbs Free Energy. For the purposes of extrasolar planet studies, Fe/FeO ratios can be inferred from either Fe-Si (as in the Method 2 above) or perhaps Fe-O mass balance in some white dwarfs (Sun-like stars have an excess of O relative to rocky planets and so place no immediate constraint on rocky planet O contents; Figure 14) .

But even here, there are significant issues in understanding Fe-O systematics in our own Solar System. Total oxygen contents appear to increase with heliocentric distance (Table 3) while Fe decreases (Table 3). The increase in O from Mercury to Vesta is expected because O is a volatile element, i.e., its 50% condensation temperature ($T_{50\%}$ = 183 K) from the Solar nebula is well below the $T_{50\%}$ (>1300 K) that characterizes the major elements that make up rocky planetary objects (Lodders 2003; Wood et al. 2019). Iron appears to behave independent of its $T_{50\%}$ value of 1334 K. McDonough and Yoshizaki (2021) hypothesize that the decrease in Fe with heliocentric distance might be controlled by the Sun's magnetic field during planetary accretion. The heliocentric trends for both O and Fe could be less dramatic for the early Solar system if Mercury has indeed lost a portion of its silicate mantle to late bombardment (Helfrich et a. 2019). In any case, gradients in Fe and O yield significant differences in planetary mantle compositions in our own Solar System: Mars and Vesta have considerably more FeO in their silicate mantles, for example, compared to Earth (Table 4), even though both planets have less Fe overall (Table 3). Similar patterns may apply elsewhere and so being able to predict Fe/O ratios amongst the inner planets, as well as in other planetary systems, is critical.

It should perhaps also be noted that the O contents in Table 3 are minimum estimates; those O contents, for the sake of a first-order comparison, are calculated assuming that all Fe exists as FeO. Correcting these is not a simple matter. Hawaiian lavas indicate that on a weight % basis the ratio $Fe_2O_3/(FeO + Fe_2O_3)$ is about 10% (Rhodes and Vollinger 2005). But Wood et al. (2006) posit that Earth's lower mantle might contain significant amounts of $Fe^{3+}$, in the form of the perovskite-structured mineral $FeAlO_3$. However, the $Fe_2O_3/(FeO + Fe_2O_3)$ ratio of the lower mantle, and even its precise bulk composition, are still unclear. Seismologists have sometimes argued that Earth's lower mantle is enriched in Fe (e.g., Anderson and Jordan 1970; van der Hilst and Karason 2000), or $SiO_2$ (e.g., Murakami et al. 2012), although later studies tend to contradict such hypotheses (e.g., Davies, 1974; Irfune et al. 2010; Hyung et al. 2016). Nevertheless, the Wood et al (2006) hypothesis is important to exoplanet studies as it implies that final planet size can affect total O contents, even without affecting the total mass of a metallic core: high-$P$ crystallization (high enough to stabilize perovskite-structured phases, at ca $P$ >23 GPa) from a deep a magma ocean could have increased the oxidation state of Earth's mantle after core formation was largely completed. Their model addresses the otherwise puzzling case of why the more O-rich Mars has a mantle with even lower $Fe_2O_3/(FeO + Fe_2O_3)$ than Earth (Herd et al. 2002; Righter et al. 2008): the answer may be that Mars is too small to allow for high-$P$ perovskite phases to crystallize to any great degree. But Mercury provides yet another confounding case: it is the most Fe-rich of the inner planets, and yet appears to have almost no FeO in its mantle (Nittler et al. 2018), despite having sufficient O to form other silicate minerals.

The first-order problem for estimating exoplanet mantle mineralogy is the extent to which Fe is partitioned into the core as Fe metal. This problem is closely followed by the effect of total O abundance (nominally $fO_2$) on the partitioning of H, C, Si, S, and other light elements into the core. One very clear prospect for future advances is to (a) better interpret spatial gradients in Fe, O and $Fe_2O_3/(FeO + Fe_2O_3)$ ratios within our own Solar System, (b) determine whether any such models could apply to other star-planet systems and (c) model such gradients, and combine these with nominal exoplanet bulk mineralogies so as to model how $fO_2$ might vary in exoplanet upper mantles (Stolper et al. 2022; Guimond et al. 2023).

***Carbon*** Carbon is heavily depleted in Earth relative to the Sun, but that pattern might not apply to all exoplanets. In a survey of 499 stars Bond et al. (2010) noted that most have C/O ratios higher than the Solar value (0.54), averaging 0.77 and ranging to > 1.0; they posit that when C/O>0.8, Si might occur largely as SiC rather than $SiO_2$ in circumstellar planet building materials. Suarez-Andres et al. (2018) re-surveyed the same 499 stars and concluded that none have C/O > 0.8, but nonetheless show that C/O ranges well above the Solar value, approaching 0.8. Kuchner and Seager (2005) also suggest that C/O can locally approach 1.0 within a circumstellar disk, even if a bulk stellar value is much lower, allowing SiC to be a major constituent of some exoplanets. The mineralogical implications are unclear. Experiments conducted by Allen-Sutter et al. (2020) show that if water is present, SiC will, at high pressures, dissociate into $SiO_2$ and diamond, and so perhaps even in high C/O planets, carbide phases might be rare, although diamond might be rather abundant. No less interesting are experiments by Hakim et al. (2018) who show that the highly reducing conditions needed to stabilize carbides could be detected through the oxidation state of Fe: detection of either $Fe^{2+}$ or $Fe^{3+}$ (as opposed to $Fe^{o}$) at a planet's surface would, in effect, be sufficient to negate a carbide-dominated mantle.

The issue of carbide and diamond contents appears to be not quite fully settled. Most of these studies also make the mistake of using Mg/Si to estimate the ratios of olivine (Mg/Si = 2 in forsterite) to pyroxene (Mg/Si = 1 in enstatite). But as the example of Mars shows, Fe can be highly abundant in silicate mantles and so the more relevant ratio is (Mg+Fe)/Si (see Putirka and Rarick 2019). Calcium contents are also frequently not so low as to be negligible in calculating Ol/Pyx ratios. New experiments on Fe-bearing, C-rich systems, with varying O contents are likely to yield additional results on the possibly mineralogical variety of exoplanetary interiors.

***Sulfur & Hydrogen*** Relative to the Sun, Earth is depleted in all three of O ($T_{50\%}$ = 183 K), C ($T_{50\%}$ = 40 K) and S ($T_{50\%}$ = 672 K) (Table 3; Fig. 14), but the surface of Mercury is intriguingly enriched in S, containing up to 4 wt. % (Nittler et al. 2011)— despite being much closer to the Sun. Moreover, because surface rocks of Mercury have almost no Fe, it is likely that S is bound up as Mg or Ca sulfides (Nittler et al. 2011) and Lark et al. (2022) suggest that Mercury's mantle may contain 7-11 wt. % sulfides. Modelling by Bourke et al. (2019) further indicates that an early magma ocean on Mercury could contain sulfide-rich layers that might be enriched in heat producing elements. And such mineral assemblages are likely to have mechanical and melting behaviors that are very distinct from S-poor, silicate mantles that are the basis of our current models. These findings also indicate that 50% condensation temperatures are an imperfect predictor of a planet's share of any particular planet-building element, and that S-rich and Fe-poor interiors are perhaps just one example of a range of possibilities for how planets may differ from Earth.

As to H, the inner planets of our Solar System have retained only a miniscule fraction relative to the Sun. But some early experiments (e.g., Fukai et al. 1982) indicated that at pressures as low as 3 GPa, significant amounts of H can partition into metallic Fe. Thus, Fe droplets sinking to form a metallic core could yield a metal core that contains H as a light-alloying element. Tagawa et al. (2016) showed that the outer core cold contain as much as 0.3 wt. % H, among other light-alloying elements. And experiments at very high P (Yang et al. 2022) indicate that FeH could be stable within Earth's inner core. It thus seems that a very wide range of core compositions are possible, including wide ranges in H (as well as S and O) in exoplanetary systems.

# ARE WE BUILDING SAND CASTLES OR CATHEDRALS?

Two different methods for estimating mineral abundances (Methods 1 and 2) accurately predict mineral abundances for natural samples, and yield good agreement when applied to exoplanets. But there are many caveats and assumptions for the latter to be accurate. If the results are so uncertain, why bother? Because it is interesting. And we have no choice. Everything we want to know about a planet – it's ability to move heat, store water, break into mobile plates, and exsolve oceans – are dependent entirely upon the minerals that make up its interior; no modeling of any use can happen absent mineralogic estimates. And for exoplanets that do have Earth-like amounts of H, C, O and S, and whose bulk compositions fall within the ultramafic ternary, our models have an excellent chance of predicting accurate upper mantle mineral abundances and core mass fractions (Figures 7-8, 9a). Perhaps our thermodynamic also extrapolate better than expected, but mineralogic estimates nevertheless point us towards the kinds of experiments that need to be conducted (e.g., Zhao et al. 2018; Brugman et al. 2021). A standard mineralogy also arranges exoplanets into classes *relative to their potential to be Earth-like*. We might never know whether a given planet has plate tectonics or granitic crust—or even its true mantle mineralogy—but numerical models that are mineralogically informed can provide a list of plausible candidates of Earth-like analogues, and should inform us of how non-analogues might evolve. Mineralogical estimates also reveal the gaps in our current understanding of Earth and the inner planets.

We have also made certain progress. Now set aside is the use of "Earth-like" to refer to any exoplanet with a rocky exterior; we must appreciate that Mercury, Venus, Earth and Mars are quite different from one another. Our next focus should perhaps concern what qualifies as "exotic", and how we decide when and if an exotic planet has been identified. For example, in a survey of >4,000 star compositions from the Hypatia catalog (Hinkel et al. 2014), Putirka and Rarick (2019) showed that elements such as Al, Ti or Cr, for example, are too low in abundance relative to Mg, Fe and Si, to allow any exoplanetary mantles to consist largely of corundum or rutile or eskolaite. Rather, familiar minerals, such as olivine and the pyroxenes, are likely to dominate, perhaps augmented by sulfides or carbides. However, Putirka and Xu (2021), argue (see also Fig. 11) that some exoplanetary mantles are not contained within the ultramafic ternary diagram (Fig. 4a); that diagram likely describes most or all the inner planets (Mercury being the most plausible exception). But even a wehrlite or a websterite mantle (which fall within but at the margins of Fig. 4a) would be exotic, let alone a mantle with abundant quartz or periclase. These could exhibit very different kinds of crust, oceans and tectonic processes than any planet in our Solar System. Errors on any given polluted white dwarf (Fig. 9-11), as well as repeat analyses (Fig. 11) show that we cannot exclude more extreme and exotic compositions.

But we need a better understanding of why our own Solar system looks the way it does. When estimating exoplanet mineralogy we are, in effect, conducting "what if" calculations: what if a planet is Earth-sized, or has an Earth-like core, or an FeO/O ratio that is like Mars, or has Mercury-like S content? But will extra-solar composition gradients mimic those of our Solar System. Or are other gradients possible? And how might we predict such? Recall also that PWDs, when passing through the Red Giant phase, obliterate their inner planetary materials; PWDs thus record the equivalent of, outer, not inner Solar system materials. How well do we know the interior compositions of Jupiter or Saturn or the moons that orbit them?

The August 1990 issue of the journal *Geophysical Research Letters* provides yet another note of caution: there, authors published predictions of what the Magellan mission might find once it arrived at Venus on August 10, 1990. The resulting Magellan data would reveal a surface vastly different than expected, illustrating the challenges in extrapolating terrestrial processes to other planets – even Earth's nominal "sister planet". The MESSENGER mission to Mercury has yielded no fewer surprises, including a S-rich, and notably Fe-free surface, on the most Fe-rich of the inner planets. If Earth-centered knowledge extrapolates poorly to Venus or Mercury, how well does our Solar system extrapolate to other stellar systems? Even if our Solar system is representative, the assumption is only useful to the extent that we understand the inner planets.

And we still debate whether Earth acquired its oceans from comet-like bombardments or degassing of its interior. The time range for the initiation of terrestrial plate tectonics covers 80% of Earth's history. The composition and mineralogy of Earth's lower mantle is only roughly understood, and debate still occurs as to the major light alloying elements of Earth's core. Current thoughts on exoplanets are clearly preliminary and contingent.

Exciting days, though, lie ahead. As explained in Guimond et al. (this issue) the James Webb Space Telescope (JWST) may reveal to us the interior compositions of exoplanets, via measurements of their atmospheres. Fortin et al. (2022), for example, show how modeling of JWST reflectance spectra should yield very useful insights regarding both planetary surface and interior behavior. Guimond et al. (2021) are also investigating how Mg/Si ratios affect the amounts of water stored in planetary interiors. And new experiments by Brugman et al. (2021) could be used to revise thermodynamic models, to better explore exoplanet-scale ranges in composition. We also still know next to nothing of Venus, and the partial melting and rheological experiments needed to predict the behavior of Mercury and Mars have mostly yet to be performed. This all leaves much exciting work to be performed, especially if we avoid being too comfortable with existing results, models and paradigms.

## ACKNOWLEDGEMENTS

Many thanks to Oliver Shorttle and Larry Nittler for their careful reading of the manuscript, their very thoughtful comments and helpful suggestions. This work was supported by NSF grant 1921182.

## REFERENCES CITED

Allen-Sutter H, Garhart E, Leinenweber K, Prakapenka V, Greenberg E, Shim S-H. (2020) Oxidation of the interiors of carbide exoplanets. The Planet Sci J 1:39

Agee CB, Li J, Shannon MC, Circone S (1995) Pressure–temperature phase diagram for the Allende meteorite. Jour Geophys Res 100:17725–17740

Agee CB, Walker D (1988) Static compression and olivine flotation in ultrabasic silicate liquid. Jour Geophys Res 93:3437-3449

Anderson DL, Jordan T (1970) The composition of the lower mantle. Phys Earth Planet Int 3:23-35

Andrault D, Bolfan-Casanova N, Nigro GL, Bouhifd MA, Garbarino G, Mezouar M (2011) Solidus and liquidus profiles of chondritic mantle: Implication for melting of the Earth across its history. Earth Planet Sci Lett 304:251–259 https://doi.org/10.1016/j.epsl.2011.02.006

Asahara Y, Kubo T, Kondo T (2004) Phase relations of a carbonaceous chondrite at lower mantle conditions. Phys Earth Planet Int 143-144:421-432

Asimow PD, Ghiorso MS (1998) Algorithmic Modifications Extending MELTS to Calculate Subsolidus Phase Relations. Am Mineral 83:1127-1131

Baedecker PA, Wasson JT (1975). Elemental fractionations among enstatite chondrites. Geochim Cosmochim Acta 39:735–765

Barrat J-A, Chaussidon M, Yamaguchi A, Beck P, Villeneuve J, Byrne DJ, Broadley MW, Marty B (2021) A 4,565-My-old andesite from an extinct chondritic protoplanet. PNAS, 118:e2026129118.

Bell DR, Ihinger PD, Rossman GR (1995) Quantitative analysis of trace OH in garnet and pyroxenes. Am Mineral 80:465-474

Birch F (1964) Density and composition of the mantle and core. J Geophys Res 69:4377–4388

Bodinier J-L, Garrido CJ, Chanefo I, Bruguier O, Gervilla F. (2008) Origin of pyroxenite-peridotite veined mantle refertilization by reactions: evidence form the Ronda Peridotite. J Petrol 49:999-1025

Bond JC, O'Brien DP, Lauretta DS (2010) The compositional diversity of extrasolar terrestrial planets I. *In situ* simulations. Astrophys J 715:1050-1070

Boukaré C-E, Parman S, Parmentier EM, Anzures BA (2019). Production and preservation of sulfide layering in Mercury's mantle. J Geophys Res Planets, 121:https://doi.org/10.1029/2019JE005942

Boyet M, Bouvier A, Frossard P, Hammouda T, Garcon M, Gannoun A (2018) Enstatite chondrites EL3 as building blocks for the Earth: the debate over the $^{146}$Sm-$^{142}$Nd systematics. Earth Planet Sci Lett 488:68-78

Brugman K, Phillips MG, Till C.B. (2021) Experimental determination of mantle solid and melt compositions for two likely rocky exoplanet compositions. J Geophys Res Planets, 126:e2020JE006371

Buckle T, Williams M, Nathwani CL, Hughes HSR (2023) WebNORM: a web application or calculating normative mineralogy. Front Eart Sci 11:1232256, doi: 10.3389/feart.2023.1232256

Campbell IH, O'Neill HS (2012) Evidence against a chondritic Earth. Nature 483:553-558

Campbell IH Taylor SR (1983) No water, no granites - no oceans, no continents. Geophys Res Lett 10:1061-1064

Cawthorn RG, Brown PA (1976) A Model for the Formation and Crystallization of Corundum-Normative Calc-Alkaline Magmas through Amphibole Fractionation. J Geol 84:467-476

Chappell BW, White AJR (1992) I- and S-type granites in the Lachlan fold belt. Earth Env Sci Trans Royal Soc Edinburgh 83:1-26. doi:10.1017/S0263593300007720

Clenet H, Jutzi M, Barrat, JA, Asphaug EI, Benz W, Gillet P (2014) A deep crust–mantle boundary in the asteroid 4 Vesta. Nature 511:303–306 https://doi.org/10.1038/nature13499

Connolly HC, Jones RH (2018) Chondrites: the canonical and non-canonical views. J Geophys Res Planets https://doi.org/10.1002/2016JE005113

Corgne A, Keshav S, Wood BJ, McDonough WF, Fei, Y (2008) Metal-silicate partitioning and constraints on core composition and oxygen fugacity during Earth accretion. Geochim Cosmochim Acta 72:574-589

Cross W, Iddings JP, Pirsson LV, Washington HS (1902) A Quantitative Chemico-Mineralogical Classification and Nomenclature of Igneous Rocks. J Geol 10:555-690

Davies GF (1974) Limits on the constitution of the lower mantle. Geophys J Astron Soc 38:479-503

Davies, G.F., Richards, M.A. (1992) Mantle convection. Journal of Geology. 100, 151-206.

Drake, M.J., and Righter, K. (2002) Determining the composition of the Earth. Nature, 416, 39-44.

Ding S, Dasgupta R, Tsuno K (2020) The solidus and melt productivity of nominally anhydrous Martian mantle constrained by new high pressure-temperature experiments—implications for crustal production and mantle source evolution. J Geophys Res Planets 123:e2019JE006078

Dorn C, Khan A, Heng K, Connolly JAD, Alibert Y, Benz W, Tackley, P. (2015) Can we constrain the interior structure of rocky exoplanets from mass and radius measurements? A&A 577, A83, doi: 10.1051/0004-6361/201424915

Doyle AE, Young ED, Klein B, Zuckerman B, Schlichting HE (2019) Oxygen fugacities of extrasolar rocks: evidence for an Earth-like geochemistry of exoplanets. Science 366:356-359

Drake MJ, Righter K (2022) Determining the composition of the Earth. Nature 416:39-44

Elardo S, Shahar A, Mock TD, Sio CK (2019) The effect of core composition on iron isotope fractionation between planetary cores and mantles. Earth Planet Sci Lett 513:124-134

Fiquet G, Auzende AL, Siebert J, Corgne A, Bureau H, Ozawa H, Garbarino G (2010). Melting of peridotite to 140 gigapascals. Science, 329:1516–1518 https://doi.org/10.1126/science.1192448.

Fitoussi C, Bourdon B (2012) Silicon isotope evidence against an enstatite chondrite Earth. Science 335:1477-1480.

Fortin M-A, Gazel E, Kaltenegger L, Holycros ME (2022) Volcanic exoplanet surfaces. Month Not Royal Astron Soc 516:4569-4575

Frost DJ, McCammon CA (2008) The redox state of Earth's mantle. Ann Rev Earth Planet Sci 36:389-420

Fukai Y, Fuzikawa A., Watanabe K, Amano M (1982) Hydrogen in Iron—its enhanced dissolution under pressure and stabilization of the γ phase. Jpn J App Phys 21, L318, 0.1143/JJAP.21.L318

Gale A, Dalton CA, Langmuir CH, Su Y, Schilling J-G (2013) The mean composition of ocean ridge basalts. Geochem Geophys Geosyst 14:489-518

Gansicke BT, Koester D, Farihi J, Girven J, Parsons SG, Breedt E (2012) The chemical diversity of exo-terrestrial planetary debris around white dwarfs. Month Not Royal Astron Soc 424:333-347

Ghiorso MS, Sack RO (1995) Chemical Mass Transfer in Magmatic Processes. IV. A Revised and Internally Consistent Thermodynamic Model for the Interpolation and Extrapolation of Liquid-Solid Equilibria in Magmatic Systems at Elevated Temperatures and Pressures. Contra Mineral Petrol 119:197-212

Grove TL, Till CB (2019) $H_2O$-rich mantle melting near the slab-wedge interface. Contrb Mineral Petrol 174:80

Guimond CM, Shorttle S, Rudge JF (2023) Mantle mineralogy limits to rocky exoplanet water inventories. Month Not Royal Astron Soc 521:2535-2552

Hakim K, van Westernen W, Dominik C (2018) Capturing the oxidation of silicon carbide in rocky exoplanetary interiors. Astron Astrophys 618, L6

Helffrich, G. (2017) Mars core structure—concise review and anticipated insights from InSight. Prog. Earth and Planet. Sci. 4:24, 10.1186/s40645-017-0139-4.

Helffrich G, Brasser R, Shahar A (2019) The chemical case for Mercury mantle stripping. Prog Earth Planet Sci 6:66, doi.org/10.1186/s40645-019-0312-z

Herd CD, Borg LE, Jones JH, Papike JJ (2002) Oxygen fugacity and geochemical variations in the Martian basalts: implications for Martian basalt petrogenesis and the oxidation state of the upper mantle of Mars. Geochim Cosmochim Acta, 66:2025-2036

Herzberg C, Raterron P, Zhang J (2000) New experimental observations on the anhydrous solidus for peridotite KLB-1. Geochem Geophys Geosyst 1:2000GC000089

Hinkel NR, Unterborn CT (2018) The star-planet connection I: using stellar composition to observationally constrain planetary mineralogy for the ten closest stars. Astrophys J 853:83

Hinkel NR, Timmes FX, Young PA, Pagano MD (2014) Turnbull, M.C. Stellar abundances in the solar neighborhood: the Hypatia Catalog. Astronom J 148: 54

Hinkel NR, Young PA, Wheeler III CH (2022) A concise treatise on converting stellar mass fractions to abundances to molar ratios. Astron J 164:256

Hirschmann, MA (2000) Mantle solidus: Experimental constraints and the effects of peridotite composition. Geochem Geophys Geosyst 1:2000GC000070

Hoffmann AW, White WM (1982) Mantle plumes form ancient oceanic crust. Earth Planet Sci Lett 57:421-436

Hollands MA, Gänsicke BT, Koester D (2018) Cool DZ white dwarfs II: compositions and evolution of old remnant planetary systems. MNRAS 477:93–111

Hollands MA, Tremblay P-E, Gänsicke BT, Koester D, Gentile-Fusillo NP (2021) Alkali metals in white dwarf atmospheres as tracers of ancient planetary crusts. Nat Astron https://doi.org/10.1038/s41550-020-01296-7.

Hauck SA III et al (2013) The curious case of Mercury's internal structure. J Geophys Res Planets 118:1204-1220

Hyung E, Huang S, Petaev MI, Jacobsen SB (2016) Is the mantle chemically stratified? Insights from sound velocity modeling and isotope evolution of an early magma ocean. Earth Planet Sci Lett 440:158-168

Irfune T, Shinmei T, McCammon CA, Miyajima N, Rubie DC, Frost DJ (2010) Iron partitioning and density changes of pyrolite in Earth's lower mantle. Science 327:193-195

Jones JH, Delano JW (1989) A three-component model for the bulk composition of the Moon. Geochim Cosmochim Acta 53, 513-527.

Jones MC, Pewsey A (2009) Sinh-Arcsinh Distributions. Biometrika 96, 761–780

Jura M, Xu S, Klein, B., Koester, D, and Zuckerman, B (2012) Two extrasolar asteroids with low volatile element mass fractions. Astrophys J 750:69

Jura M, Dufour P, Xu S, Zuckerman B, Klein B, Young E, Melis C (2015) Evidence for an anhydrous carbonaceous extrasolar minor planet. Astrophys J 799:109

Kaminsky F (2012) Mineralogy of the lower mantle: a review of 'super-deep' mineral inclusions in diamond. Earth-Sci Rev 110:127-147

Khan, A, Connolly JAD (2008) Constraining the composition and thermal state of Mars from inversion of geophysical data. J Geophys Res Planets 113:E07003

Khan A, Pommier A, Neumann GA, Mosegaard K (2013) The lunar moho and the internal structure of the Moon: a geophysical perspective. Tectonophys 609:331-352

Khan A, Connolly JAD, Taylor SR (2008) Inversion of seismic and geodetic data for the major element chemistry and temperature of the Earth's mantle. J Geophys Res 113:B09308, doi:10.1029/2007JB005239.

Kim T, Ko B, Greenberg E, Prakapenka V, Shim S-H, Lee Y (2020). Low melting temperature of anhydrous mantle materials at the core-mantle boundary. Geophys Res Lett 47:e2020GL089345, https://doi.org/10.1029/2020GL089345

King SD (2018) Ceres internal structure from geophysical constraints. Meteor Planet Sci 53: 1999-2007

Kinzler RJ (1997) Melting of mantle peridotite at pressure approaching the spinel to garnet transition: application to mid-ocean ridge basalt petrogenesis. J Geophys Res 102:853-874

Klein BZ, Jagoutz O Behn MD (2017) Archean crustal compositions promote full mantle convection. Earth Planet Sci Lett 474:516-526

Klein B, Jura M, Koester D, Zuckerman B (2011) Rocky extrasolar planetary compositions derived from externally polluted white dwarfs. Astrophys J 741:64

Kaminsky F (2012) Mineralogy of the lower mantle: A review of 'super-deep' mineral inclusions in diamond. Earth-Sci Rev 110:127-147

Knapmeyer-Endrun B et al. (2021) Thickness and structure of the Martian crust from InSight seismic data. Science 373:438-443

Knibbe JS et al. (2021) Mercury's interior structure constrained by density and P-wave velocity measurements of liquid Fe-Si-C alloys. J. Geophys Planets 126:10.1029/2020JE006651

Kombayashi T, Pesce G, Morard G, Anotonangeli D, Sinmyo R, Mezouar M (2019) Phase transition boundary between fcc and hcp structures in Fe-Si alloy and its implications for terrestrial planetary cores. Am Mineral 104:94-99

Kuchner MJ, Seager S (2005) Extrasolar carbon planets. arXiv:Astro-ph/0504214v2

Kuskov OL, Kronrod EV, Kronrod VA (2019) Effect of thermal state on the mantle composition and core sizes of the Moon. Geochem Int 57:605-620

Kuwayama Y, Hirose K, Cobden L, Kusakabe M, Tateno S, Oishi Y (2021) Post-Perovskite Phase Transition in the Pyrolitic Lowermost Mantle: Implications for Ubiquitous Occurrence of Post-Perovskite Above CMB. Geophys Res Lett 49:e2021GL096219

Lambart S, Baker MB, Stolper EM (2016) The role of pyroxenite in basalt genesis: melt-PX, a melting parameterization for mantle pyroxenites between 0.9 and 5 GPa. J Geophys Res 10.1002/2015JB012762

Lark L, Parman S, Huber C, Parmentier E, Head J (2022) Sulfides in Mercury's mantle: implications for Mercury's interior as interpreted from moment of inertia. Geophys Res Letters 49:e2021GL096713

Le Bas MJ, Streckeisen AL (1991) The IUGS systematics of igneous rocks. Journal of the Geol Soc London 148:825-833

Le Bas MJ, Le Maitre RW, Streckeisen A, Zanettin B (1986) A chemical classification of volcanic rocks based on the total alkali-silica diagram. J Petrol 27:745-750

Lewis GN (1901) The law of physio-chemical change. Proc Am Acad Arts Sci, 37:49-69

Lodders K (2003) Solar system abundances and condensation temperatures of the elements. Astrophys J 591:1220-1247

Lodders K, Fegley B Jr (1998) The Planetary Scientist's Companion, Oxford University Press, Oxford

Lodders K, Fegley B Jr (2018) Chemistry of the Solar System, RSCPublishing, Royal Society of Chemistry, Cambridge

Lodders K, Palme H, Gail HP (2009) Abundances of the elements in the solar system. *In* Landolt-Börnstein, Vol. VI/4B, Trümper JE (ed.), Springer-Verlag, Berlin, p 560-630

Longhi J (2006) Petrogenesis of picritic mare magmas: constraints on the extent of lunar differentiation. Geochim Cosmochim Acta 70:5919-5934

Lyubetskaya T, Korenaga J (2007) Composition of Earth's primitive mantle and its variance 1. Methods and Results. J Geophys Res 112:B03211, doi:10.1029/2005JB004223

Magg E et al. (2022) Observational constraints on the origin of the elements. IV: The standard composition of the Sun. A&A 661:A140, https://doi.org/10.1051/0004-6361/202142971

McDonough WF, Sun S-s (1995) The composition of the Earth Chem Geol 120:223-253

McDonough WF, Yoshizaki T (2021) Terrestrial planet compositions controlled by accretion disk magnetic field. Prog Earth Planet Sci 8:39, https://doi.org/10.1186/s40645-021-00429-4

Mattern E, Matas J, Ricard Y, Bass J (2005) Lower mantle composition and temperature form mineral physics and thermodynamic modeling. Geophys J Int 160, 973-990

Mezger K, Schonbachler M, Bouvier A (2020) Accretion of the Earth - Missing components? Space Sci Rev 216:27, https://doi.org/10.1007/s11214-020-00649-y

Mittlefehldt DW (2015) Asteroid (4) Vesta: 1. The howardite-eucrite-diogenite (HED) clan of meteorites. Geochem 75:155-183

Morse SA (1980) Basalts and Phase Diagrams: An Introduction to the Quantitative use of Phase Diagrams in Igneous Petrology. Springer-Verlag, Berlin.

Murakami M, Ohishi Y, Hirao N, Hirose K (2012) A perovskitic lower mantle inferred from high-pressure, high temperature sound velocity data. Nature 485:90-94

Namur O, Collinet M, Charlier B, Grove TL, Holtz F, McCammon C (2014) Melting processes and mantle sources of lavas on Mercury. Earth Planet Sci Lett 439:117-128

Nitttler LR, Chabot NL, Grove TL, Peplowski PN (2018) Chapter 2. The Chemical Composition of Mercury. *In* Mercury, The View after MESSENGER, Solomon, SC, Nittler LR, Anderson B (eds), Cambridge University Press, Cambridge p 30 - 51

Nittler LR, McCoy TJ, Clark PE, Murphy ME, Trombka JI, Jarosewich E (2004) Bulk element compositions of meteorites: a guide for interpreting remote-sensing geochemical measurements of planets and asteroids. Antarctic Met Res 17:231-251

Nittler LR, Starr RD, Weider SZ, McCoy TJ, Boynton WV, Ebel DS, Sprague AL (2011) The major-element composition of mercury's surface from messenger x-ray spectrometry. Science 333:1847–1850

Nomura R, Hirose K, Uesugi K, Ohishi Y, Tsuchiyama A, Miyake A, Ueno Y (2014). Low core-mantle boundary temperature inferred from the solidus of pyrolite. Science 343:522–525

Ogawa M, Schubert G, Abdelfattah M (1991) Numerical simulations of three-dimensional thermal convection in a fluid with strongly temperature-dependent viscosity. J Fluid Mech 233:299-328

O'Neil C (2021) End member Venusian core scenarios: does Venus have an inner core? Geophys Res Lett 48:10.1029/2021GL095499

Palme H, O'Neill HStC (2003) Cosmochemical estimates of mantle composition. *In* The Mantle and Core, Vol. 2 Treatise on Geochemistry, Carlson RW (ed), Elsevier–Pergamon, Oxford, p 1-38

Pottasch SR (1964) A comparison of the chemical composition of the solar atmosphere with meteorites. Annales Astrophys, 27:163-169.

Palme H, Zipfel J (2022) The composition of CI chondrites and their contents of chlorine and bromine: results from instrumental neutron activation analysis. Met Planet Sci 57:317-333

Peterson BT, DePaolo DJ (2007) Mass and Composition of the Continental Crust Estimated Using the CRUST2.0 Model. Am Geophys Union, Fall Meeting, V33A-1161

Prytulak J, Elliott T (2007) $TiO_2$ enrichment in ocean island basalts. Earth Planet Sci Lett 263:388–403

Putirka KD, Dorn C, Hinkel NR, Unterborn CT (2021) Compositional diversity of rocky exoplanets. Elements 17:235-240

Putirka K, Perfit M, Ryerson FJ, Jackson MG (2007) Ambient and excess mantle temperatures, olivine thermometry, and active vs. passive upwelling, Chem Geol 241:177-206

Putirka KD, Rarick J (2019) The composition and mineralogy of rocky exoplanets: a survey of >4000 stars from the Hypatia Catalog. Am Mineral 104:817-829

Putirka K, Tao Y, Hari KR, Perfit M, Jackson M, Arevalo Jr R (2018) The mantle source of thermal plumes: minor elements in olivine and major oxides of primitive liquids (and why the olivine compositions don't matter). Am Mineral 103:1253-1270

Putirka, K.D., Xu, S. (2021) Polluted white dwarfs reveal exotic mantle rock types on exoplanets in our solar neighborhood. Nat Comm 12, 6168, https://doi.org/10.1038/s41467-021-26403-8

Ransford GA (1979) A comparison of two accretional heating models. Proc Lunar Planet Sci Conf 10:1867-1879

Raymond CA, Park RS, Asmar SW, Konopliv AS, Buczkowski DL, De Sanctis MC, McSween HY, Russell CT, Jaumann R, Preusker F, and the Dawn Team (2013) The Crust and Mantle of Vesta's Southern Hemisphere. European Planet Sci Congr Abstracts. 8:EPSC2013-1002

Reach WT, Lisse C, von Hippel T, Mullally F (2009) The dust cloud around white dwarf G 29-38. II Spectrum from 5 to 40 μm and mid-infrared photometric variability. Astrophys J 693: 697-712.

Rhodes JM, Vollinger MJ (2005) Ferric/ferrous ratios in 1984 Mauna Loa lavas: a contribution to understanding the oxidation state of Hawaiian magmas. Contra Mineral Petrol 149:666-674

Righter K, Drake MJ (1996) Core formation in Earth's Moon, Mars and Vesta. Icarus 124:513-529.

Righter K, Yang H, Costin CG, Downs RT (2008) Oxygen fugacity in the Martian mantle controlled by carbon: new constraints from the nakhlite MIL 03346. Met Planet Sci 43:1709-1723

Rogers LK et al. (2024) Seven white dwarfs with circumstellar gas discs I: white dwarf parameters and accreted planetary abundances. MNRAS 527, 6038-6054

Rubie DC, Frost DJ, Mann U, Asahara Y, Nimmo, F., Tsuno, K., Kegler, P., Holzheid, A.& Palme, H. (2011) Heterogeneous accretion, composition and core-mantle differentiation of the Earth. Earth and Planetary Science Letters, 301, 31-42.

Rudnick RL, Gao S (2014) 4.1 Composition of the Continental Crust. *In* Treatise of Geochemistry 2nd ed., Holland HD, Turekian KK (eds), Elsevier, Amsterdam, p 1-51

Safranov VS (1978) The heating of the earth during its formation. Icarus 33:3-12

Salters ,VJM, Stracke A (2004) Composition of the depleted mantle. Geochem Geophys Geosyst 5:Q05B07, doi:10.1029/2003GC000597

Sanloup C, Jambon A, Gillet P (1999) A simple chondritic model of Mars. Phys Earth Planet Int 112:43-54

Shah O, Helled R, Alibert Y, Mezger K (2022) Possible chemical composition and interior structure models of Venus inferred from numerical modeling. Astrophys J 926:217

Shea T, Hammer JE, Hellegrand E, Mourney AJ, Costa F, First EC, Lynn KJ, Melnik O (2019) Phosphorous and aluminum zoning in olivine: contrasting behavior of two nominally incompatible trace elements. Contra Mineral Petrol 174:85

Sobolev AV, Hoffmann AW, Sobolev SV, Nikogosian IK (2005) An olivine-free mantle source of Hawaiian shield basalts. Nature 434:590-597

Solomotov VS, Moresi L-N (2000) Scaling of time-dependent stagnant-lid convection: application to small-scale convection on Earth and other terrestrial planets. J Geophys Res, 105:21795-21817

Sori MM (2018) A thin, dense crust for Mercury. Earth Planet Sci Lett 490:92-99

Steenstra ES, Knibbe JS, Rai N, van Westrenen W (2016) Constraints on core formation in Vesta from metal-silicate partitioning of siderophile elements. Geochim Cosmochim Acta 177:48-61

Stevenson D (2003) Styles of mantle convection and their influence on planetary evolution. CR Geosci 335:99–111

Stolper E (1980) A phase diagram for mid-ocean ridge basalts: preliminary results and implications for petrogenesis. Contrib Mineral Petrol 74:13-27

Stolper E, Shorttle O, Antoshechkina PM, Asimow PD (2020) The effects of solid-solid phase equilibria on the oxygen fugacity of the upper mantle. Amer Mineral 105, 1445-1471

Suarez-Andres L, Israelian G, Gonzalez Hernandez JI, Adibekyan VZh, Delgado Mena E, Santos NC, Sousa SG (2018) C/O vs. Mg/Si ratios in Solar type stars: the HARPS sample. Astron Astrophys 614:A84

Szurgot M (2015) Core mass fraction and mean atomic weight of terrestrial planets, moon and protoplanet Vesta. *In* Comparative Tectonics Geodynamics of Venus, Earth and Rocky Exoplanets, Contribution 5001, Lunar and Planetary Institute, Houston, p 35.

Takahashi E (1986) Melting of a dry peridotite KLB-1 up to 14 GPa: implications on the origin of peridotitic upper mantle. J Geophys Res 91:9367-9382

Tagawa S, Ohta K, Hirose K, Kato C, Ohishi Y (2016). Compression of Fe-Si-H alloys to core pressures. Geophys Res Lett, 43, 3686–3692

Taylor GJ (2013) The bulk composition of Mars. Chemi der Erde, 73:401-420

Teraskai H et al. (2019) Pressure and composition effects on sound velocity and density of core-forming liquids: implication to core compositions of terrestrial planets. J Geophys Res Planets 10.1029/2019JE005936

Thompson JB Jr (1982) Chapter 2. Ration space: an algebraic and geometric approach. *In* Characterization of Metamorphism through Mineral Equilibria, Ferry JM (ed), De Gruyter, Berlin, p 33-52

Till CB, Elkins-Tanton LT, Fischer KM (2010) A mechanism for low-extent melts at the lithosphere-asthenosphere boundary. Geochem Geophys Geosyst 11:Q10015

Till CB, Grove TL, Krawczynzki MJ (2012) A melting model for variably depleted and enriched lherzolite in the plagioclase and spinel stability fields. J Geophys Res 117:B06206

Togashi S, Kita N, Tomiya A, Morshita Y (2017) Magmatic evolution of the lunar highland rocks estimated from trace elements in plagioclase: a new bulk silicate Moon model with sub-chondritic Ti/Ba, Sr/Ba and Sr/Al ratios. Geochim Cosmochim Acta 210:152-183

Trierweiler IL, Doyle AE, Young ED (2023) A chondritic solar neighborhood. arXiv:2306.03743v1 [Astro-ph.EP]

Tronnes RG, Baron MA, Eigenmann KR, Guren MG, Heyn BH, Loken A, Mohn CE (2019) Core formation, mantle differentiation and core-mantle interaction within Earth and the terrestrial planets. Tectonophys 760:165-198

Unterborn CT, Panero WR (2017) The effects of Mg/Si on the exoplanetary refractory oxygen budget. Astrophys J 845:61

Unterborn CT, Panero WR (2019) The pressure and temperature limits of likely rocky exoplanets. J Geophys Rese Planets 124:1704-1716

van der Hilst RD, Karason H (2000) Compositional heterogeneity in the bottom 1000 kilometers of Earth's mantle: toward a hybrid convection model. Science 283:1885-1888

Vander Kaaden KE, McCubbin FM (2016) The origin of boninites on Mercury: an experimental study of the northern volcanic plains lavas. Geochim Cosmochim Acta, 173:246-263

Vander Kaaden KE, McCubbin FM, Nittler LR, Peplowski PN, Weider, SZ, Frank, EA, McCoy TJ (2017) Geochemistry, mineralogy, and petrology of boninitic and komatiitic rocks on the Mercurian surface: Insights into the Mercurian mantle. Icarus 285:155-168

Verma SP, Torres-Alvarado IS, Velasco-Tapia F (2003). A revised CIPW norm. Schweiz Mineral Petrogr Mittl 83, 197–216, doi:10.5169/seals-63145

Viswanathan V, Rambaux N, Fienga A, Laskar J, Gastineau M. (2019) Observational constraint on the radius and oblateness of the Lunar core-mantle boundary. Geophys Res Lett 10.1029/2019GL082677

Vocaldo L (2010) New views of the Earth's inner core from computational mineral physics. *In* New Frontiers in Integrated Solid Earth Sciences, International Year of Planet Earth, Cloetingh S, Negendank J (eds), Springer Science+Business Media, Berlin, p 397-412

Wade J, Wood BJ (2005) Core formation and the oxidation state of the Earth. Earth Planet Sci Lett 236:78-95

Walker D (1986) Melting equilibria in multicomponent systems and liquidus/solidus convergence in mantle peridotite. Contra Mineral Petrol 92:303-307

Walter MJ (1998) Melting of garnet peridotite and the origin of komatiite and depleted lithosphere. J Petrol 39:29–60

Wang J, Takahashi E, Xiong X, Chen L, Li L, Suzuki T, Walter MJ (2020) The water-saturated solidus and second critical endpoint of peridotite: implications for magma genesis within the mantle wedge. J Geophys Res Sol Earth 10.1029/2020JB019452

Wanke H, Dreibus G (1994) Chemistry and accretion history of Mars. Phil Trans Royal Soc A, 349:545-557

Warren JM (2016) Global variation in abyssal peridotite compositions. Lithos 248-251:193-219

Weller MB, Lenardic A (2018) On the evolution of terrestrial planets: bi-stability, stochastic effects, and the non-uniqueness of tectonic states. Geoscience Frontiers 9:91-102

Whitney DL, Evans BW (2010) Abbreviations for names of rock-forming minerals. Amer Mineral 95, 185-187

Wicks JK, Duffy TS (2016) Crystal structures of minerals in the lower mantle. *In* Deep Earth: Physics and Chemistry of the Lower Mantle and Core, Geophysical Monograph 217, Terasaki H, Fischer RA (eds), American Geophysical Union, John Wiley & Sons, Inc., New York, p 69-87

Wood BJ, Kiseeva ES, Mirolo FJ (2014) Accretion and core formation: the effects of sulfur on metal-silicate partition coefficients. Geochim Cosmochim Acta 145:248-267

Wood BJ, Smythe DJ, Harrison T (2019) The condensation temperatures of the elements: a reappraisal. Am Mineral 104:844-856

Wood BJ, Walter MJ, Wade J (2006) Accretion of the Earth and segregation of its core. Nature 441:825-833

Workman RK, Hart SR (2005) Major and trace element composition of the depleted MORB mantle (DMM). Earth Planet Sci Lett 231:53-72

Xu S, Bonsor A (2021) Exogeology from polluted white dwarfs. Elements 17:241-244

Xu S, Dufour P, Klein B, Melis C, Monson NN, Zuckerman B, Young ED, Jura M (2019) Compositions of planetary debris around dusty white dwarfs. Astronom J 158:242

Xu S, Jura M, Klein B, Koester D, Zuckerman B (2013) Two beyond-primitive extrasolar planetesimals. Astrophys J 766:132

Xu S, Zuckerman B, Dufour P, Young ED, Klein B, Jura M (2017) The chemical composition of an extrasolar Kuiper-belt-object. Astrophys J Lett 836:L7

Yang H, Muir JM, Zhang F (2022) Iron hydride in the Earth's inner core and its geophysical implications. Geochem Geophys Geosyst 23, e2022GC010620, 10.1029/2022GC010620

Yoshizaki T, McDonough WF (2020) The composition of Mars. Geochim Cosmochim Acta 273:137-162

Zhao Y-H, Zimmerman ME, Kohlstedt DL (2018) Effect of iron content on the creep behavior of olivine: 2 hydrous conditions. Phys Earth Planet Int, 278:26-33

Zuckerman B. et al. (2011) An aluminum/calcium-rich, iron-poor, white dwarf star: evidence for an extrasolar planetary lithosphere. Astrophys J, 739:101

# Figure Captions

**Figure 1.** Cross section of Earth illustrating the core, mantle and crust (crust is not to scale), mantle plumes, and plate tectonics.

**Figure 2.** Cross sectional view of Earth's mantle mineralogy (Adapted from Kaminsky, 2012); horizontal widths show relative mineral abundances (bottom scale); Earth's geotherm (upper horizontal scale) is also shown. Along the geotherm is a proposed standard state (2 GPa, 1350$^o$C) that is recommended for comparing planets due to the high abundance of and low uncertainties of experiments that bracket that state, at 1-3 GPa and 1300-1400$^o$C, and because diverse compositions should equilibrate rapidly at these conditions.

**Figure 3.** (a) Ternary eutectic for the case of pure diopside (Di; $CaMgSi_2O_6$), forsterite (Fo; $Mg_2SiO_4$) and anorthite (An; $CaAl_2Si_2O_8$), adapted from Morse (1980). Temperature contours are shown as dashed lines; the three heavy dark lines are "cotectic" curves that indicate the temperature and compositions of co-saturation of An + Ol, Di + Ol and An + Di; these intersect at the eutectic composition. For any combination of the three minerals, Di, Fo and An, melting will begin at the eutectic, at 1270$^o$C. (b) A more realistic ternary, the plagioclase-saturated ("+ plag") Di-Ol-Qzz ($SiO_2$) system; melting will begin at a red dot, the precise position of which depends upon pressure (adapted from Stolper 1980); curves show the positions of the cotectic curves at 1 atmosphere and 10, 15 and 20 kbar.

**Figure 4.** (a) Ultramafic rock classification diagram (Le Bas and Streckeisen 1991). Shaded green region shows the location of terrestrial mantle rocks, (b) Many ternary diagrams, such as in panel (a) or Figure 3, represent a "projection" from a 3D space of four minerals (b) into a 2D plane, that illustrates the abundances of three minerals. In (b) the rock contains a combination of olivine (Ol) + orthopyroxene (Opx) + clinopyroxene (Cpx) + garnet (Gar) and lies just above the Ol + Cpx + Opx plane. The projection, (c) is obtained by renormalizing so that the sum Ol + Cpx + Opx = 1 (or 100). The "+ gar" in (c) shows that the projection of the indicated sample is from the mineral garnet.

**Figure 5.** Solidus (T at which melting beings) and liquidus (T at which melting is completed) curves for Earth, Mars, Mercury and Chondrite meteorites.

**Figure 6.** (a) Chemical classification diagram for volcanic rocks. "Basalts" are the dominant crust type of the inner planets and differentiated meteorites. Andesites are common above subduction zones and represent the average composition of Earth's continental crust. (b) Comparison of the compositions of mantle and crust rock types.

**Figure 7.** Test of Model 2 (Table 9a) for its ability to predict mineral proportions in natural rock samples, where mineral proportions are measured directly, or calculated by mass balance from measured mineral and whole rock compositions.

**Figure 8.** (a) shows a test of Model 2's ability to predict mineral abundances for the same data as shown in Figure 7. (b) shows Model 2's ability to predict core mass fractions for the inner planets, Moon and Vesta, from Fe-Si mass balance and resulting estimates of total metallic Fe.

**Figure 9 (a)** 1,000 Monte Carlo simulations with 1$\sigma$, 2 $\sigma$ and 3 $\sigma$ contours, of the PWD composition Ton 345, and its reported errors, along with 23 PWD compositions (Putirka and Xu, 2021; see also Jura et al. 2015), projected into the Fe-free ternary system Ol + Cpx + Opx. (**b-e**) Error analysis using mean measurements uncertainties for PWDs, but projected into the ternary

Ol+Pyx+Fe Metal. (**f-g**) Using the same measurement uncertainties as in Fig. 9a, panels f and g compare how errors vary depending upon the particular bulk composition that is plotted. **(h-i).** Using the same measurement uncertainties as in Fig. 9a, panels h and i show how the projected error depends upon the precise mineral formulas used to express the ternary mineral compositions.

**Figure 10.** (**a**) PWD mineral abundances (from bulk compositions in Putirka and Xu (2021) are plotted using Method 1, utilizing Model 2 of Table 9a. (b) PWD mineral abundances are calculated using Method 2. Both (**a**) and (**b**) show mineral estimates for the inner planets, using compositions from Table 3 and compositions from McDonough and Yoshizaki (2021). Also shown are results from 1,000 Monte Carlo simulations using a man PWD composition and uncertainties as input. In (**b**) Most Monte Carlo simulations plot on the boundaries of the ternary diagram. (**c**) Comparison of the observed distribution of bulk PWD Fe metal contents (green bars), obtained using Model 4 of Table 9a (green bars) with 1,000 Monte Carlo simulations (red curve) assuming a Gaussian distribution, using the mean PWD composition and error as input. (**d-f**) mineral abundances for PWD BSPs (green bars) using Model 1 of Table 9a, as in Figure 10a, with Monte Carlo simulations of mean PWD (red curve). Values in green are mean and standard deviations of PWD mineral abundance estimates; values in red are the same for the Monte Carlo simulations. Gray shaded areas of (**d-f**) are mineral estimates that would fall within the ultramafic ternary diagram (Fig. 4a). Blue bars indicate Earth-like compositions.

**Figure 11**. Panel j compares multiple reported compositions for several PWDs. Average and median uncertainties for the seven PWDs are shown as black filled symbols. Standard deviations of the observed PWD compositions would be smaller than the mean and median uncertainties, except for PW 0843 and PG 1015. Note that with the projection from the wüstite (FeO) apex, the data in Fig. 10 are not sensitive to Fe contents; variations in Fe within the Ol-Opx-Cpx-Wus tetrahedron occur along a line connecting a given point to the Wus apex, as in Fig. 4d-e, if one replaces garnet with wüstite). Standard deviations of the multiple estimates for most of the PWDs shown in Fig. 10 will also yield even smaller uncertainties than indicated by the mean and median compositional ranges. Data sources are as follows: G241-6, Jura et al. (2012), Zuckerman et al. (2011); G40 Jura et al. (2012); HS 2243+8023, Klein et al. (2021); PG 0843+517 and PG 1015+161, Xu et al. (2019), Gansicke et al. (2012); PG 1225-079, Xu et al. (2013), Klein et al. (2011); WD 1425+540, Xu et al. (2017). Note: Gansicke et al. (2012) do not report Ca for PG 0843, so we substitute the mean of 23 PWDs where Ca = -6.94 dex.

**Figure 12.** (**a**) olivine structure model showing the crystallographic sites where Fe, Mg and Si occur; (**b**) Gibbs Free Energy vs. Composition (G-X diagrams, panels T1-T4) and (**c**) binary temperature vs composition phase diagram (T-X diagram, bottom panel) showing the complete solid solution between Fa and Fo in olivine.

**Figure 13.** Hypothetical G-X diagrams for the cases when (**a**) SiO2 and (**b**) MgO vary within ranges defined by Earth and the inner planets and cases that might obtain in exoplanets, where (**c**) $SiO_2$ and (**d**) MgO might vary more broadly, and unconventional or unknown phases with relatively low Gibbs Free Energies may be lurking.

**Figure 14.** A comparison of O and S contents for Earth, Mercury, Mars and the Sun.

Figure 1

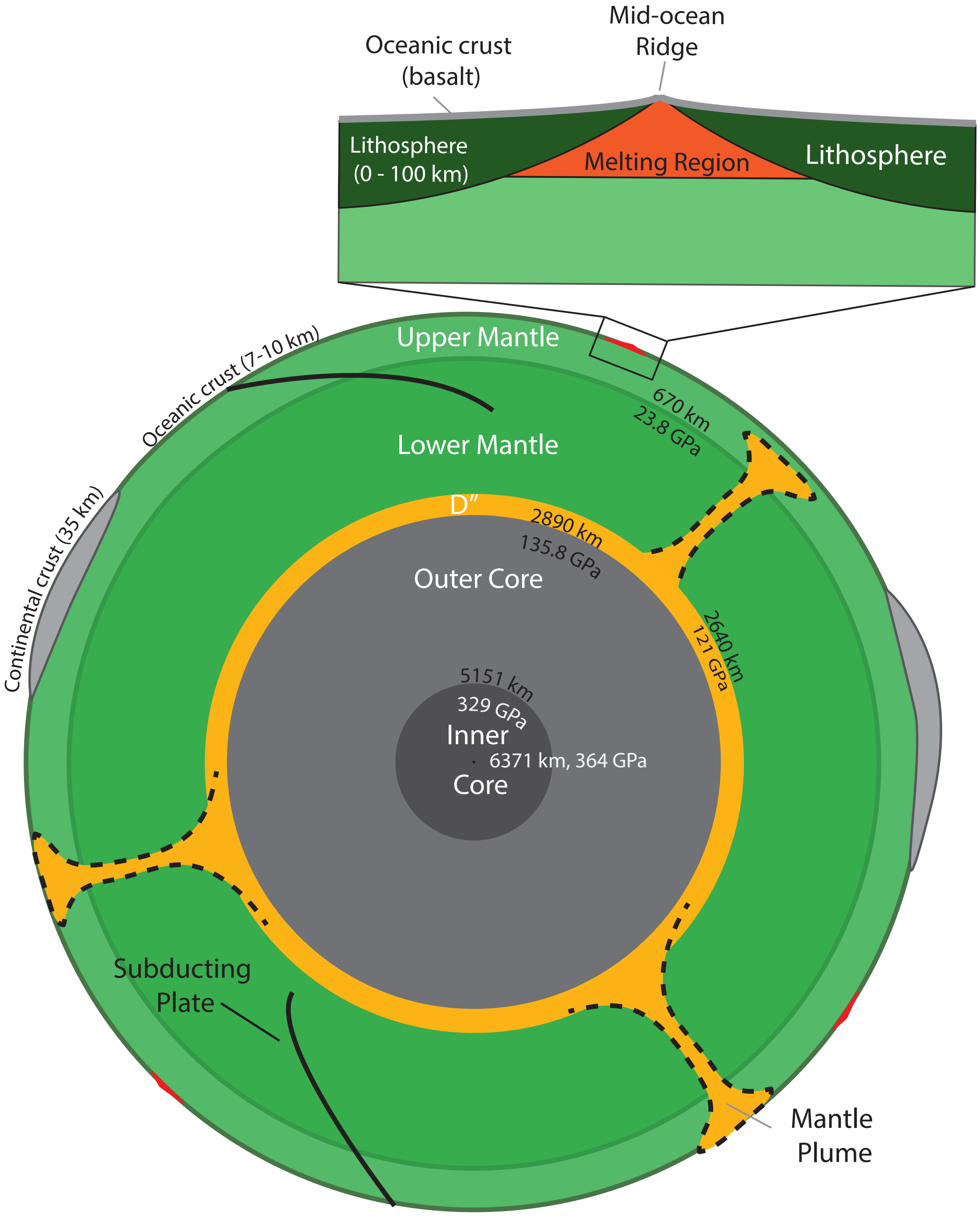

Figure 2

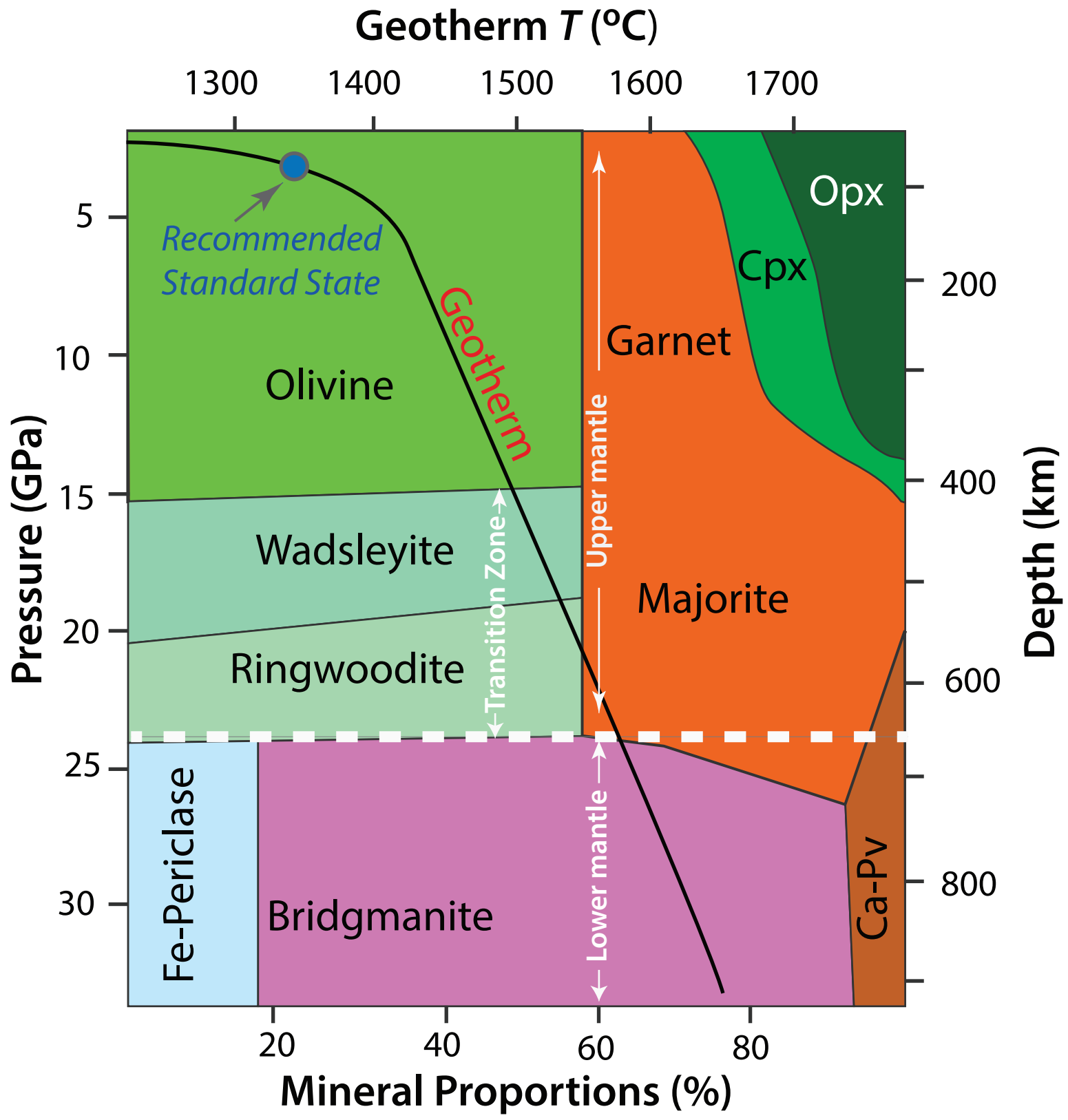

Figure 3

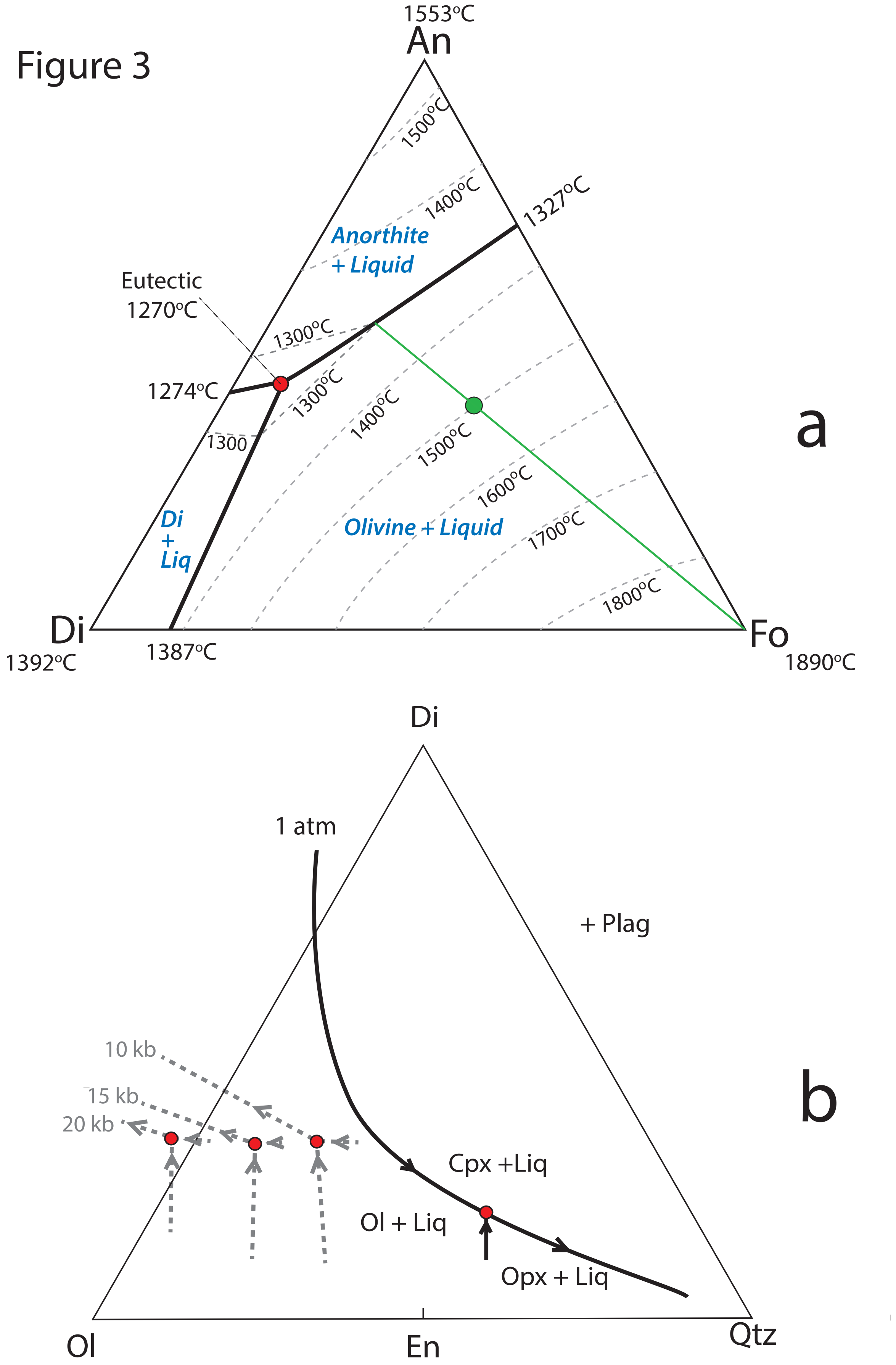

Figure 4

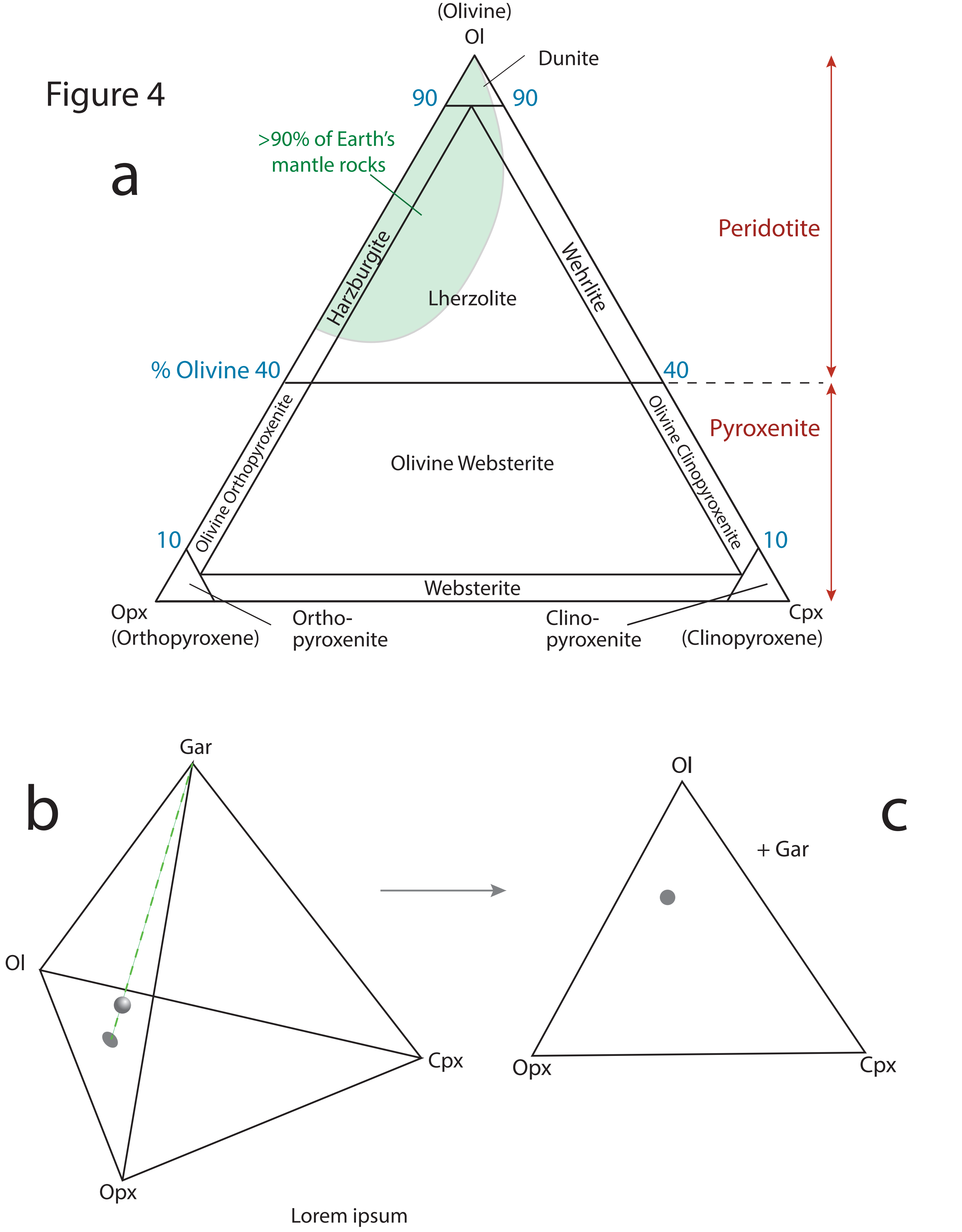

Figure 5

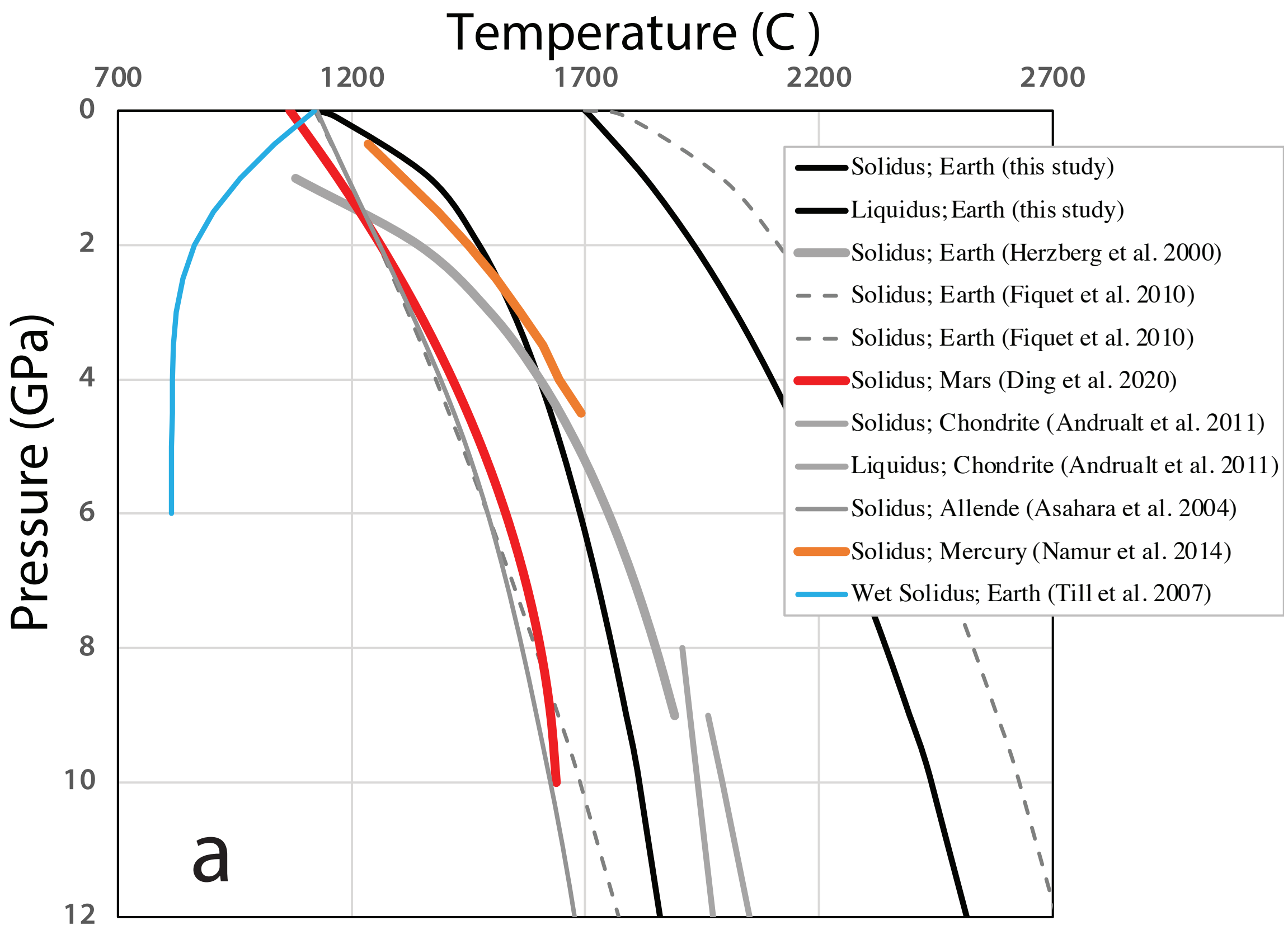
Temperature (C )
700
1200
1700
2200
2700
0
2
4
6
8
10
12
Pressure (GPa)
a
Solidus; Earth (this study)
Liquidus; Earth (this study)
Solidus; Earth (Herzberg et al. 2000)
Solidus; Earth (Fiquet et al. 2010)
Solidus; Earth (Fiquet et al. 2010)
Solidus; Mars (Ding et al. 2020)
Solidus; Chondrite (Andrualt et al. 2011)
Liquidus; Chondrite (Andrualt et al. 2011)
Solidus; Allende (Asahara et al. 2004)
Solidus; Mercury (Namur et al. 2014)
Wet Solidus; Earth (Till et al. 2007)

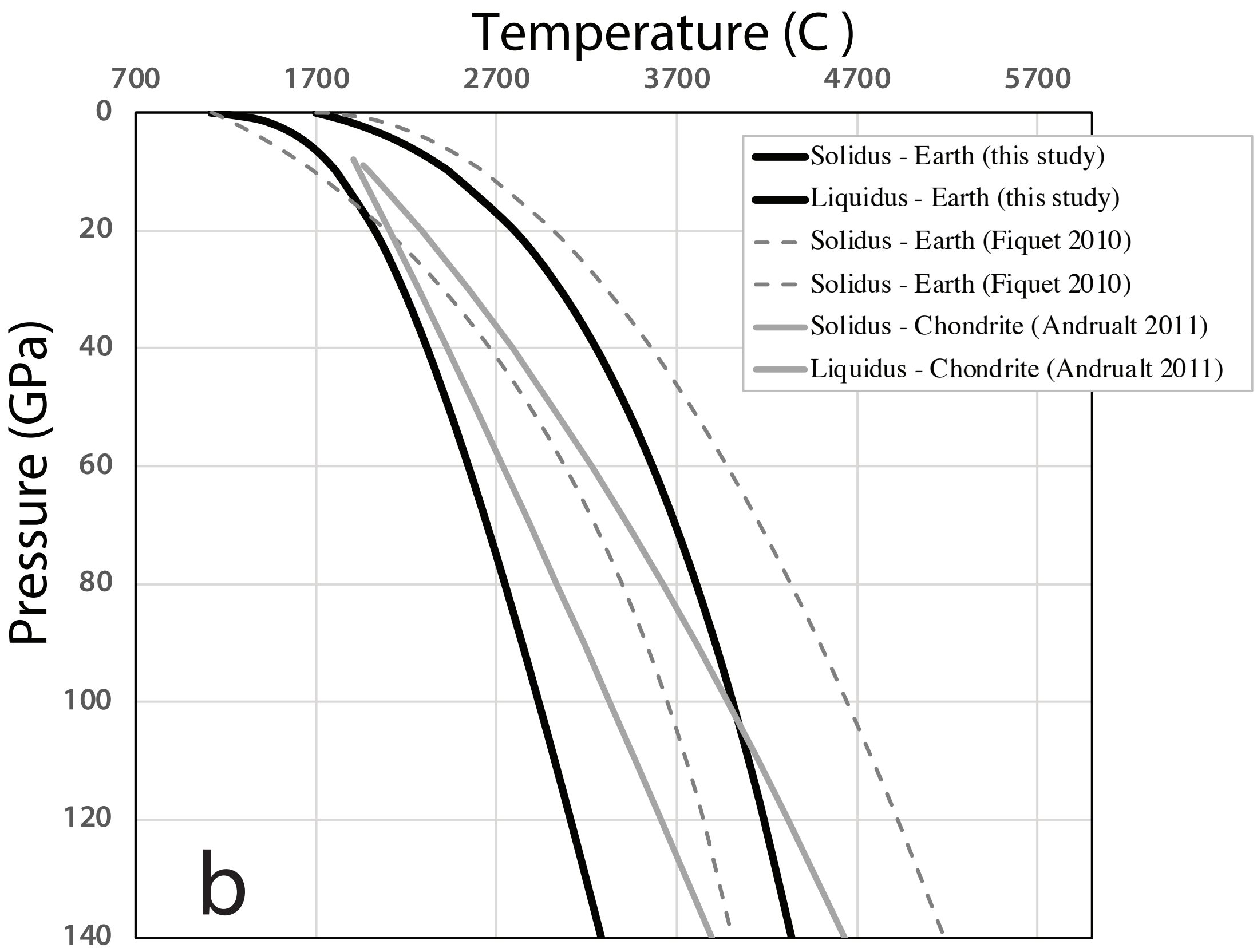
Temperature (C )
700
1700
2700
3700
4700
5700
0
20
40
60
80
100
120
140
Pressure (GPa)
b
Solidus - Earth (this study)
Liquidus - Earth (this study)
Solidus - Earth (Fiquet 2010)
Solidus - Earth (Fiquet 2010)
Solidus - Chondrite (Andrualt 2011)
Liquidus - Chondrite (Andrualt 2011)

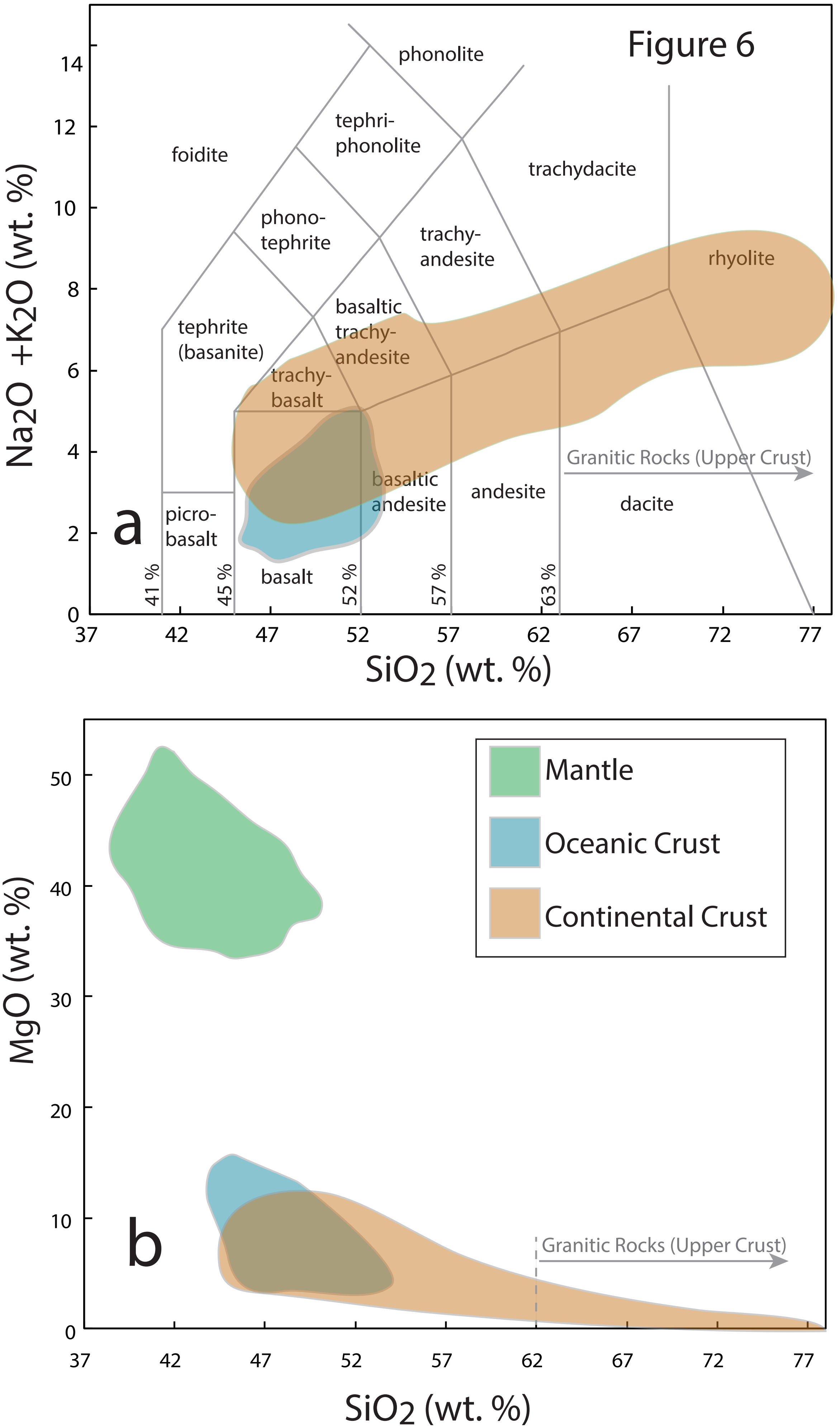

Figure 7

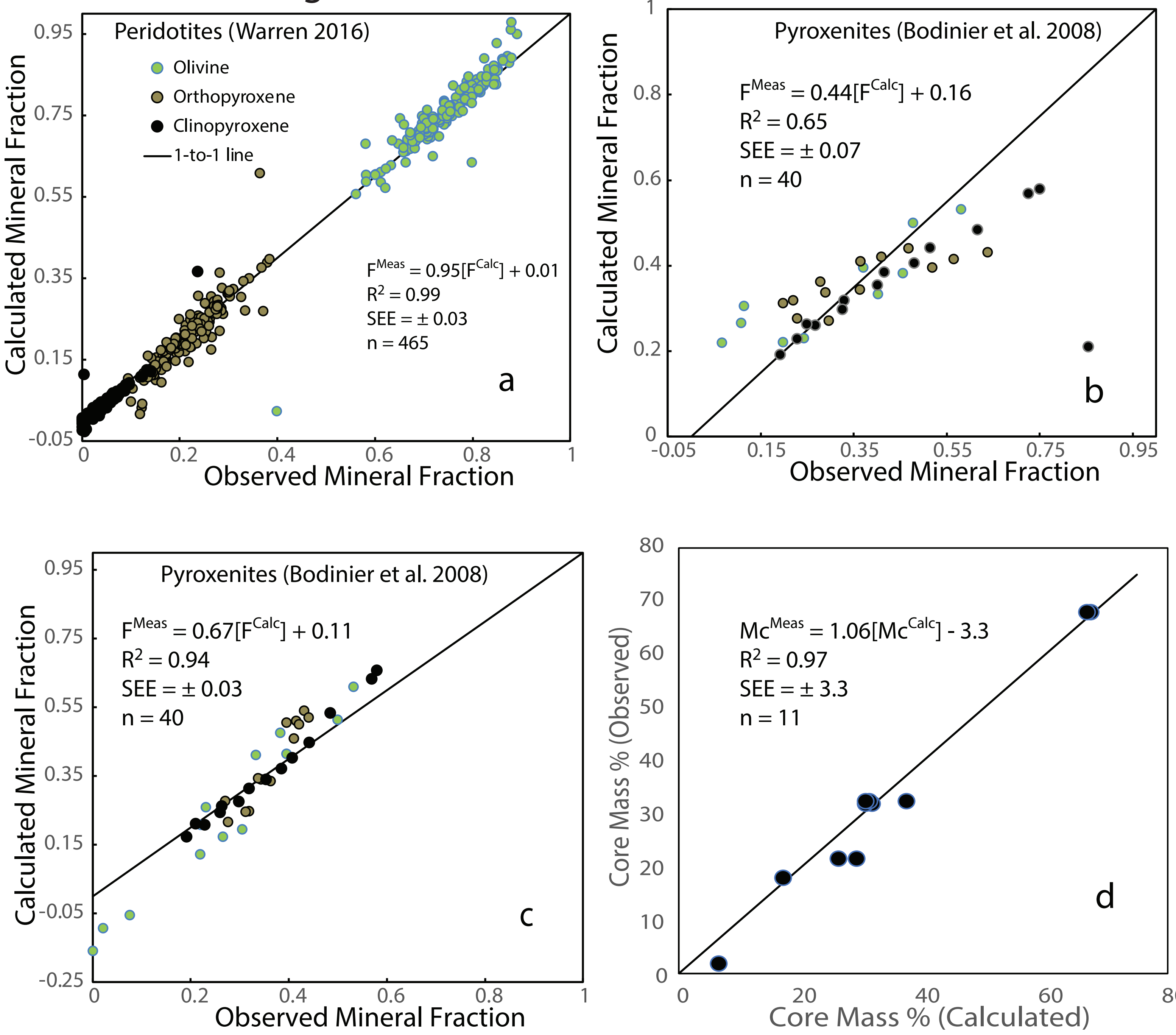

Peridotites (Warren 2016)
Olivine
Orthopyroxene
Clinopyroxene
1-to-1 line
F^Meas = 0.95[F^Calc] + 0.01
R^2 = 0.99
SEE = ± 0.03
n = 465
a
Calculated Mineral Fraction
Observed Mineral Fraction
Pyroxenites (Bodinier et al. 2008)
F^Meas = 0.44[F^Calc] + 0.16
R^2 = 0.65
SEE = ± 0.07
n = 40
b
Pyroxenites (Bodinier et al. 2008)
F^Meas = 0.67[F^Calc] + 0.11
R^2 = 0.94
SEE = ± 0.03
n = 40
c
Mc^Meas = 1.06[Mc^Calc] - 3.3
R^2 = 0.97
SEE = ± 3.3
n = 11
d
Core Mass % (Observed)
Core Mass % (Calculated)

# Figure 8

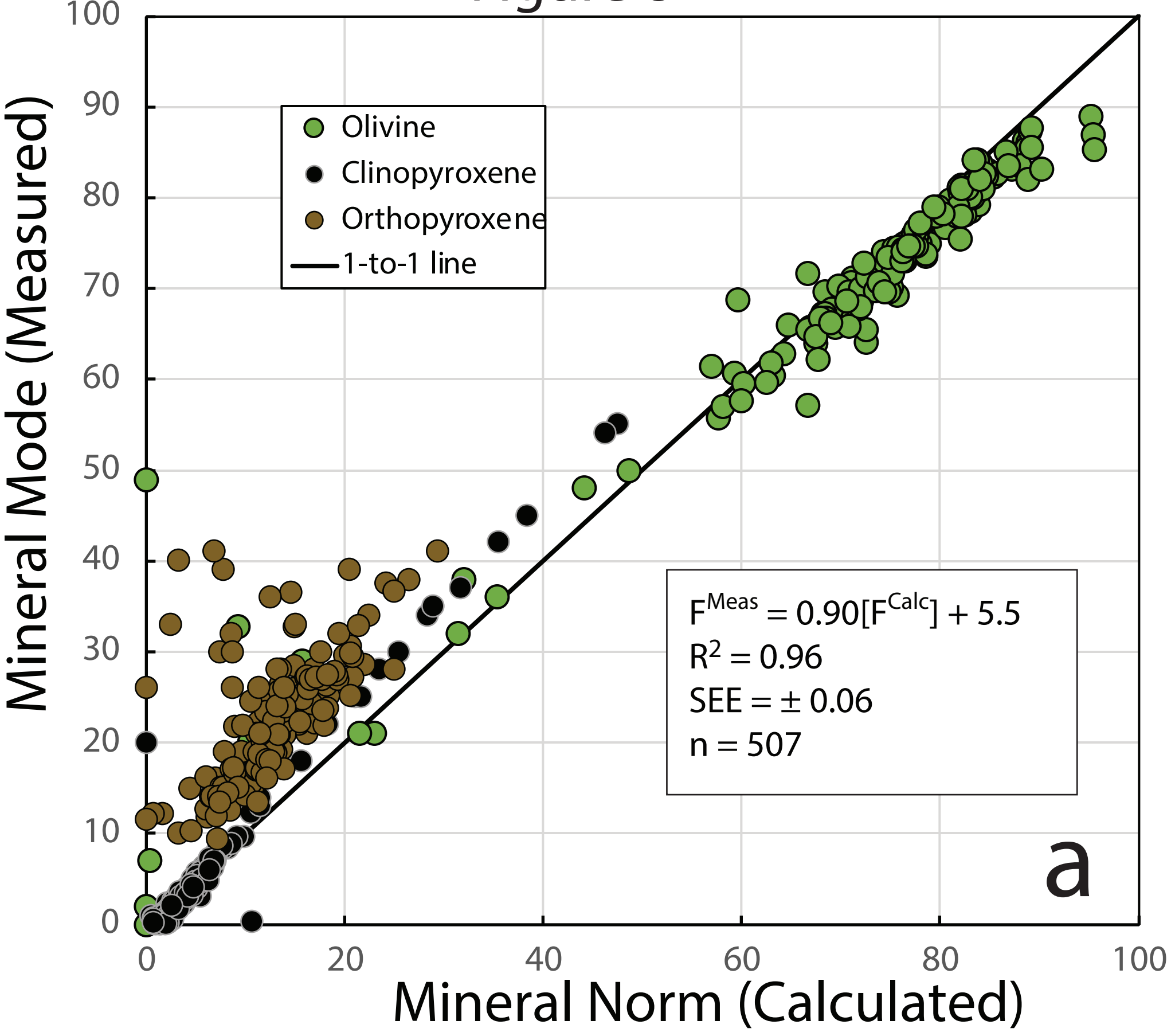

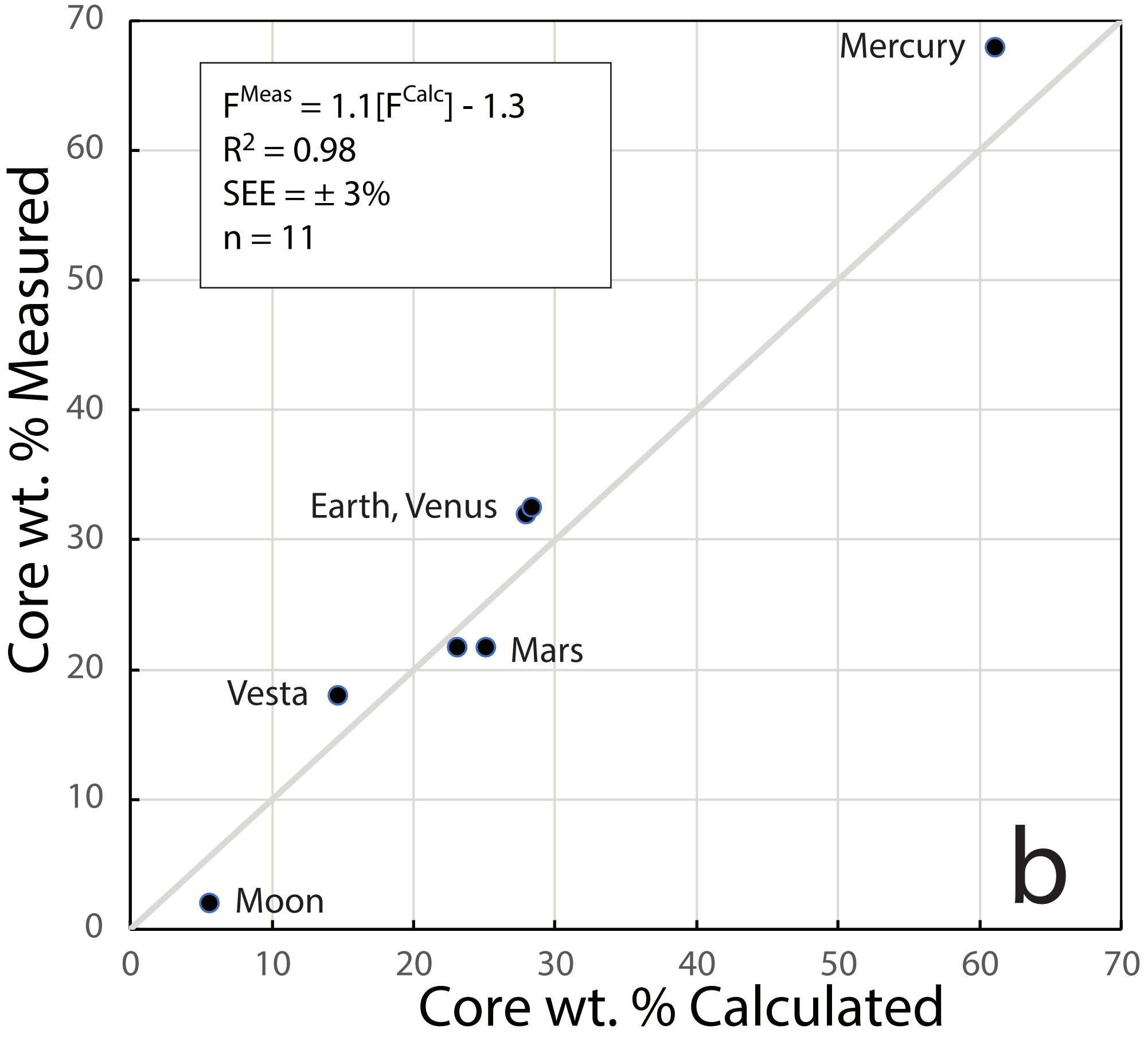

Core wt. % Measured
Core wt. % Calculated
F^Meas = 1.1[F^Calc] - 1.3
R² = 0.98
SEE = ± 3%
n = 11
Mercury
Earth, Venus
Mars
Vesta
Moon
b

Figure 9

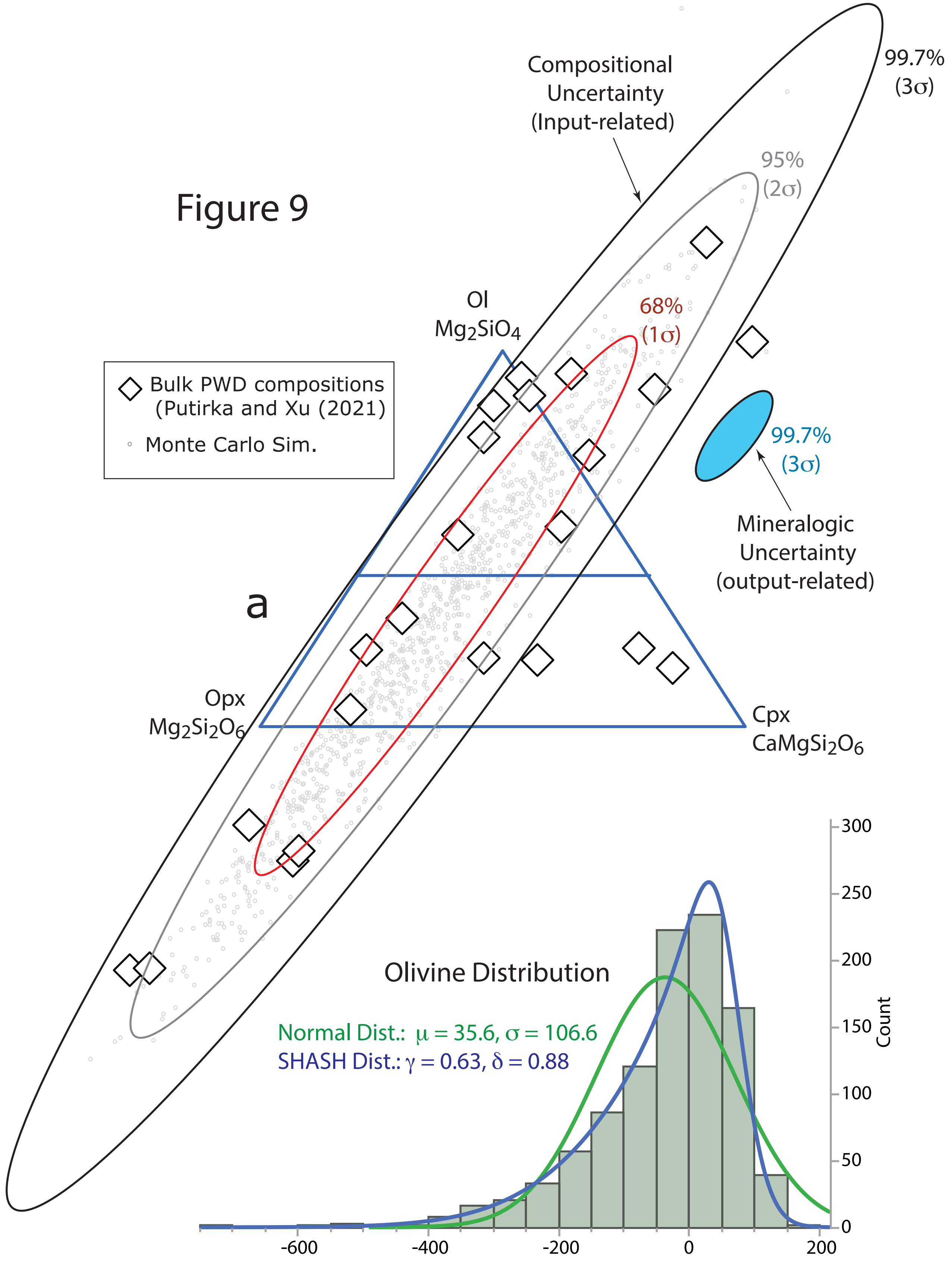

Compositional
Uncertainty
(Input-related)
99.7%
(3σ)
95%
(2σ)
68%
(1σ)
Ol
Mg2SiO4
Bulk PWD compositions
(Putirka and Xu (2021)
Monte Carlo Sim.
99.7%
(3σ)
Mineralogic
Uncertainty
(output-related)
a
Opx
Mg2Si2O6
Cpx
CaMgSi2O6
Olivine Distribution
Normal Dist.: μ = 35.6, σ = 106.6
SHASH Dist.: γ = 0.63, δ = 0.88
Count
300
250
200
150
100
50
0
-600
-400
-200
0
200

# Figure 9

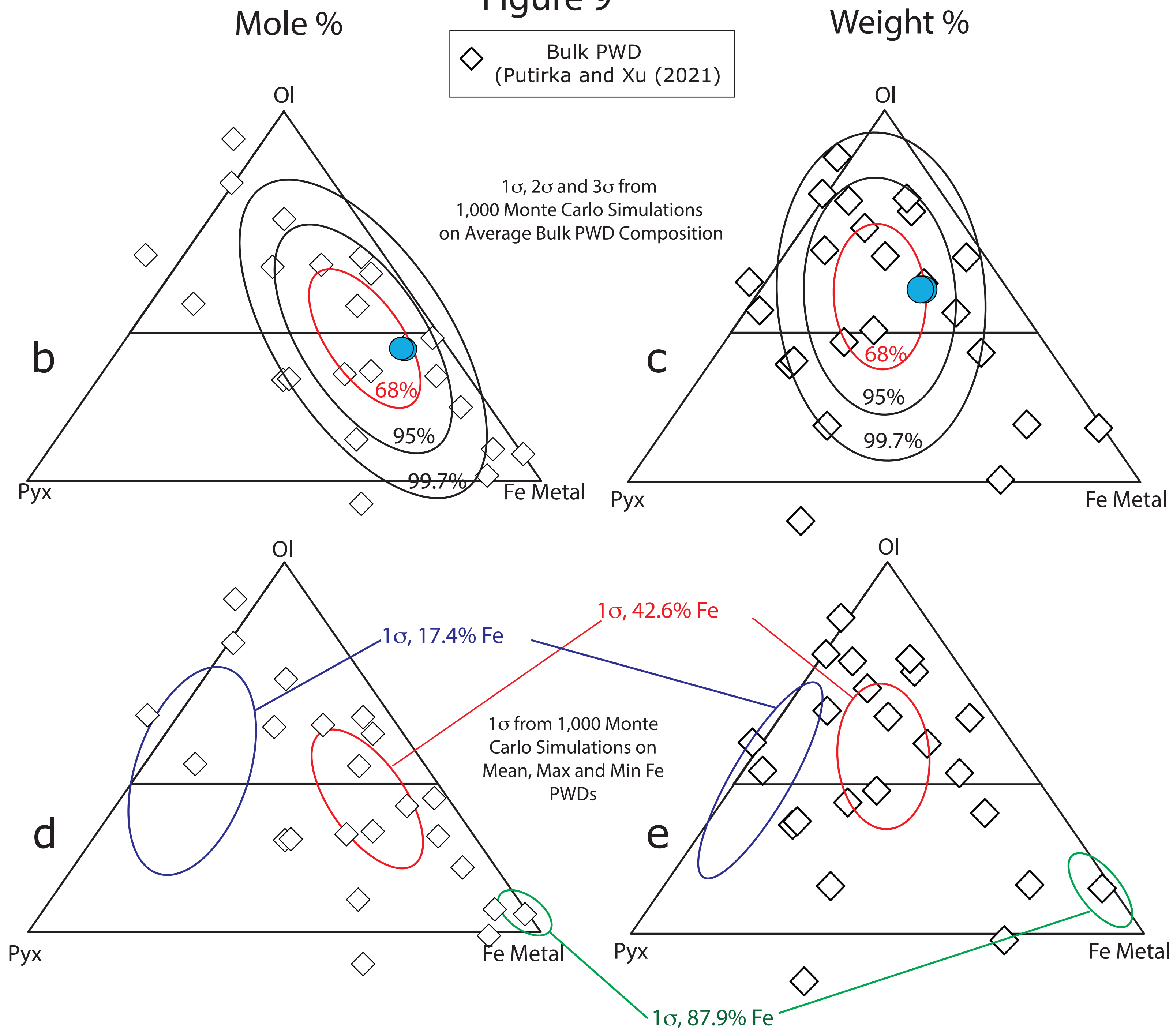

Mole %
Weight %
Bulk PWD
(Putirka and Xu (2021)
Ol
Pyx
Fe Metal
b
c
68%
95%
99.7%
1σ, 2σ and 3σ from
1,000 Monte Carlo Simulations
on Average Bulk PWD Composition
d
e
1σ, 17.4% Fe
1σ, 42.6% Fe
1σ, 87.9% Fe
1σ from 1,000 Monte
Carlo Simulations on
Mean, Max and Min Fe
PWDs

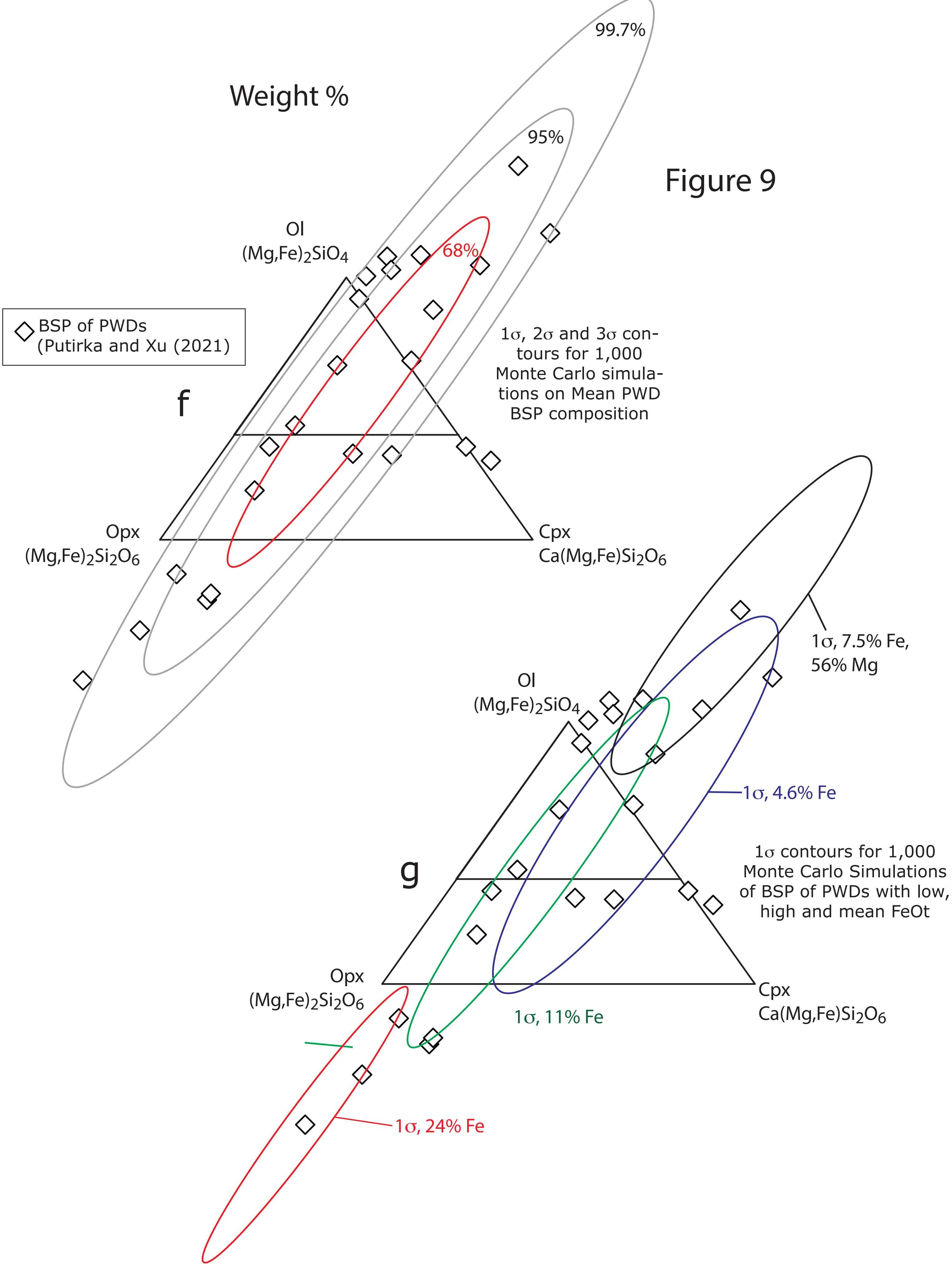
Weight %
Figure 9
99.7%
95%
68%
Ol
$(Mg,Fe)_2SiO_4$
BSP of PWDs
(Putirka and Xu (2021)
1σ, 2σ and 3σ contours for 1,000 Monte Carlo simulations on Mean PWD BSP composition
f
Opx
$(Mg,Fe)_2Si_2O_6$
Cpx
$Ca(Mg,Fe)Si_2O_6$
1σ, 7.5% Fe, 56% Mg
Ol
$(Mg,Fe)_2SiO_4$
1σ, 4.6% Fe
1σ contours for 1,000 Monte Carlo Simulations of BSP of PWDs with low, high and mean FeOt
g
Opx
$(Mg,Fe)_2Si_2O_6$
Cpx
$Ca(Mg,Fe)Si_2O_6$
1σ, 11% Fe
1σ, 24% Fe

Figure 9

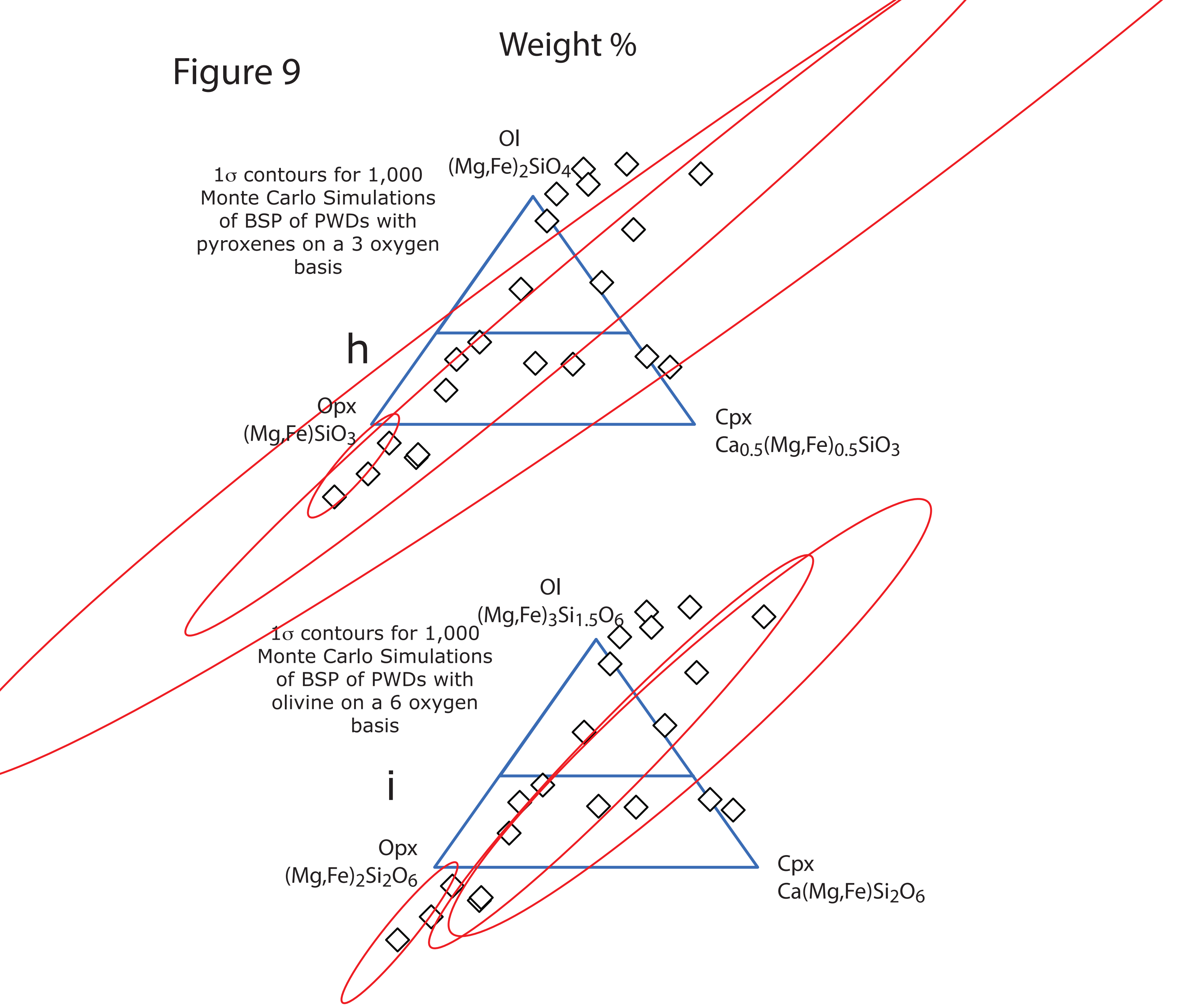
Weight %
h
Ol
(Mg,Fe)$_2$SiO$_4$
1σ contours for 1,000 Monte Carlo Simulations of BSP of PWDs with pyroxenes on a 3 oxygen basis
Opx
(Mg,Fe)SiO$_3$
Cpx
Ca$_{0.5}$(Mg,Fe)$_{0.5}$SiO$_3$
i
Ol
(Mg,Fe)$_3$Si$_{1.5}$O$_6$
1σ contours for 1,000 Monte Carlo Simulations of BSP of PWDs with olivine on a 6 oxygen basis
Opx
(Mg,Fe)$_2$Si$_2$O$_6$
Cpx
Ca(Mg,Fe)Si$_2$O$_6$

Figure 10

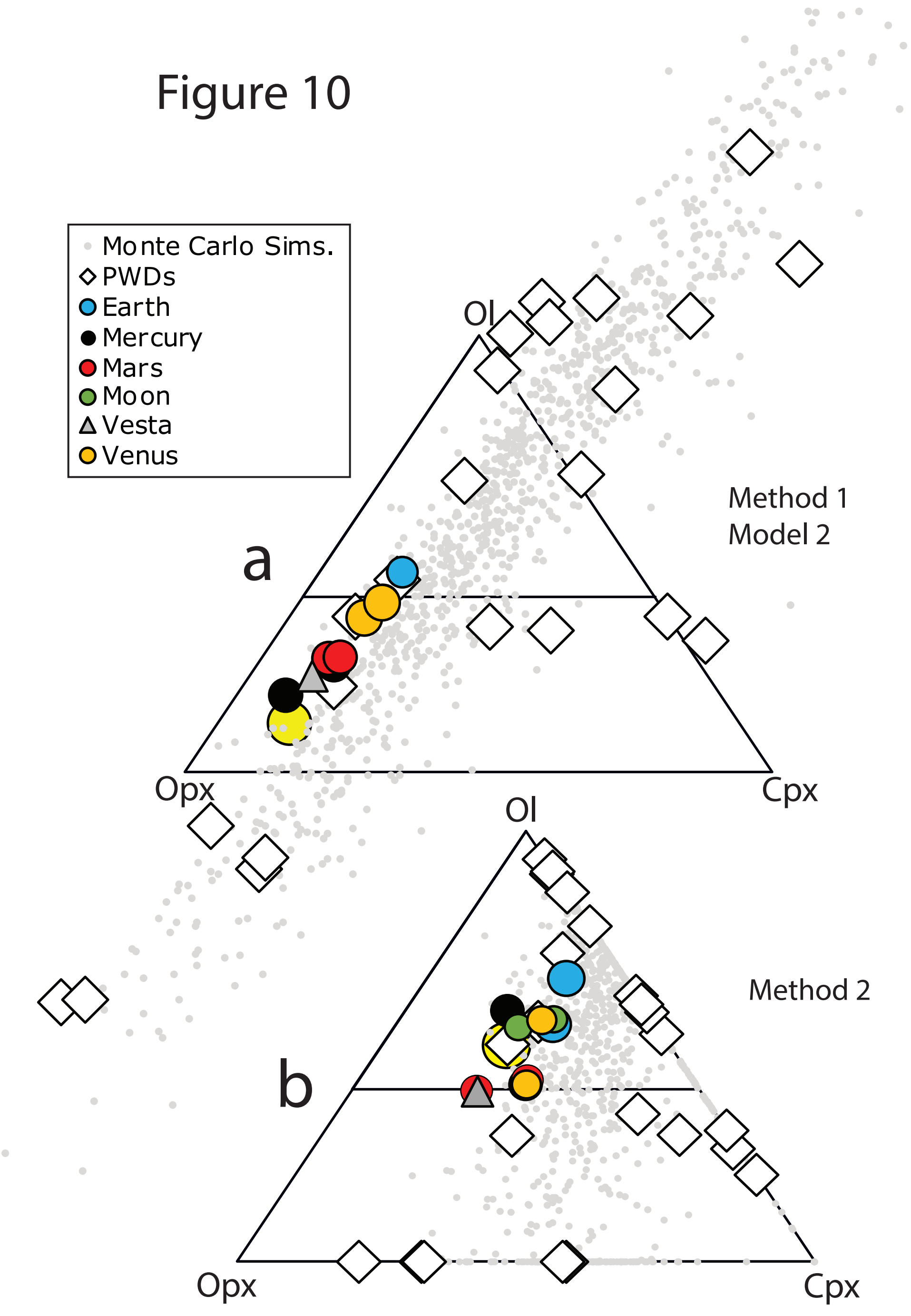

# Figure 10

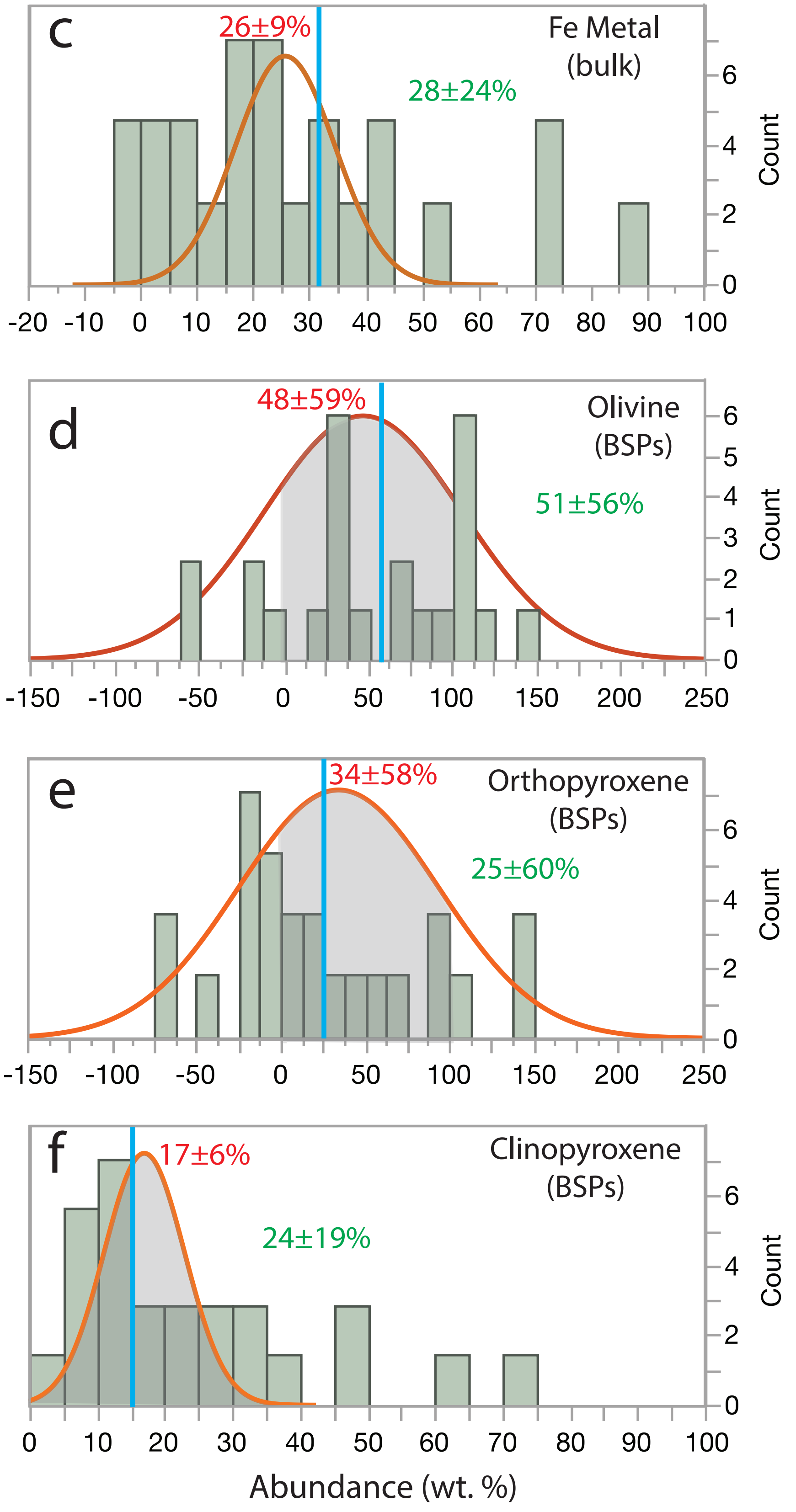

c
26±9%
Fe Metal
(bulk)
28±24%
Count
-20 -10 0 10 20 30 40 50 60 70 80 90 100
d
48±59%
Olivine
(BSPs)
51±56%
Count
-150 -100 -50 0 50 100 150 200 250
e
34±58%
Orthopyroxene
(BSPs)
25±60%
Count
-150 -100 -50 0 50 100 150 200 250
f
17±6%
Clinopyroxene
(BSPs)
24±19%
Count
0 10 20 30 40 50 60 70 80 90 100
Abundance (wt. %)

Figure 11

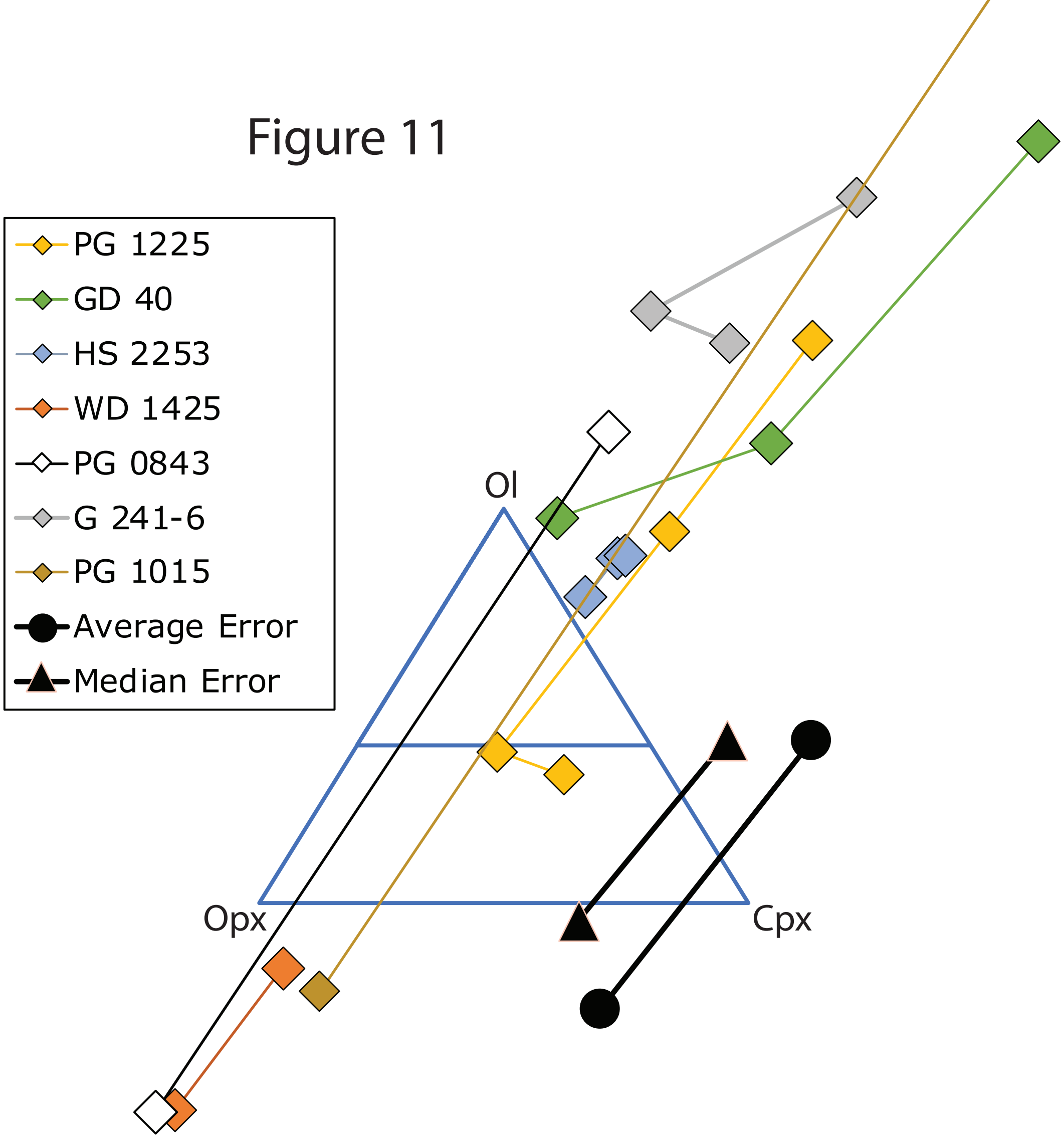

Figure 12 **Olivine**

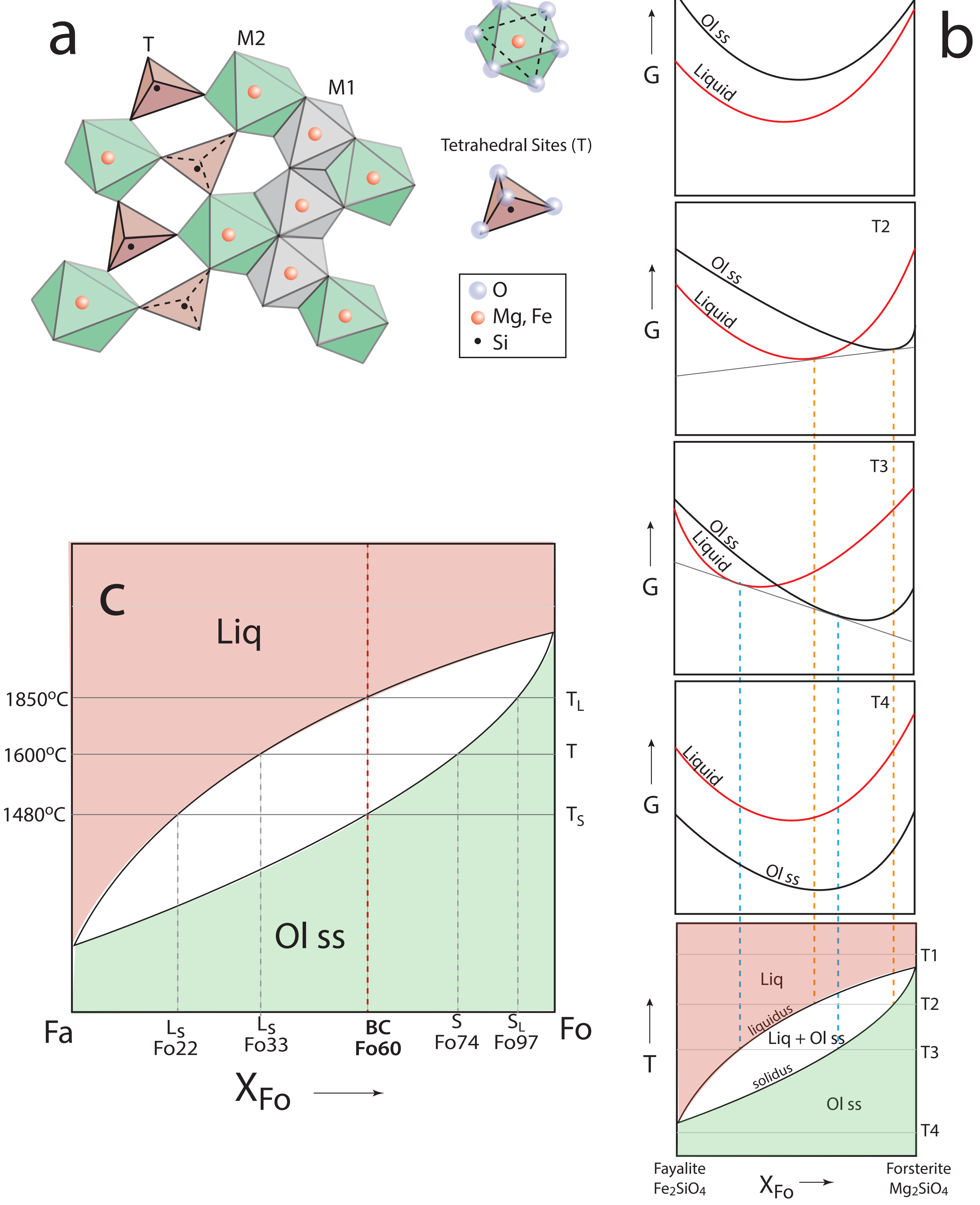

# Figure 13

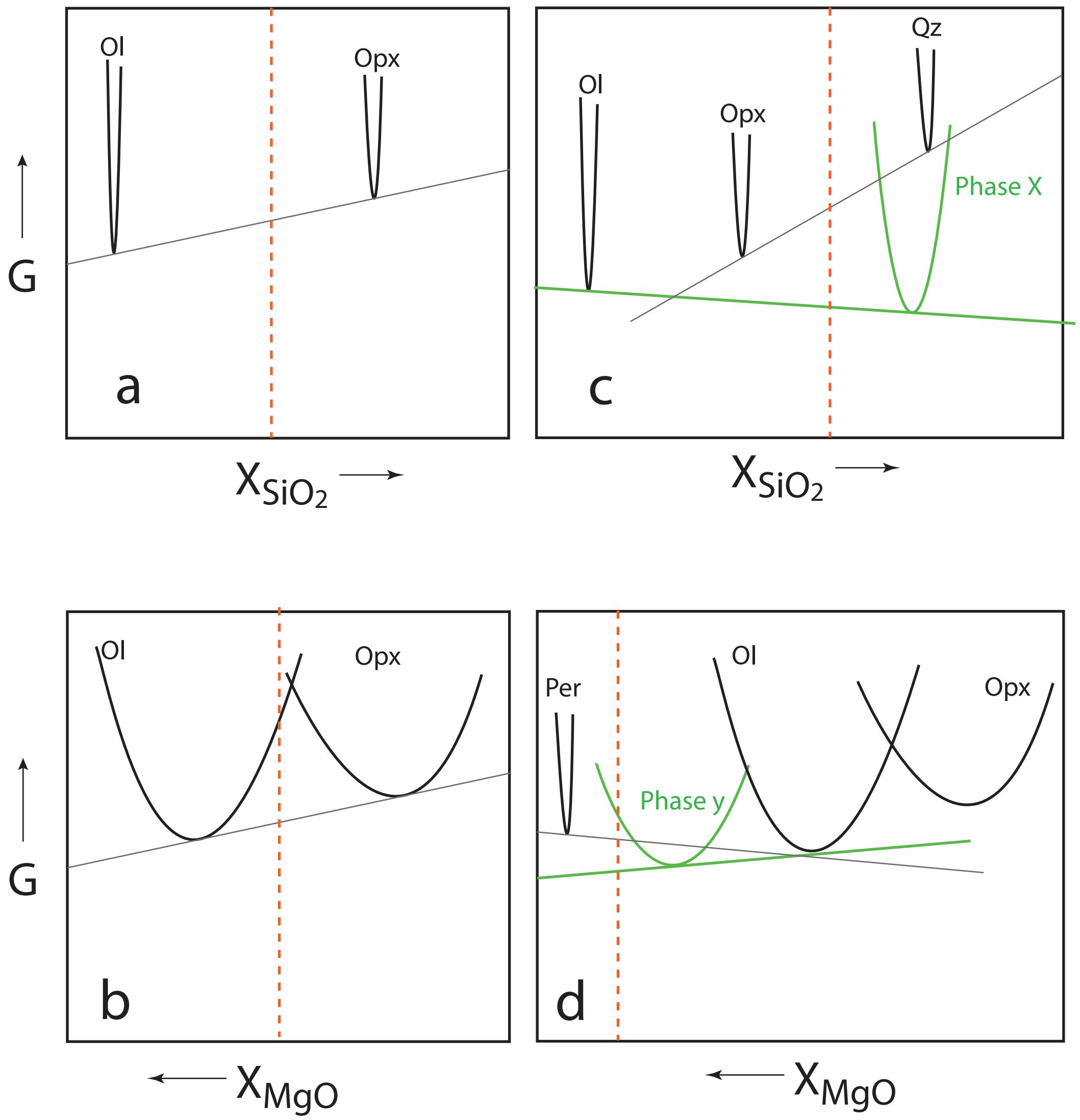

Figure 14

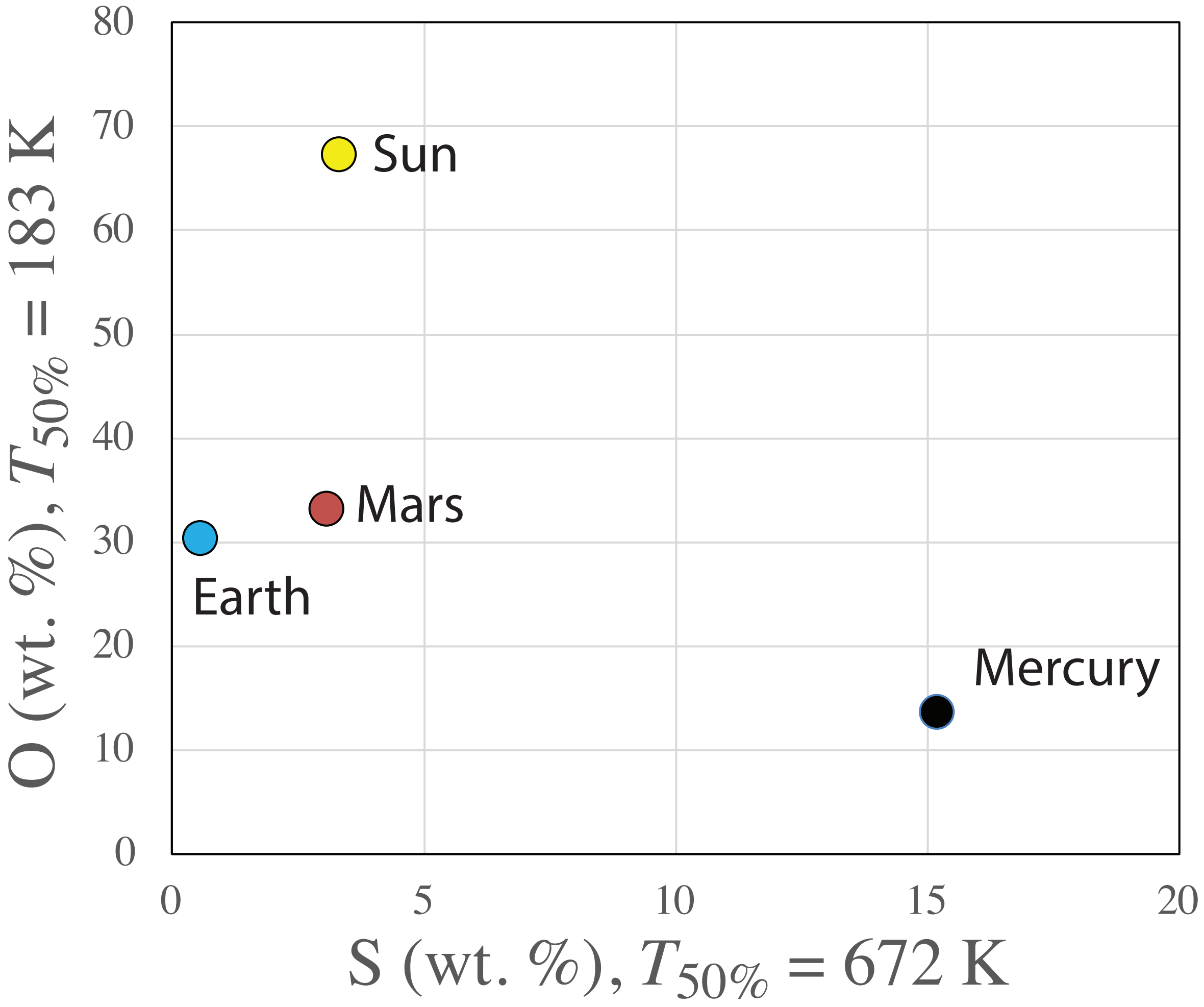